\documentclass[aps,nofootinbib,showkeys,noshowpacs,preprintnumbers,amsmath,amssymb]{revtex4}
\pdfoutput=1

\def\be{\begin{equation}}
\def\ee{\end{equation}}
\def\ba{\begin{eqnarray}}
\def\ea{\end{eqnarray}}
\def\bea{\begin{eqnarray}}
\def\eea{\end{eqnarray}}
\def\bes{\begin{subequations}}
\def\ees{\end{subequations}}
\newcommand{\A}{{\mathcal{A}}}
\newcommand{\tA}{{\widetilde {\mathcal{A}}}}
\newcommand{\ta}{{\widetilde a}}
\newcommand{\tal}{{\widetilde \alpha}}

\newcommand{\td}{{\widetilde d}}

\newcommand{\MSbar}{\overline{\rm MS}}  
\newcommand{\tk}{{\widetilde k}}
\usepackage{graphics}
\usepackage{graphicx}
\usepackage{dcolumn}
\usepackage{bm}
\usepackage{epsfig}
\usepackage{graphicx,color}
\usepackage{multirow}

\begin{document}

\preprint{USM-TH-351}

\title{Nearly perturbative lattice-motivated QCD coupling with zero IR limit}

\author{C\'esar Ayala$^1$}
\author{Gorazd Cveti\v{c}$^1$}
\author{Reinhart K\"ogerler$^2$}
\author{Igor Kondrashuk$^3$}

\affiliation{$^1$Department of Physics, Universidad T{\'e}cnica Federico Santa Mar{\'\i}a, Casilla 110-V, Valpara{\'\i}so, Chile\\ 
$^2$Department of Physics, Universit\"at Bielefeld, 33501 Bielefeld, Germany\\ 
  $^3$ Grupo de Matem\'atica Aplicada {\rm \&} Grupo de F\'isica de Altas Energ\'ias, Departamento de Ciencias B\'asicas, Universidad del B\'io-B\'io,  Campus Fernando May, 
  Casilla 447, Chill\'an, Chile}

\date{\today}

\begin{abstract}
The product of the gluon dressing function and the square of the ghost dressing function in the Landau gauge can be regarded to represent, apart from the inverse power corrections $1/Q^{2n}$, a nonperturbative generalization $\A(Q^2)$ of the perturbative QCD running coupling $a(Q^2)$ ($\equiv \alpha_s(Q^2)/\pi$). Recent large volume lattice calculations for these dressing functions \textcolor{black}{indicate that the coupling defined in such a way} goes to zero as $\A(Q^2) \sim Q^2$ when the squared momenta $Q^2$ go to zero ($Q^2 \ll 1 \ {\rm GeV}^2$). In this work we construct such a QCD coupling $\A(Q^2)$ which fulfills also various other physically motivated conditions. At high momenta it becomes the underlying perturbative coupling $a(Q^2)$ to a very high precision. And at intermediate low squared momenta $Q^2 \sim 1 \ {\rm GeV}^2$ it gives results consistent with the data of the semihadronic $\tau$ lepton decays as measured by OPAL and ALEPH. The coupling is constructed in a dispersive way, \textcolor{black}{resulting as a byproduct in} the holomorphic behavior of $\A(Q^2)$ in the complex $Q^2$-plane which reflects the holomorphic behavior of the spacelike QCD observables. Application of the Borel sum rules to $\tau$-decay $V+A$ spectral functions allows us to obtain values for the gluon (dimension-4) condensate and the dimension-6 condensate, which reproduce the measured OPAL and ALEPH data to a significantly better precision than the perturbative $\MSbar$ coupling approach. 
\end{abstract}

\maketitle

\section{Introduction}
\label{sec:intr}
In QCD, the extension of the perturbative QCD (pQCD) to a low energy regime of squared momenta $|Q^2| \lesssim 1 \ {\rm GeV}^2$ remains an open question.\footnote{
  We use here the usual convention, $Q^2 \equiv -q^2 = -(q^0)^2 + {\vec q}^2$ where $q$ is the 4-momentum of a physical process characterizing the running coupling.} This problem is related with a host of other unsolved problems. The present world average value for the pQCD coupling $a(Q^2) \equiv \alpha_s(Q^2)/\pi$, namely  $a(M_Z^2,\MSbar) = (0.1185 \pm 0.0006)/\pi$ \cite{PDG2014} or  $a(M_Z^2,\MSbar) = (0.1181 \pm 0.0011)/\pi$ \cite{PDG2016}, appears to be, to a certain degree, in tension with the well measured low-energy physics of $\tau$ lepton semihadronic decays which would require a higher value of the coupling in most cases of analyses within the contour improved perturbation theory (CIPT) \cite{Pivovarov:1991rh,LeDiberder:1992te}. In such evaluations, $\MSbar$ pQCD is applied together with the Operator Product Expansion approach (OPE) in the form of QCD sum rules.
Another \textcolor{black}{inconvenience of the pQCD coupling $a(Q^2)$ in widely used renormalization schemes, such as $\MSbar$ and similar schemes, is that $a(Q^2)$} does not reflect the analyticity properties of spacelike observables ${\cal D}(Q^2)$ in the complex $Q^2$-plane, required by the general principles of Quantum Field Theory \cite{BS,Oehme}. Namely, spacelike physical quantities, such as current correlators and \textcolor{black}{differential cross sections for deep inelastic lepton-hadron scattering (DIS)}, are required to be holomorphic (analytic) functions in the complex $Q^2$-plane, with the exception of a part of the negative semiaxis, i.e., for $Q^2 \in \mathbb{C} \backslash (-\infty, -M_{\rm thr}^2]$, where $M_{\rm thr} \sim 0.1$ GeV is a threshold scale of the order of the light meson mass. For example, $\MSbar$ pQCD coupling $a(Q^2;\MSbar)$ has a cut stretching over the entire negative semiaxis plus a part of the positive axis, $(-\infty,+\Lambda_{\rm Lan.}^2)$ where $\Lambda_{\rm Lan.}^2 \sim 0.1 \ {\rm GeV}^2$ is the Landau pole (or branching point). This nonholomorphic behavior at low positive $Q^2$, $0 < Q^2 \lesssim 0.1\ {\rm GeV}^2$  \textcolor{black}{is a drawback} because it means that the infrared (IR) regime $|Q^2| < 1 \ {\rm GeV}^2$ cannot be described reliably  \textcolor{black}{by truncated perturbation series in} pQCD in such schemes. \textcolor{black}{However, we wish to stress that the running coupling itself  is, in general, not an observable.\footnote{\textcolor{black}{However, it could be defined to be equal to a specific physical observable [effective charge (ECH)] in the pQCD sense \cite{Grunberg} and in a more general nonperturbative sense \cite{DBCK} (cf.~also Refs.~\cite{MSS1,MSS2,MagrGl,mes2,DeRafael,MagrTau,Nest3a,Nest3b,NestBook} for general dispersive ECH approaches).}} Therefore, a good holomorphic behavior of $a(Q^2)$ is not a necessary requirement for the correct holomorphic nature of observables. But as long as one believes that a perturbation expansion makes sense at all (at least for part of the observable), one would expect a connection between the analyticity of $a(Q^2)$ and of the spacelike physical observables ${\cal D}(Q^2)$. Such a connection can be conveniently realized by replacing the pQCD coupling $a(Q^2)$ by a coupling $\A(Q^2)$ which reflects qualitatively the holomorphic behavior of spacelike physical observables.}

  Within the present paper we aim at finding an effective running QCD coupling $\A(Q^2)$ which, \textcolor{black}{although not being an effective charge}, is an extension of the perturbative QCD coupling $a(Q^2)$ to the infrared (IR) regime  $|Q^2| \lesssim 1 \ {\rm GeV}^2$, accounting there for the recent lattice results and simultaneously avoiding the problems with unphysical nonholomorphic behavior. A reasonable definition of such a coupling is $\A(Q^2) = \A_{\rm latt.}(Q^2) - \Delta \A_{\rm NP}(Q^2)$, where $\A_{\rm latt.}(Q^2)$ denotes the product of the gluon dressing function and the square of the ghost dressing function as obtained in lattice calculations in the Landau gauge and in the lattice MiniMOM scheme \cite{MiniMOM} (cf.~also \cite{BoucaudMM,CheRet}), and $-\Delta \A_{\rm NP}(Q^2)$ represents subtraction of the main part of the nonperturbative (NP) contributions manifesting themselves at $|Q^2| > \Lambda_{\rm QCD}^2$ as  $\sim (\Lambda_{\rm QCD}^2/Q^2)^n$ ($n \geq 1$) \textcolor{black}{and not diverging at $|Q^2| < \Lambda_{\rm QCD}^2$.}  Recent lattice results for the dressing functions \textcolor{black}{indicate that the coupling $\A_{\rm latt.}(Q^2)$ defined in the mentioned way} goes to zero, as $\sim Q^2$, in the deep IR regime $0< Q^2 < 0.1 \ {\rm GeV}^2$. Assuming that no finetuning occurs when the mentioned nonperturbative contributions are subtracted from the product of dressing functions, the resulting extended coupling $\A(Q^2)$, in the mentioned MOM scheme, is also going to zero as  $\A(Q^2) \sim Q^2$ in the deep IR regime. Such a coupling is then respecting qualitatively the lattice results for the low-momentum dressing functions in the deep IR, and in the ultraviolet (UV) regime ($|Q^2| \gg \Lambda_{\rm QCD}^2$) it tends fast to the pQCD coupling $a(Q^2)$ in the same renormalization scheme, reproducing the high-energy QCD phenomenology correctly. Such $\A(Q^2)$ is constructed here within a dispersive approach, \textcolor{black}{and it turns out to be} a holomorphic coupling reflecting the analytic behavior of spacelike observables. This property makes the application of OPE and QCD sum rules a consistent approach, because the coupling $\A(Q^2)$ and the analogs of the truncated perturbation series [in terms of $\A(Q^2)$] for spacelike observables reflect correctly the holomorphic behavior of those observables, in contrast to the case of QCD sum rules in $\MSbar$ pQCD. Further, this holomorphic property turns out to solve the numerical problem of the renormalon-related asymptotic divergence of the truncated perturbation series of physical quantities in pQCD \cite{BGA,Techn}. 
The coupling $\A(Q^2)$ also has to reproduce correctly the well-measured results of the $\tau$-lepton semihadronic decays, which represents the physics of the moderate IR regime.

In this work we construct such a coupling $\A(Q^2)$. It turns out that it is generically nonperturbative, i.e., it differs from its underlying pQCD coupling by nonperturbative terms\footnote{
  It is possible to show that pQCD renormalization schemes exist in which pQCD coupling $a(Q^2)$ is holomorphic for $Q^2 \in \mathbb{C} \backslash (-\infty, -M_{\rm thr}^2]$ and at the same time reproduces the high-energy QCD phenomenology as well as the semihadronic $\tau$-lepton decay physics \cite{anpQCD1a,anpQCD1b,anpQCD2}.}
such as powers of $1/(Q^2+M^2)$ where $M^2 \lesssim 1 \ {\rm GeV}^2$.  
In our previous, shorter, work \cite{AQCDprev} we presented a construction of such a coupling in a renormalization scheme which agrees up to three-loops with the lattice MiniMOM scheme, and we applied the QCD Borel sum rules  to the semihadronic $\tau$-decay data by fitting the theoretical results to the experimental results of the OPAL Collaboration. Here we extend the work \cite{AQCDprev}, by working in a renormalization scheme which agrees with the lattice MiniMOM scheme up to four-loops,\footnote{MiniMOM scheme is known at present to four loops \cite{MiniMOM,BoucaudMM,CheRet}.} we include the experimental data of ALEPH Collaboration, and we investigate in the Borel sum rule analysis of the OPAL Collaboration data also the quality of the results when the upper bound of integration $\sigma_{\rm max}$ is gradually reduced. We stress that for defining our coupling $\A(Q^2)$ we use exclusively the lattice MiniMOM (MM) scheme, in which the results of the lattice calculations for the dressing functions were obtained.\footnote{
  In this scheme, however, we rescale $Q^2$ from the $\Lambda_{\rm MM}$ to the usual $\Lambda_{\MSbar}$ convention.} The reason for this is that we do not know \textcolor{black}{in advance how} the dressing functions, and our coupling $\A(Q^2)$, change in the deep IR regime when the renormalization scheme is changed. The changes of a running coupling induced by the changes of the renormalization scheme in the UV (perturbative) regime are known \cite{Stevenson}; \textcolor{black}{these changes then affect the values of the NP parameters of the coupling $\A(Q^2)$ such as the effective gluon mass \cite{Cornwall} or the values (normalization) of $\A(Q^2)$ at low $Q^2$ \cite{Brod2}, for example via a matching procedure.}\footnote{\textcolor{black}{\label{ftnBrod2} In Ref.~\cite{Brod2}, the matching of $\A(Q^2)$ and $d \A(Q^2)/d \ln Q^2$ at an IR/UV transition scale $Q_0^2 \sim 1 \ {\rm GeV}^2$ is imposed, fixing the values of $\A(0) > 0$ and $Q_0^2$. On the other hand, our coupling $\A(Q^2)$ will be holomorphic, no explicit IR/UV matching scale will exist. Instead of the matching, we will impose various physically motivated conditions which will affect simultaneously the behavior of $\A(Q^2)$ in the UV and IR regimes.}}

In Sec.~\ref{sec:latt} we present lattice results for the mentioned product of dressing functions, and give arguments why the running coupling $\A(Q^2)$ should have qualitatively the same behavior, even in the deep IR regime. In Sec.~\ref{sec:pQCD} we describe the construction of the underlying pQCD coupling, and of its discontinuity function along the cut, in the renormalization scheme which agrees up to four loops with the mentioned MOM scheme. In Sec.~\ref{sec:constr} we then construct, by dispersive approach, the coupling $\A(Q^2)$ which fulfills a host of restrictions: it has the mentioned behavior in the deep IR regime; it merges fast with the underlying pQCD coupling $a(Q^2)$ in the UV regime; and it reproduces the correct decay ratio $r_{\tau} \approx 0.20$ for the $\tau$-lepton semihadronic strangeless decays. We require that the high-energy QCD value of $\A(M_Z^2) = a(M_Z^2)$ (in MiniMOM) corresponds to the modern central world average values of the $\MSbar$ pQCD coupling $a(M_Z^2,\MSbar) = 0.1185/\pi$ \cite{PDG2014} or $0.1181/\pi$ \cite{PDG2016}. In Appendix \ref{app:tcp} we summarize the transformation of perturbation coefficients under the change of scheme, and in Appendix \ref{app:An} we summarize the construction of $\A_n$ which are the analogs of the higher powers $a^n$ in the $\A$-coupling framework. In Sec.~\ref{sec:BSR} we apply Borel sum rules to the obtained coupling for the $\tau$ spectral functions,  and  thus extract the values of the gluon (dimension $D=4$) condensate and $D=6$ condensate by fitting the theoretical results to the experimental results of the OPAL and ALEPH Collaborations. We also compare the extracted results with those obtained \textcolor{black}{when $\MSbar$ pQCD coupling is} used. In Appendix \ref{app:fit} we explain the fitting procedures used. In Sec.~\ref{sec:pred} we present some predictions of the $\A$-coupling framework, and discuss several aspects of the framework. In Appendix \ref{app:mAQCD} we present the related ``mass-modified'' couplings. In Sec.~\ref{sec:diff} we compare the obtained coupling with the related coupling of Ref.~\cite{AQCDprev}, and compare the extracted values of the gluon condensate with those in other works in the literature. In Sec.~\ref{sec:summ} we summarize our results.

The programs for calculation of the coupling $\A(Q^2)$ and of its higher power analogs $\A_n(Q^2)$, in Mathematica \cite{Math}, are available freely \cite{prgs}.

\section{Lattice coupling}
\label{sec:latt}

In Ref.~\cite{LattcoupNf0}, lattice calculation of the Landau gauge gluon and ghost dressing functions, $Z_{\rm gl}(Q^2)$ and $Z_{\rm gh}(Q^2)$, were performed with large physical volume and high statistics, giving presumably reliable results in the IR regime $0<Q^2 < 1 \ {\rm GeV}^2$. On the other hand, in pQCD, a Slavnov-Taylor identity yields for  the running coupling $a(Q^2)$ the following combination of the gluon and ghost dressing functions $Z_{\rm gl}$ and $Z_{\rm gh}$ and the ghost-gluon vertex function ${\widetilde Z}_1(Q^2)$:
\be
a(Q^2)  =  a(\Lambda^2) \frac{Z_{\rm gl}^{(\Lambda)}(Q^2) Z_{\rm gh}^{(\Lambda)}(Q^2)^2}{{\widetilde Z}_1 ^{(\Lambda)}(Q^2)^2} 
\label{alatt}                   
\ee  
in any gauge. The scale $\Lambda$ in the superscript indicates that the theory has UV cutoff squared scale $\Lambda^2$ ($\gg Q^2$).
However, in the Landau gauge, the ghost-gluon vertex function ${\widetilde Z}_1(Q^2)$, with symmetric momentum points, is not affected by renormalization to any loop order \cite{Taylor,Slavnov:1972fg,Slavnov:1974dg,Faddeev:1980be} (cf.~also \cite{DSEscale,Kondrashuk:2000qb,Cvetic:2002in}). \textcolor{black}{It is equal to one in the Landau gauge in the $\MSbar$ scheme and in a MOM scheme called MiniMOM (MM) \cite{MiniMOM} (cf.~also \cite{BoucaudMM,CheRet}).}\footnote{\textcolor{black}{Various QCD couplings coexist in MOM schemes, depending on which QCD vertex is considered \cite{CelGon}; they converge to each other at high $|Q^2|$, but may differ significantly in the IR regime. The coupling (\ref{alatt}) corresponds to the coupling $(g^{\prime \prime}_{\rm mom}/2 \pi)^2$ of Ref.~\cite{CelGon} in the Landau gauge, but now with ${\widetilde Z}_1$ not renormalized by MOM but rather ${\widetilde Z}_1=$${\widetilde Z}_1( \MSbar )$ $= {\widetilde Z}_1({\rm MM})=1$.}}  
Therefore, it appears natural to extend the definition of the QCD running coupling to the following combination of the Landau gauge dressing functions:
\be
\A_{\rm latt.}(Q^2)  =  \A_{\rm latt.}(\Lambda^2)  \frac{Z_{\rm gl}^{(\Lambda)}(Q^2) Z_{\rm gh}^{(\Lambda)}(Q^2)^2}{{\widetilde Z}_1 ^{(\Lambda)}(Q^2)^2} \ ,
\label{Alatt}
\ee
\begin{figure}[htb] 
  \centering\includegraphics[width=100mm, height=6.5cm]{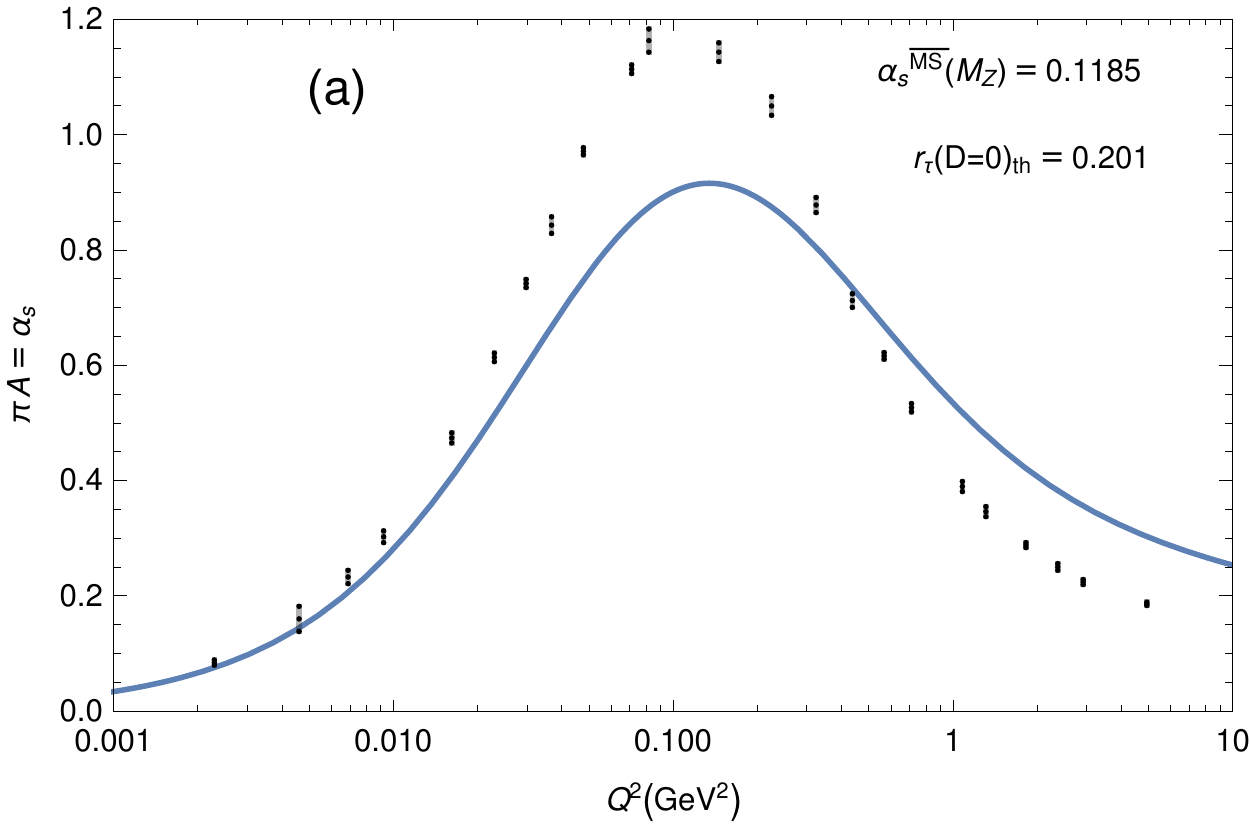}
  \vspace{-0.4cm}
 \caption{\footnotesize  The $N_f=0$ lattice values $\pi \A_{\rm latt.}(Q^2)$ at low $Q^2$, from Ref.~\cite{LattcoupNf0}: the points with calculational uncertainties included. The squared momenta are rescaled, from the MiniMOM (MM) lattice scheme scale to the usual $\MSbar$-like scale: $Q^2 = Q^2_{\rm latt.} ({\Lambda}_{\MSbar}/{\Lambda_{\rm MM}})^2 \approx Q^2_{\rm latt.}/1.9^2$ at $N_f=0$. The solid curve is the ($N_f=3$) theoretical coupling in the same IR regime (see later, Sec.~\ref{sec:constr}).}
\label{FigAlatt}
\end{figure}
where ${\widetilde Z}_1 ^{(\Lambda)}(Q^2)^2=1$ (Landau gauge and MiniMOM scheme) and it is assumed that $\Lambda$ is determined by the lattice spacing. The results for the combination (\ref{Alatt}) in the low-$Q^2$ regime, as obtained by the mentioned lattice $N_f=0$ calculations \cite{LattcoupNf0}, are presented as points in Fig.~\ref{FigAlatt}.
One striking fact is that this lattice coupling results go to zero as $\A_{\rm latt.}(Q^2) \sim Q^2$ when $Q^2 \to 0$. Very similar results were obtained later by another group \cite{LattcoupNf0b}, also for $N_f=0$, who used different physical volumes and lattice spacings.
\textcolor{black}{The lattice results for $N_f=2$ \cite{LattcoupNf2} and $N_f=2+1+1$ \cite{LattcoupNf4} are also available, they give for the values of the scale $Q^2_{\rm max}$ at which $\A_{\rm latt.}(Q^2)$ has maximum, $Q^2_{\rm max} \approx (0.15 \pm 0.05) \ {\rm GeV}^2$ [after rescaling from the MiniMOM scale convention ($\Lambda_{\rm MM}$) to $\Lambda_{\MSbar}$-scale convention, cf.~Eq.~(\ref{LambdaMMMSbar}) in Sec.~\ref{sec:pQCD}]. This is similar to the $N_f=0$ case where $Q^2_{\rm max} \approx 0.135 \ {\rm GeV}^2$ (cf.~footnote \ref{ftnQ2max} in Sec.~\ref{sec:constr}). Further, the mentioned $N_f=2$ and $N_f=2+1+1$ lattice results also indicate that $\A_{\rm latt.}(Q^2) \to 0$ when $Q^2 \to 0$. So the two basic features of $\A_{\rm latt.}$, which we are going to take into account in the construction of our ($N_f=3$) coupling $\A(Q^2)$, hold for $N_f=0, 2$ and $2+1+1$. Nonetheless, the statistics and the volume of the mentioned $N_f=2$ and $N_f=2+1+1$ lattice calculations \cite{LattcoupNf2,LattcoupNf4} are lower than those of the $N_f=0$ results \cite{LattcoupNf0,LattcoupNf0b}.  In the $N_f=2$ and $N_f=2+1+1$ cases \cite{LattcoupNf2,LattcoupNf4}, only few lattice results (points) for $\A_{\rm latt.}(Q^2)$ are available at $Q^2 < Q^2_{\rm max}$, and they do not reach very low scales $Q^2 < 0.01 \ {\rm GeV}^2$, in contrast to the $N_f=0$ case \cite{LattcoupNf0,LattcoupNf0b}. Therefore, in our comparisons of $\A_{\rm latt.}$ and $\A$, cf.~Fig.~\ref{FigAlatt} here and Fig.~\ref{FigComb} in Sec.~\ref{sec:constr}, we use the $N_f=0$ lattice results of Ref.~\cite{LattcoupNf0}. The actual values of the Landau gauge gluon and ghost propagators [and thus of $\A_{\rm latt.}(Q^2)$ of Eq.~(\ref{Alatt})], though, do have nonnegligible dependence on $N_f$ \cite{Nfprops1} (cf.~also \cite{Nfprops2} for a related study).}

We should take into account that the lattice coupling (\ref{Alatt}), at $|Q^2| > \Lambda_{\rm QCD}^2$, differs from the perturbative coupling by higher-dimension (higher-twist) terms of the form $\sim {\Lambda_{\rm QCD}^2}/(Q^2)^n$ ($n=1,\ldots$; for $|Q^2| \gg \Lambda_{\rm QCD}^2$) \cite{LattcoupNf4}, which are of nonperturbative (NP) origin.\footnote{
NP terms are formally those which are functions $F(a)$ of the pQCD coupling $a(Q^2)$ such that $F(a)$ is a function nonanalytic in $a=0$. Since a function $F(a) \sim \exp(-B/a)$ (with $B$ a constant) is such a function, and the power terms ${\Lambda_{\rm QCD}^2}/(Q^2)^n$ can be represented as $\exp(-B/a(Q^2))$ at large $|Q^2| \gg {\Lambda}^2_{\rm QCD}$ [we note that there $a(Q^2) \sim 1/\ln(Q^2/ {\Lambda}^2_{\rm QCD})$ is small], the contributions $\sim {\Lambda_{\rm QCD}^2}/(Q^2)^n$  ($n=1,\ldots$; for $|Q^2| \gg \Lambda_{\rm QCD}^2$) fall clearly into the category of such NP contributions.}
To be in accordance with this fact, we take the following position. The major part of the NP contributions can be separated in $\A_{\rm latt.}(Q^2)$ for all $Q^2$ (not just for $|Q^2| \gg \Lambda_{\rm QCD}^2$) as an additive term
\be
\A_{\rm latt.}(Q^2) =  \A(Q^2) + \Delta \A_{\rm NP}(Q^2) \ .
\label{AlattA}
\ee
Here, $\Delta \A_{\rm NP}(Q^2)$ is the general NP contribution, which will be considered to be a correction to the basic $\A(Q^2)$, the latter being nearly perturbative at large $|Q^2|$, in the sense that
\be
\A(Q^2) - a(Q^2)   \sim \left( \frac{\Lambda^2_{\rm QCD}}{Q^2} \right)^{N_{\rm max}} \qquad (|Q^2| > \Lambda^2_{\rm QCD}) \ ,
\label{Aadiff1}
\ee
where $a(Q^2)$ is the underlying pQCD coupling, i.e., the pQCD coupling in the same renormalization scheme (MiniMOM) as $\A(Q^2)$, and $N_{\rm max}$ is a relatively large integer (in our case it will be $N_{\rm max} = 5$). The correction $\Delta \A_{\rm NP}(Q^2)$ will be evidently very small in the UV regime, \textcolor{black}{and this is supported by experimental evidence as pQCD has been shown to describe well QCD phenomena at high $|Q^2| \gg \Lambda^2_{\rm QCD}$.}\footnote{\textcolor{black}{Later we will see that the MiniMOM scheme, rescaled to $\Lambda_{\MSbar}$-scale convention, is in the perturbative regime not very far from the $\MSbar$ renormalization scheme widely used in the high-$|Q^2|$ QCD, cf.~Eqs.~(\ref{c2c3}).}}  Crucially, we will assume that the  additive correction $\Delta \A_{\rm NP}(Q^2)$ cannot lead to finetuning in deep IR regime $|Q^2| \lesssim 0.1 {\rm GeV}^2$. In view of the lattice result $\A_{\rm latt.}(Q^2) \sim Q^2$ when $Q^2 \to 0$, the mentioned assumption of no finetuning implies that in the deep IR regime we cannot have
$|\Delta \A_{\rm NP}(Q^2)| > |Q^2|$ when $Q^2 \to 0$. Stated otherwise, we will have simultaneously
\be
\Delta \A_{\rm NP}(Q^2) \sim Q^2 \qquad {\rm and} \quad \A(Q^2) \sim Q^2
\qquad (Q^2 \to 0).
\label{noft}
\ee
The important consequence of the no-finetuning assumption is thus that the nearly perturbative $\A(Q^2)$ coupling, which we intend to construct here, has the behavior $\A(Q^2) \sim Q^2$ when $Q^2 \to 0$, qualitatively the same as $\A_{\rm latt.}(Q^2)$.

On the other hand, the authors of Refs.~\cite{DSEdecoupFreez,PTBMF} defined their (nearly perturbative) coupling $\A(Q^2)$ in a way different from Eqs.~(\ref{Alatt})-(\ref{noft}). Their definition involves a dynamical gluon mass $M(Q^2)$, and is multiplicative in nature instead of additive [cf.~Eq.~(\ref{AlattA})]
\be
\A_{\rm latt.}(Q^2)  = \A(Q^2) \frac{Q^2}{( Q^2 + M(Q^2)^2 ) } \ ,
\label{AFreez}
\ee
where $\A_{\rm latt.}(Q^2)$ is the product of the Landau gauge dressing functions, Eq.~(\ref{Alatt}). \textcolor{black}{These works employ a DSE approach, using a combination of the pinch technique (PT) and the background field method (BFM). A comparatively large positive value $\A(0) \approx 1$ is obtained, and the coupling $\A(Q^2)$ is close to the Bjorken polarized sum rule effective charge \cite{PTBMF}.}
  
Due to the multiplicative nature of the relation (\ref{AFreez}), the additive finetuning is not evident in this form. Further, since $M(0) > 0$, this definition implies a freezing of $\A(Q^2)$ at a finite positive value in the deep IR regime, $\A(0) = a_0 > 0$, \textcolor{black}{this holding even when the decoupling solutions of the DSE approach were used for the dressing functions.} There is some ambiguity in the definition of the running dynamical mass of the gluon. The index $N_{\rm max}$ in the relation (\ref{Aadiff1}) depends in the case of such coupling on the behavior of $M(Q^2)$ in the UV regime.  We will not follow this line of reasoning here, but will adopt the reasoning leading to Eqs.~(\ref{AlattA})-(\ref{noft}).

When using the product of the dressing functions Eq.~(\ref{Alatt}) and taking $\A_{\rm latt.}(Q^2) \sim \A(Q^2)$, then the behavior $\A(Q^2) \sim Q^2$ as $Q^2 \to 0$ is obtained with the decoupling solutions \cite{DSEdecoup} for the gluon and ghost dressing functions in the Dyson-Schwinger equations (DSE) approach, with the modified Gribov-Zwanziger approach \cite{Gribovdecoup}, and with some solutions of the functional renormalization group (FRG) approach \cite{FRG}. On the other hand, when again using the expression (\ref{Alatt}) and $\A_{\rm latt.}(Q^2) \sim \A(Q^2)$, then the behavior of $\A(Q^2)$ with $\A(0)>0$ is obtained with the scaling solution \cite{DSEscale} of the DSE approach, with the Gribov-Zwanziger approach \cite{Gribovscale}, and with some solutions of the FRG approach \cite{FRG}. However, as seen earlier, this appears not to be consistent with the lattice results \cite{LattcoupNf0,LattcoupNf0b,LattcoupNf4} $\A(0)=0$. 

\textcolor{black}{Further, in Ref.~\cite{DSE4glMM} the DSE approach for the four-gluon vertex in the Landau gauge and the MiniMOM scheme (${\widetilde Z}_1=1$) was applied, and it gave for the decoupling solution the value $\A(0)=0$, but for the scaling solution the obtained value of $\A(0)$ was a small positive number; qualitatively the same conclusion was reached from the three-gluon vertex. There could be uncontrolled approximations in lattice calculations; application of stochastic quantization approach to DSE may indicate the possible existence of an appropriate boundary condition which would restrict the lattice configurations in the Landau gauge so as to give possible scaling solutions to the gluon and ghost dressing functions \cite{Llanes} and thus a positive value of $\A(0)$. }  

We recall that in our coupling $\A(Q^2)$, Eqs.~(\ref{AlattA})-(\ref{noft}), we will take as input from the lattice coupling (\ref{Alatt}) only the two main features of the latter: $\A_{\rm latt}(Q^2) \sim Q^2 \to 0$ when $Q^2 \to 0$, and that the coupling has its maximum at $Q^2 \sim 0.1 \ {\rm GeV}^2$. One may ask whether these main qualitative features survive when the lattice coupling is calculated in other gauges, not just the Landau gauge. Large volume lattice calculations of the coupling Eq.~(\ref{Alatt}) in other gauges have not been performed yet. Nonetheless, application of DSE-like methods in the Coulomb gauge shows, Ref.~\cite{RW} (see also \cite{RQ}), that the decoupling solution exists also in this gauge, namely the solution which in the limit $Q^2 \to 0$ gives: $Z_{\rm gl}^{(\Lambda)} \sim Q^2 \to 0$, $Z_{\rm gh}^{(\Lambda)} \sim {\rm const}$ and ${\widetilde Z}_1^{(\Lambda)} = {\rm const}$. This indicates that the lattice coupling (\ref{Alatt}) in the Coulomb gauge probably also has the same mentioned qualitative features as in the Landau gauge. Furthermore, the author of \cite{RQ} argues that in general we should expect that the decoupled solutions of the DSE and DSE-like methods get realized in true QCD, and that the scaling solutions are exceptional cases at certain critical values of parameters. In this work, we will assume that the mentioned two main features of the lattice coupling are valid in any gauge.

\textcolor{black}{We wish to stress that the condition (\ref{Aadiff1}) only means that the NP contributions in the coupling $\A(Q^2)$ are very suppressed in the UV regime $|Q^2| > 1 \ {\rm GeV}^2$, but are expected to be significant in the regime $|Q^2| \lesssim 1 \ {\rm GeV}^2$. Further, we point out that the applications of our coupling $\A(Q^2)$ will be made only in the intermediate and high-$Q^2$ regime ($|Q^2| \gtrsim 1 \ {\rm GeV}^2$), and will depend only indirectly on the vanishing of $\A(Q^2)$ in the deep IR regime (via holomorphic behavior, etc.). We do not claim the physical reality of the vanishing of the coupling, $\A(Q^2) \sim Q^2$ as $Q^2 \to 0$, and the meaning of this behavior is at the moment not clear, as can be seen from parts of the discussion of this Section.} 

\section{The underlying pQCD coupling}
\label{sec:pQCD}

The running coupling $\A(Q^2)$ that we intend to construct coincides at high $|Q^2|$ with the usual perturbative coupling $a(Q^2)$ in the same renormalization scheme. We will call this coupling $a(Q^2)$ the underlying pQCD coupling. In order to be able to impose the ``lattice condition'' (\ref{noft}), i.e., $\A(Q^2) \sim Q^2$ when $Q^2 \to 0$ , we choose to work in the same renormalization scheme in which the lattice coupling $\A_{\rm latt}(Q^2)$ was calculated, i.e., in the MiniMOM scheme \cite{MiniMOM,BoucaudMM,CheRet}), which is known analytically only perturbatively, up to four loops. The reason for this choice of scheme is that we do not know \textcolor{black}{in advance how} the change of the renormalization scheme would affect the coupling $\A(Q^2)$ in the deep IR regime; we know how to calculate the effects of this change in the perturbative regime \cite{Stevenson}.\footnote{
  In principle, we could construct $\A$ in any other scheme, e.g., in $\MSbar$ scheme, but then it would not be clear how such a coupling compares with $\A_{\rm latt}$ of Ref.~\cite{LattcoupNf0} in the deep IR regime. For an application and discussion of the MiniMOM scheme in pQCD, see Ref.~\cite{KatMol}.} \textcolor{black}{Nonetheless, the change of the (perturbative) renormalization scheme definitely influences the behavior of $\A(Q^2)$ in the IR regime. Specifically, it is expected that scheme changes affect in our coupling the values of the UV/IR transition scale (pQCD-onset scale $M_0^2$, see later) and the NP parameters  (such as $M_j^2$ and ${\cal F}_j$, see later). For similar effects of scheme changes, we refer to Ref.~\cite{Brod2} and footnote \ref{ftnBrod2} in the present work.} 
On the other hand, we will use the number of active quark flavors to be $N_f=3$
(the lattice results of Ref.~\cite{LattcoupNf0,LattcoupNf0b} are for $N_f=0$). The variation of $N_f$ does not seem to change significantly the lattice results, not even the location of the maximum of $\A(Q^2)$, cf.~Refs.~\cite{LattcoupNf4,LattcoupNf2} where the lattice coupling was calculated for $N_f=4, 2$, respectively, but with smaller lattice volumes.

Since the MiniMOM scheme is known perturbatively up to four loops \cite{MiniMOM}, this means that in the corresponding perturbative Renormalization Group Equation (pRGE)
\be
\frac{d a(Q^2)}{d \ln Q^2} = - \beta_0 a(Q^2)^2 \left[ 1 + c_1 a(Q^2) + c_2 a(Q^2)^2 + c_3 a(Q^2)^3 + \ldots \right]
\label{RGE4l}
\ee
only the coefficients $\beta_0, c_1, c_2, c_3$ on the right-hand side are known. The first two coefficients, $\beta_0=(1/4) (11 - 2 N_f/3)$ and $c_1\equiv \beta_1/\beta_0=(1/4)(102 - 38 N_f/3)/(11 - 2 N_f/3)$, are universal in the class of mass independent schemes, while $c_j \equiv \beta_j/\beta_0$ ($j\geq 2$) together with the scale parameter $\Lambda$ characterize the renormalization scheme \cite{Stevenson}. Hence, the underlying MiniMOM coupling could be taken such as determined by an initial value and by the RGE-running with a $\beta$-function being a polynomial truncated at the $c_3$-term. This turns out to be impractical for numerical calculations (using Mathematica \cite{Math}), because, as we will see, the coupling $\A(Q^2)$ will be constructed dispersively and for that we will need to have a high precision of the spectral (discontinuity) function $\rho^{\rm (pt)}(\sigma) = {\rm Im} \; a(Q^2=-\sigma - i \epsilon)$ at large positive $\sigma$. If using the running coupling $a(Q^2)$ determined by RGE-running (\ref{RGE4l}) with the $\beta$-function truncated at four loops (at the $c_3$-term), such an RGE has no explicit solution in terms of known functions,\footnote{The same is true also for the three-loop version of the RGE (\ref{RGE4l}), i.e., truncated at the $c_2$-term, cf.~Ref.~\cite{Gardi:1998qr}.} and the integration of such an RGE would have to be performed numerically in the entire complex $Q^2$-plane, leading to large uncertainties near the cut region $Q^2 < 0$ and thus to uncertain values of $\rho^{\rm (pt)}(\sigma) = {\rm Im} \; a(Q^2=-\sigma - i \epsilon)$. Therefore, we will use for our $\beta$-function a specific Pad\'e form \cite{GCIK} whose expansion gives the known MiniMOM coefficients $c_2$ and $c_3$ and the solution of the corresponding RGE is explicitly known, in terms of Lambert functions (the latter being well known by Mathematica). The RGE is in this case
\be
\frac{d a(Q^2)}{d \ln Q^2}  =  \beta(a(Q^2))
\equiv - \beta_0 a(Q^2)^2 \frac{ \left[ 1 + a_0 c_1 a(Q^2) + a_1 c_1^2 a(Q^2)^2 \right]}{\left[ 1 - a_1 c_1^2 a(Q^2)^2 \right] \left[ 1 + (a_0-1) c_1 a(Q^2) + a_1 c_1^2 a(Q^2)^2 \right]} \ ,
\label{beta}
\ee
where
\be
a_0  =  1 + \sqrt{c_3/c_1^3}, \quad 
a_1 = c_2/c_1^2 +   \sqrt{c_3/c_1^3} .
\label{a0a1}
\ee
It is straightforward to check that the expansion of the $\beta$-function of the RGE (\ref{beta}) up to $\sim a(Q^2)^5$ reproduces the four-loop polynomial $\beta$-function on the right-hand side of RGE (\ref{RGE4l}). The RGE (\ref{beta}) has explicit solution \cite{GCIK} in terms of Lambert functions, namely
\be
 a(Q^2) = 
\frac{2}{c_1} \left[ - \sqrt{\omega_2} - 1 - W_{\mp 1}(z) + 
\sqrt{(\sqrt{\omega_2} + 1 + W_{\mp 1}(z))^2 
- 4(\omega_1 + \sqrt{\omega_2})} \right]^{-1},
\label{a4l}
\ee
where $\omega_1= c_2/c_1^2$, $\omega_2=c_3/c_1^3$,  $Q^2 = |Q^2| \exp(i \phi)$. Here, the upper index for the Lambert function $W$, i.e., $W_{-1}$, is used when $0 \leq \phi < \pi$, and the lower index $W_{+1}$ when $-\pi \leq \phi < 0$, and the argument $z = z(Q^2)$ in the Lambert functions $W_{\pm 1}(z)$ is
\be
z \equiv z(Q^2) = -\frac{1}{c_1 e} \left( \frac{\Lambda_L^2}{Q^2} \right)^{\beta_0/c_1} \ ,
\label{zexpr}
\ee
where $\Lambda_L(N_f)$ we call the Lambert scale ($\Lambda_L \sim \Lambda_{\rm QCD}$). In Ref.~\cite{GCIK} we used a slightly different expression for $z$, namely without the factor $1/( c_1 e)$, 
which just redefines the Lambert scale $\Lambda_L$. Here we decide to keep this convention factor, i.e., Eq.~(\ref{zexpr}), as it is kept also in the three-loop \cite{Gardi:1998qr} and two-loop solutions 
\cite{Gardi:1998qr,Magradze:1998ng} involving the Lambert functions. Further, there exists another solution  \cite{GCIK} to the RGE (\ref{beta}),  but it turns out to have the Landau branching point of the cut, 
$Q_{\rm br}^2$, at a higher positive value than the solution (\ref{a4l}). We consider this to be an unattractive feature and, therefore, we do not consider this other solution. We point out that the considered
pQCD coupling (\ref{a4l}) has a cut along the semiaxis $Q^2 \leq Q_{\rm br}^2$, and thus includes also a part of the positive axis $0 \leq Q^2 \leq Q_{\rm br}^2$ (where $Q_{\rm br}^2 \sim 1 \ {\rm GeV}^2$). 
This part of the cut is called the Landau (ghost) cut, since it does not reflect the analyticity properties of the spacelike QCD observables as dictated by general principles of Quantum Field Theory \cite{BS,Oehme}.
The result for the running coupling given in Eq. (\ref{a4l}) has been obtained in Ref.~\cite{GCIK} by deforming a solution of the RGE of a particular form given in Refs.~\cite{Jones:1983ip} and \cite{Novikov:1983uc},  to a solution of the RGE of the form whose $\beta$-function gives, when expanded, the expansion (\ref{RGE4l}) with chosen coefficients $c_j$ ($j \leq 3$). The RGE of Refs.~\cite{Jones:1983ip} and  \cite{Novikov:1983uc} is called NSVZ $\beta$-function.\footnote{
This NSVZ $\beta$-function is for bare coupling constant cases. For renormalized couplings the full scheme should be specified; for $\mathcal{N} =$ 1 supersymmetric QED see Refs.~\cite{SQED}.}

The lattice MiniMOM (MM) scheme was determined to 4-loop in Refs.~\cite{MiniMOM,BoucaudMM,CheRet}, where, for $N_f=3$, the two $c_j$ scheme parameters are
\be
c_2({\rm MM},N_f=3)=9.2970, \quad c_3({\rm MM},N_f=3)=71.4538 \ .
\label{c2c3}
\ee
This can be compared with the corresponding coefficients in $\MSbar$, which are\footnote{Expansion of the $\beta$-function (\ref{beta}) gives also an estimated value for the MiniMOM $c_4$ coefficient, namely $c_4({\rm MM},N_f=3)=201.84$, to be compared with the recently obtained $\MSbar$ value $c_4(\MSbar,N_f=3)=56.588$ \cite{5lMSbarbeta}.}
$c_2(\MSbar,N_f=3)=4.4711$ and $c_3(\MSbar,N_f=3)=20.9902$. 

In order to fix the only parameter left free in the explicit solution (\ref{a4l}), namely the Lambert scale $\Lambda_L$ appearing in Eq.~(\ref{zexpr}), we proceed in the following way. Due to experimental uncertainty of the value $a(M_Z^2,\MSbar) \equiv {\overline a}(M_Z^2)$, and because several of the results of our analysis depend crucially on the value of ${\overline a}(M_Z^2)$, we will consider in the following the three values ${\overline a}(M_Z^2) = 0.1185/\pi$ \cite{PDG2014}, $0.1181/\pi$ \cite{PDG2016} and $0.1189/\pi$, which has $N_f=5$ in all cases.\footnote{We recall that the world average given by Particle Data Group in 2014 is ${\overline a}(M_Z^2)=(0.1185 \pm 0.0006)/\pi$ \cite{PDG2014}, and in 2016 is ${\overline a}(M_Z^2)=(0.1181 \pm 0.0011)/\pi$ \cite{PDG2016}. \label{PDGfootnote}}
We RGE-evolve this value down into the $N_f=3$ regime of positive $Q^2$, with four-loop $\MSbar$ beta function \cite{4lMSbarbeta}, and take into account the corresponding three-loop quark threshold relations \cite{CKS} at $Q^2_{\rm thr.} = (\kappa {\overline m}_q)^2$ ($q=b, c$) with $\kappa=2$ and with the $\MSbar$ quark masses
${\overline m}_q \equiv {\overline m}_q({\overline m}_q^2)$ set equal to ${\overline m}_b=4.20$ GeV and ${\overline m}_c=1.27$ GeV. This then gives, at the upper edge $Q_c^2=(\kappa {\overline m}_c)^2$ ($=6.452 \ {\rm GeV}^2$) of the $N_f=3$ regime, in $\MSbar$, the value ${\overline a}(Q_c^2,N_f=3) (\equiv {\overline a}_c) = 0.26588/\pi$ if ${\overline a}(M_Z^2)=0.1185/\pi$, and accordingly in the other two cases, cf.~Eqs.~(\ref{aQc}) below. The last step remaining is to change this value ${\overline a}_c$ from $\MSbar$ to MiniMOM scheme Eq.~(\ref{beta}), but with the same scale scheme parameter $\Lambda_{\MSbar}$  as used in $\MSbar$; we will call this scheme: Lambert-MiniMOM, shorthand LMM, to distinguish it from the lattice MiniMOM, shorthand MM. 
This change is performed by solving numerically for $a_c \equiv  a(Q_c^2,{\rm LMM})$ the integrated form of RGE in its subtracted
form, cf.~App.~A of \cite{Stevenson} and App.~A of \cite{CK}
\bea
\lefteqn{
\frac{1}{a_{c}} + c_1 \ln \left( \frac{c_1 a_{c}}{1\!+\!c_1 a_{c}} \right)
+ \int_0^{a_{c}} dx \left[ \frac{ \beta(x) + \beta_0 x^2 (1\!+\!c_1 x)}
{x^2 (1\!+\!c_1 x) \beta(x) } \right] =}
\nonumber\\
&& \frac{1}{{\overline a}_{c}} + c_1 \ln \left( \frac{c_1 {\overline a}_{c}}{1\!+\!c_1 {\overline a}_{c}} \right)
+ \int_0^{{\overline a}_{c}} dx \left[ \frac{ {\overline \beta}(x) + \beta_0 x^2 (1\!+\!c_1 x)}
{x^2 (1\!+\!c_1 x) {\overline \beta}(x) } \right],
\label{match}
\eea
where $\beta(x)$ is the LMM beta function appearing in the RGE (\ref{beta}), and ${\overline \beta}(x)$ is the $\MSbar$ beta function, and ${\overline a}_{c} \equiv a(Q_c^2,\MSbar)$ (with $N_f=3$).

This change of scheme gives the following results, at $Q_c^2 = (2 {\overline m}_c)^2 - 0$ ($=6.452 \ {\rm GeV}^2$) of the $N_f=3$ regime:
\bes
\label{aQc}
\bea
a(M_Z^2,\MSbar) & = & \frac{0.1185}{\pi} \Rightarrow a(Q_c^2, \MSbar) \equiv {\overline a}_c = \frac{0.26588}{\pi} \Rightarrow a(Q_c^2,{\rm LMM}) \equiv a_c = \frac{0.28209}{\pi} \ ,
\label{ain01185}
\\
a(M_Z^2,\MSbar) & = & \frac{0.1181}{\pi} \Rightarrow a(Q_c^2, \MSbar) = \frac{0.26375}{\pi} \Rightarrow a(Q_c^2,{\rm LMM})= \frac{0.27946}{\pi} \ ,
\label{ain01181}
\\
a(M_Z^2,\MSbar) & = & \frac{0.1189}{\pi} \Rightarrow a(Q_c^2, \MSbar) \equiv {\overline a}_c = \frac{0.26805}{\pi} \Rightarrow a(Q_c^2,{\rm LMM}) \equiv a_c = \frac{0.28476}{\pi} \ ,
\label{ain01189}
\eea
\ees
Using these results, and requiring that the solution (\ref{a4l}) of the LMM scheme agree with these results, we can fix the Lambert scale of Eq.~(\ref{zexpr}), at $N_f=3$
\bes
\label{LambdaL}
\bea
\Lambda_L&=&0.11564 \ {\rm GeV} \quad (\pi {\overline a}(M_Z^2)=0.1185) \ ,
\label{LambdaL01185}
\\
\Lambda_L&=&0.11360 \ {\rm GeV} \quad (\pi {\overline a}(M_Z^2)=0.1181) \ ,
\label{LambdaL01181}
\\
\Lambda_L&=&0.11771 \ {\rm GeV} \quad (\pi {\overline a}(M_Z^2)=0.1189) \ ,
\label{LambdaL01189}
\eea
\ees
When comparing with lattice results, we must take into account that the momentum scale parameter $\Lambda$ is different in the lattice MiniMOM (MM), \cite{MiniMOM}
\be
\frac{\Lambda_{\rm MM}}{\Lambda_{\MSbar}} = 1.8968 \; ({\rm for} \; N_f=0); \; 1.8171 \; ({\rm for} \; N_f=3);
\label{LambdaMMMSbar}
\ee
This demonstrates that the meaning of the momentum scales is different in the lattice MiniMOM (MM, using $\Lambda_{\rm MM}$) and in the Lambert-MiniMOM (LMM, using $\Lambda_{\MSbar}$ instead), respectively, 
although the same $c_2$ and $c_3$ scheme parameters are used. For example, the scale $Q^2=0.45 \ {\rm GeV}^2$ in the lattice MM with $N_f=0$ or $N_f=3$ corresponds to 
the scale $Q^2 = 0.45 \times (\Lambda_{\MSbar}/\Lambda_{\rm MM})^2 \approx 0.125 \ {\rm GeV}^2$ in $N_f=0$ LMM and to $Q^2 \approx 0.136 \ {\rm GeV}^2$ in $N_f=3$ LMM scheme. \textcolor{black}{Due to this aspect, we will be able to apply the couplings $a(Q^2)$ and $\A(Q^2)$ in the LMM scheme in general at lower values of $Q^2$ than in the MM scheme. Another important aspect which will allow us to use the coupling $\A(Q^2)$ in evaluations of physical quantities at lower positive $Q^2$ values than usual will be the holomorphic behavior of $\A(Q^2)$, cf.~the next Section.}

If we repeat this calculation, but use instead the five-loop $\MSbar$ $\beta$-function \cite{5lMSbarbeta} and the corresponding four-loop quark thresholds \cite{4lquarkthresh} (i.e., the ``5+4'' approach), 
the resulting values do not change significantly enough for our purposes and precisions. For example, in Eq.~(\ref{ain01185}) the value $\pi a_c =0.28209$ was obtained by the ``4+3'' 
approach for $\pi {\overline a}(M_Z^2) = 0.1185$, and this same value  $\pi a_c$ is obtained in the ``5+4'' approach with only slightly different high-energy initial value, n
amely  $\pi {\overline a}(M_Z^2) = 0.1186$. Further, in this context we recall that MiniMOM scheme is known only to four loops \cite{MiniMOM,BoucaudMM,CheRet}.

The pQCD running coupling (\ref{a4l})-(\ref{zexpr}), with the MiniMOM scheme parameters $c_2$ and $c_3$ Eqs.~(\ref{c2c3}) and with the momentum scale $\Lambda_L$ Eqs.~(\ref{LambdaL}), 
is thus in the scheme which agrees up to (known) four-loop level with the Lambert-MiniMOM scheme. Moreover, the expression (\ref{a4l}) involves the Lambert functions which are efficiently evaluated
with Mathematica \cite{Math}, and this expression can be evaluated efficiently also along the cut axis of the pQCD coupling, allowing for a fast and precise evaluation of the cut 
discontinuity function $\rho_{a}(\sigma) \equiv {\rm Im} a(Q^2=-\sigma - i \epsilon)$, even at very high values of $\sigma$. This will be of practical importance in the next Section where we will 
construct a holomorphic coupling $\A(Q^2)$ dispersively, using the discontinuity function $\rho_{a}(\sigma)$.

We point out that in our previous work \cite{AQCDprev}, the underlying coupling was constructed according to an analogous but considerably simpler formula \cite{Gardi:1998qr} involving Lambert functions, 
such that it reproduced the three-loop MiniMOM scheme parameter $c_2$, but not the four-loop MiniMOM parameter $c_3$ Eq.~(\ref{c2c3}).

\section{Construction of the coupling $\A(Q^2)$}
\label{sec:constr}

\begin{figure}[htb]
\centering\includegraphics[width=130mm]{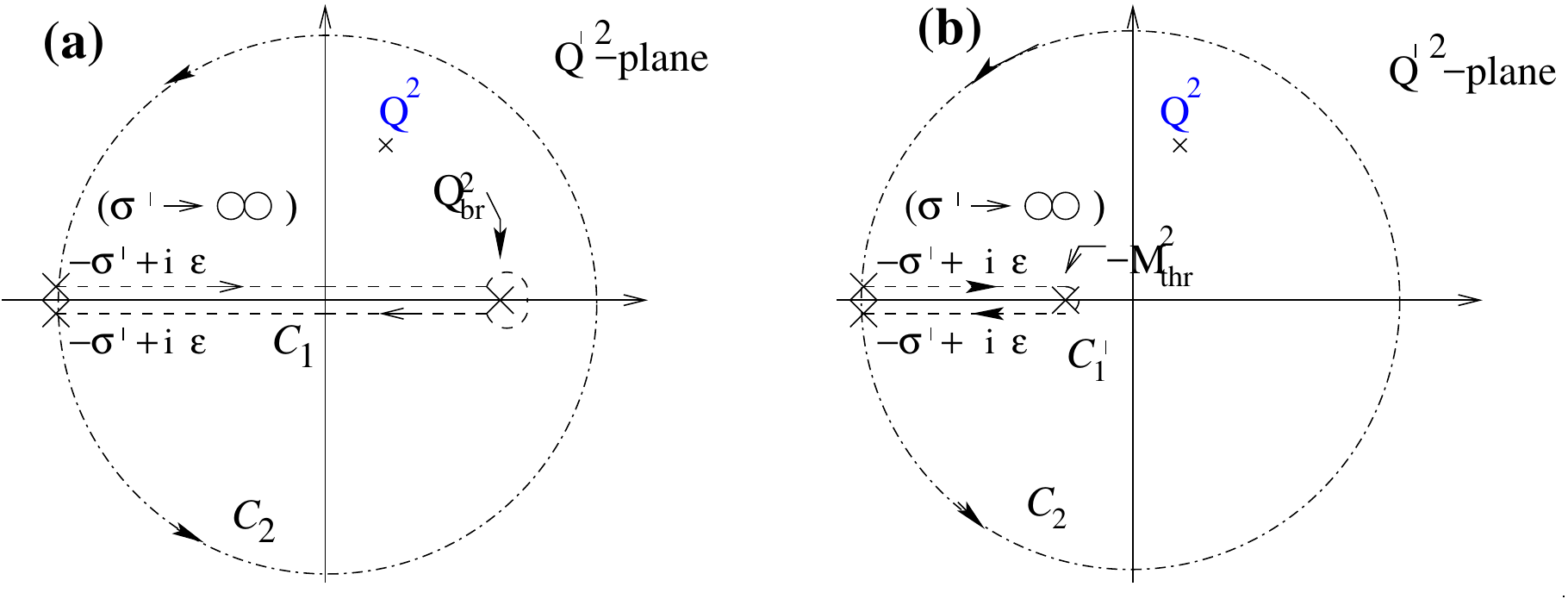}
\caption{\footnotesize (a) The contour of integration for the integrand $a(Q'^2)/(Q'^2 - Q^2)$ leading to the dispersion relation (\ref{adisp}) for  $a(Q^2)$; (b) the contour of integration for the integrand  $\A(Q'^2)/(Q'^2 - Q^2)$ leading to the dispersion relation (\ref{Adisp}). The radius $\sigma'$ of the circular part tends to infinity.}
\label{Figintpath}
\end{figure}

The starting point in our construction of the coupling $\A(Q^2)$ will be the requirement that, for consistency reasons, the analytic properties of $\A(Q^2)$ in the complex $Q^2$-plane reflect the corresponding analytic properties of the spacelike QCD physical quantities ${\cal D}(Q^2)$ \cite{BS,Oehme}, such as current correlators (Adler function), \textcolor{black}{DIS differential cross sections}, and various amplitudes for physical processes. Stated otherwise, we will require that $\A(Q^2)$ be an analytic function of $Q^2$ for $Q^2 \in \mathbb{C} \backslash (-\infty, -M_{\rm thr}^2]$, where $M_{\rm thr} \sim 0.1$ GeV is a threshold scale. This implies that the only singularity structure of $\A(Q^2)$ is the cut along the negative semiaxis $(-\infty, -M_{\rm thr}^2]$.
Using this property and the asymptotic freedom of $\A(Q^2)$, application of the Cauchy integral formula to the integrand  $\A(Q'^2)/(Q'^2 - Q^2)$ along the path in Fig.~\ref{Figintpath}(b) in the complex $Q^{'2}$-plane leads directly to the following dispersive relation:
\be
\A(Q^2) = \frac{1}{\pi} \int_{\sigma=M^2_{\rm thr}-\eta}^{\infty} \frac{d \sigma \rho_{\A}(\sigma)}{(\sigma + Q^2)} 
\qquad (\eta \to +0),
\label{Adisp}
\ee
where $\rho_{\A}(\sigma) \equiv {\rm Im} \mathcal{A}(-\sigma - i \varepsilon)$ is the discontinuity function (spectral function) of $\A$ along the cut.

The idea for this approach goes back to the perturbation theory analytic in its ``minimal'' form, as described in Refs.~\cite{ShS,MS,Sh1Sh2}, where the authors used for $\rho_{\A}(\sigma)$ the pQCD spectral function $\rho_a(\sigma) \equiv {\rm Im} a(Q^2=-\sigma - i \epsilon)$ and setting $M^2_{\rm thr}=0$.\footnote{
  This approach was called Analytic Perturbation Theory (APT). It was extended to any physical quantity in Ref.~\cite{KS}, and to analogs of noninteger powers of the coupling \cite{BMS} (Fractional Analytic Perturbation Theory - FAPT). We refer to Refs.~\cite{Prosperi,Shirkov,Bakulev,Stefanis} for reviews of these approaches. In Refs.~\cite{APTappl1,APTappl2} are given some of the applications of these works to (low-energy) QCD phenomenology of nucleon structure function sum rules and of nucleon structure functions, respectively, and in Ref.~\cite{APTappl3} to the gluon propagator.}
Many other models leading to holomorphic couplings have been constructed and applied since then, among them those of Refs.~\cite{Nest2,Webber,Boucaud,Alekseev,CV12,1danQCD,2danQCD,anOPE,anOPE2,Brod,Brod2,ArbZaits,Shirkovmass,KKS,Luna,Nest1,NestBook}. Most of these couplings have finite values $\A(0) < \infty$. The construction of Refs.~\cite{Nest1} gives a holomorphic coupling infinite at the origin $\A(0) = \infty$. Some of the couplings constructed and motivated in this literature fulfill the condition $\A(0)=0$, namely Refs.~\cite{ArbZaits,Boucaud,mes2}. The latter were obtained independently of the lattice results \cite{LattcoupNf0,LattcoupNf0b,LattcoupNf4,LattcoupNf2,Latt3gluon} and of the Gribov-related and DSE approaches  \cite{DSEdecoup,Latt3gluon,Gribovdecoup} which also give $\A(0)=0$ if using the relation (\ref{Alatt}) and $\A \sim \A_{\rm latt.}$ or analogous relations \cite{Latt3gluon}.

Reviews of a variety of such models can be found, cf.~Refs.~\cite{GCrev,Brodrev}. Further, mathematical packages for numerical evaluation of various holomorphic couplings and of their power analogs are available, Refs.~\cite{BK,ACprogr}.
Most, but not all, of the models use constructions with dispersion integrals (i.e., Cauchy integral formula) applied to the couplings, automatically ensuring their holomorphic behavior. However, there exist also related dispersive approaches which are applied directly to spacelike QCD quantities \cite{MSS1,MSS2,MagrGl,mes2,DeRafael,MagrTau,Nest3a,Nest3b,NestBook}. In the course of all such constructions, nonperturbative terms are generated, either inside the couplings and/or in the physical quantities. 

In our dispersive construction of the coupling $\A(Q^2)$, Eq.~(\ref{Adisp}), the dispersion relation for the underlying pQCD coupling $a(Q^2)$, Eq.~(\ref{a4l}), will play an important role. This coupling $a(Q^2)$ has singularity structure in the complex $Q^2$-plane represented by a larger  cut $ (-\infty, +Q_{\rm br}^2]$, which involves also the Landau (ghost) cut $(0,+Q_{\rm br}^2]$ (where $0 < Q_{\rm br}^2 \sim 1 \ {\rm GeV}^2$). As mentioned, this Landau cut of pQCD coupling $a(Q^2)$ does not reflect the general properties of spacelike physical QCD quantities, and prevents us from evaluating the coupling at low momenta $0 < Q^2 < Q_{\rm br}^2$. Analogously to the dispersive relation (\ref{Adisp}), 
the Cauchy integral formula applied in this case to the function  $a(Q'^2)/(Q'^2 - Q^2)$ gives the following dispersion integral:
\begin{equation}
a(Q^2) = \frac{1}{\pi} \int_{\sigma= - {Q^2_{\rm br}} - \eta}^{\infty}
\frac{d \sigma {\rho}_a(\sigma) }{(\sigma + Q^2)}
   \qquad (\eta \to +0),
\label{adisp}
\end{equation}
where $\rho_a(\sigma) \equiv {\rm Im} a(Q^2=-\sigma - i \epsilon)$ is the discontinuity function of $a$ along its cut. The contours of the Cauchy theorems in the complex $Q^{'2}$-plane, leading to the relations (\ref{adisp}) and (\ref{Adisp}) are presented in Figs.~\ref{Figintpath}(a) and (b).

In the UV regime of large positive $\sigma = - Q^2$, the discontinuity function $\rho_{\A}(\sigma)$ tends to the corresponding (underlying) pQCD function $\rho_a(\sigma)$ as dictated by the asymptotic freedom of QCD. Therefore, we will set equal $\rho_{\A}(\sigma) = \rho_a(\sigma)$  for $\sigma > M_0^2$ where $M_0^2 \sim 1$-$10 \ {\rm GeV}^2$ is a ``pQCD-onset'' scale. On the other hand, in the IR regime, $0 < \sigma < M_0^2$, the spectral function $\rho_{\A}(\sigma)$ is a priori unknown, and contributes to the part $\Delta \A_{\rm IR}(Q^2)$ of $\A(Q^2)$
\bes
\label{ADA}
\bea
 \A(Q^2) &=& \frac{1}{\pi} \int_{\sigma=M^2_0}^{\infty} \frac{d \sigma \rho_{a}(\sigma)}{(\sigma + Q^2)} + \Delta \A_{\rm IR}(Q^2) ,
\label{ADAa}
\\
\Delta \A_{\rm IR}(Q^2) & = & \frac{1}{\pi} \int_{\sigma=M_{\rm thr}^2}^{M_0^2} \frac{d \sigma \rho_{\A}(\sigma)}{(\sigma + Q^2)}
\eea
\label{ADAb}
\ees
We will parametrize the (unknown) function $\Delta \A_{\rm IR}(Q^2)$ as a Pad\'e of the type $[M-1/M]$, i.e., as a ratio of a polynomial in $Q^2$ of power $M-1$ divided by a polynomial of power $M$: 
\bes
\label{DAPD}
\bea
\Delta \A_{\rm IR}(Q^2) &=& [M-1/M](Q^2) = \frac{\sum_{n=1}^{M-1} A_n Q^{2n}}{\sum_{n=1}^M B_n Q^{2n}}
\label{M1M}
\\
& = & \sum_{j=1}^{M} \frac{{\cal F}_j}{Q^2 + M_j^2} \ .
\label{PFM1M}
\eea
\ees
In the second line, we rewrote the mentioned Pad\'e as a sum of partial fractions, as can always be done, and we assume $M_j^2 > 0$ to maintain the holomorphic behavior of $\A(Q^2)$. If the spectral function $\rho_{\A}(\sigma)$ were a nonnegative function, the coupling $ \Delta \A_{\rm IR}(Q^2)$ [and $\A(Q^2)$] would be a Stieltjes function. In such a case, a mathematical theorem \cite{BakerMorris} guarantees that a sequence of Pad\'e's $[M-1/M](Q^2)$ exists which converges to $\Delta \A_{\rm IR}(Q^2)$ when $M \to \infty$, for any $Q^2 \in \mathbb{C} \backslash (-\infty, -M_{\rm thr}^2]$ (cf.~also \cite{Peris}). In our case the spectral function is negative for some (low) $\sigma$ values, as can be concluded from the lattice results of Fig.~\ref{FigAlatt}. Namely, since $\A(Q^2)$ is not monotonic at low positive $Q^2$, the slope
\be
\frac{d \A(Q^2)}{d \ln Q^2} = - \frac{Q^2}{\pi}  \int_{\sigma=M^2_{\rm thr}-\eta}^{\infty} \frac{d \sigma \rho_{\A}(\sigma)}{(\sigma + Q^2)^2}
\label{slope}
\ee
changes sign at low $Q^2 \sim 0.1 \ {\rm GeV}^2$, and therefore the spectral function $\rho_{\A}(\sigma)$ cannot be nonnegative at all $\sigma$. Nonetheless, although in the considered case (where $\A$ and $\Delta A_{\rm IR}$ are not Stieltjes) no mathematical theorem is known which would guarantee the convergence $[M-1/M](Q^2) \to  \Delta A_{\rm IR}(Q^2)$, we will assume that there is such a convergence.\footnote{
As pointed out in the book \cite{BakerMorris},  the mathematical theory of Pad\'e approximants is still very incomplete. Note that in the mentioned theorem, the Stieltjes nature is not a necessary condition for the convergence behavior (it is a sufficient condition). However, Pad\'e approximants are in general regarded as efficient analytic continuations of a function $f(z)$ into the complex $z$-plane when we have only limited information available for $f(z)$, such as the first few derivatives at a point $z_0$, or the structure of poles and/or zeros of the function, etc., cf.~Ref.\cite{BakerMorris}. As a consequence, Pad\'e approximants are widely used in natural sciences, including various areas of physics. 
In QCD, in the limiting case $M \rightarrow \infty$, successful Pad\'e approximants exist for any range of $Q^2$. An example for a Pad\'e approximant form $[M-1/M](a(Q^2))$ is the expression for $\beta(a(Q^2))/a(Q^2)$ in Eq. (\ref{beta}), with $M=4$.}
In the sequence $[M-1/M](Q^2) \to  \Delta A_{\rm IR}(Q^2)$, we will make the approximation $M=3$, i.e., we will assume that $\Delta \A_{\rm IR}(Q^2)$ can be sufficiently well approximated by $[2/3](Q^2)$ Pad\'e approximant. It is straightforward to see that this implies for the spectral function $\rho_{\A}(\sigma)$ in the IR regime ($0 < \sigma < M_0^2$) the following expression:
\bes
\label{DADrho}
\bea
  \Delta \A_{\rm IR}(Q^2) &=& [2/3](Q^2) = \sum_{j=1}^{3} \frac{{\cal F}_j}{Q^2 + M_j^2}
\label{DAIR}
\\
\Leftrightarrow
\rho_{\A}(\sigma) & = & \pi \sum_{j=1}^{3} {\cal F}_j \; \delta(\sigma - M_j^2) \qquad (0 < \sigma < M_0^2),
\label{Drho}
\eea
\ees
i.e., the spectral function in the IR regime is approximated (parametrized) by three delta functions. On physical grounds, we expect the squared masses $M_j^2$ to lie in the IR regime, i.e., $0 < M_j^2 < M_0^2$ ($j=1,2,3$). The total discontinuity function is then
\be
\rho_{\A}(\sigma) =  \pi \sum_{j=1}^{3} {\cal F}_j \; \delta(\sigma - M_j^2)  + \Theta(\sigma - M_0^2) \rho_a(\sigma) \ ,
\label{rhoA}
\ee
where $\Theta$ is the Heaviside step function.\footnote{The parametrization of spectral functions in the unknown regime as a linear combination of delta functions has been used in the literature, principally for spectral functions ${\rm Im} {\cal D}(-\sigma - i \epsilon)$ of spacelike QCD observables ${\cal D}(Q^2)$ such as the current correlation functions and the related Adler function, in the context of the large-$N_c$ QCD approximation, cf.~Refs.~\cite{DeRafael,MagrTau,Peris}. We assume that a different ansatz, e.g., with finite width Breit-Wigner resonance forms, would in general give similar results for the coupling $\A(Q^2)$.}
As a result, the considered coupling $\A(Q^2)$ is parametrized as 
\bea
\A(Q^2) & = &  \sum_{j=1}^3 \frac{{\cal F}_j}{(Q^2 + M_j^2)} + \frac{1}{\pi} \int_{M_0^2}^{\infty} d \sigma \frac{ \rho_a(\sigma) }{(Q^2 + \sigma)} \ .
\label{AQ2}
\eea
The coupling (\ref{AQ2}) has seven free parameters ${\cal F}_j$, $M^2_j$ ($j=1,2,3$) and $M_0^2$. We will order the squared mass parameters in the following way: $0 < M_1^2 < M_2^2 < M_3^2$ ($< M_0^2$). We recall that, since $\A(Q^2)$ is not Stieltjes and $\rho_a(\sigma)>0$, at least one of the scale parameters ${\cal F}_j$ ($j=1,2,3$) will be negative.

We note that the scale $\Lambda_L$ appearing in the pQCD spectral function $\rho_a(\sigma)$ was fixed by the central value of the world average $\alpha_s(M_Z^2; \MSbar) \equiv \pi {\overline a}(M_Z^2)$, Eqs.~(\ref{LambdaL}). We therefore need seven conditions to eliminate the free parameters.  Four of these conditions will come from the imposed requirement that the coupling $\A$ effectively merges with the pQCD coupling at high momenta, namely the condition Eq.~(\ref{Aadiff1}) with $N_{\rm max}=5$,
\be
\A(Q^2) - a(Q^2) \sim \left( \frac{\Lambda_L^2}{Q^2} \right)^5 \quad
(|Q^2| > \Lambda_L^2) \ .
\label{Aadiff2}
\ee
This condition together with the use of the world average values $\alpha_s(M_Z^2;\MSbar) \approx 0.1185$ \cite{PDG2014} for fixing of the $\Lambda_L$ scale in the underlying pQCD coupling $a(Q^2)$ [cf.~Eqs.~(\ref{aQc})-(\ref{LambdaL})] means that we assume that the mentioned world average values can be extracted from the high-energy QCD phenomenology only, $|Q^2| > 1 \ {\rm GeV}^2$. This may not be exactly true, as the world average values were extracted, by application of pQCD(+OPE), from a large set of QCD processes, some of them being low-energy processes ($|Q^2| \sim 1 {\rm GeV}^2$) such as the semihadronic decays of the $\tau$ lepton. 

If in Eq.~(\ref{Aadiff2}) we increased the power index, i.e., $N_{\rm max} > 5$, the numerical results for $\A(Q^2)$ would change insignificantly: $\A$ would merge even slightly better with $a$ in the UV regime, but the number of conditions and parameters in $\A$ would increase. On the other hand, $N_{\rm max}=5$ is sufficiently high for the application of OPE. Namely, in Sec.~\ref{sec:BSR} we apply sum rules with OPE up to terms $\sim (\Lambda_L^2/Q^2)^3$, therefore the condition (\ref{Aadiff2}) is safely sufficient to ensure that the OPE approach with $\A$ or with $a$ gives in principle\footnote{But not in practice, due to truncation of series and fitting.} the same OPE terms for inclusive physical quantities.    
  
The fifth condition will be Eq.~(\ref{noft}), i.e., $\A(Q^2) \sim Q^2$ at $Q^2 \to 0$ as suggested by lattice results, Fig.~\ref{FigAlatt}.

Altogether, these five conditions can be written in a more explicit manner, by using the expressions (\ref{AQ2}) and (\ref{adisp}) for $\A$ and $a$ and applying them for $Q^2 \to 0$ and for $|Q^2|>\Lambda_L^2$
\bes
\label{cons}
\bea
- \frac{1}{\pi} \int_{M_0^2}^{\infty} d \sigma \frac{\rho_a(\sigma)}{\sigma} &=& 
\sum_{j=1}^3  \frac{{\cal F}_j}{M_j^2} \ ;
\label{Q2to0}
\\
\frac{1}{\pi} \int_{-Q_{\rm br}^2}^{M_0^2} d \sigma \sigma^k \rho_a(\sigma) &=& \sum_{j=1}^3 {\cal F}_j M_j^{2 k}  \quad (k=0,1,2,3) \ .
\label{1u}
\eea
\ees
The first identity represents the condition $\A(Q^2) \sim Q^2$ when $Q^2 \to 0$. The second identity with $k=0$ means that $\A(Q^2) - a(Q^2)$ contains no term $\sim (\Lambda_L^2/Q^2)^1$ (when $|Q^2| > \Lambda_L^2$); etc.; with $k=3$ means that $\A(Q^2) - a(Q^2)$ contains no term $\sim (\Lambda_L^2/Q^2)^4$. Thus, Eqs.~(\ref{1u}) mean that the relation (\ref{Aadiff2}) is fulfilled.

The sixth condition will be the requirement that $\A(Q^2)$ for positive $Q^2$ have its maximum at about the same squared momentum $Q^2_{\rm max}$ as $\A_{\rm latt.}(Q^2)$, i.e., $Q^2_{\rm max} \approx 0.135 \ {\rm GeV}^2$, cf.~Fig.~\ref{FigAlatt}.\footnote{\label{ftnQ2max}
  We note that in MiniMOM (MM) scheme, $\A_{\rm latt.}(Q^2)$ has maximum value at about $Q^2 \approx 0.45 \ {\rm GeV}^2$, cf.~\cite{LattcoupNf0} for $N_f=0$ and \cite{LattcoupNf2} for $N_f=2$. We use the Lambert-MiniMOM scheme, i.e., MiniMOM by rescaling of $Q^2$ by a factor $\Lambda_{\MSbar}^2/\Lambda_{\rm MM}^2$, Eq.~(\ref{LambdaMMMSbar}). Using for this factor the $N_f=3$ value $1/1.8171^2$ (we work in $N_f=3$ Lambert-MiniMOM), we obtain the estimate $Q^2_{\rm max} \approx 0.45/1.8171^2 \ {\rm GeV}^2 \approx 0.135 \ {\rm GeV}^2$.}
We recall that our $\A(Q^2)$ is expected to reproduce qualitatively the main features of $\A_{\rm latt.}(Q^2)$ for low positive $Q^2$, and in particular is expected to reproduce also approximately the location of the local maximum.

The last, seventh, condition comes from requiring that the main features of the semihadronic $\tau$-lepton decay physics be reproduced. Stated otherwise, we will require that the approach with the coupling $\A(Q^2)$ reproduce the experimentally suggested value of the V+A semihadronic $\tau$-decay ratio parameter $r^{(D=0)}_{\tau} \approx 0.20$ \cite{ALEPH2,DDHMZ} (cf.~also App.~B of \cite{anpQCD1b}). This is the QCD part of the V+A $\tau$-decay ratio $R_{\tau} = \Gamma(\tau^- \to \nu_{\tau}{\rm hadrons}(\gamma))/\Gamma(\tau^- \to \nu_{\tau} e^- {\bar \nu}_e (\gamma))$, where the hadrons are strangeless ($\Delta S=0$) and the quark mass effects and other higher-twist effects are subtracted, i.e., it is the dimension $D=0$ strangeless and massless part. It is canonical in the sense that its pQCD expansion begins with $r^{(D=0)}_{\tau, {\rm pt}} = a + {\cal O}(a^2)$.
More explicitly, $r^{(D=0)}_{\tau}$ is defined by the relations
\bes
\label{Rtaudef}
\bea
R_{\tau} (\Delta S = 0) & \equiv & \frac{\Gamma(\tau^- \to \nu_{\tau}{\rm hadrons}(\gamma))}{\Gamma(\tau^- \to \nu_{\tau} e^- {\bar \nu}_e (\gamma))} -  R_{\tau} (\Delta S \not= 0) \ ,
\label{Rtau1}
\\
R_{\tau} (\Delta S = 0) & = & 3 |V_{u d}|^2 (1 + \delta_{\rm EW}) \left[ 1 + \delta^{\prime}_{\rm EW} + r_{\tau} + \Delta r_{\tau}(m_{u,d} \not= 0) \right] ,
\label{Rtau2}
\eea
\ees
where $\delta_{\rm EW} \approx 1.0198 \pm 0.0006$ \cite{ALEPH2} and $\delta_{\rm EW}^{\prime} = 0.0010$ \cite{Braaten:1990ef} are EW correction parameters, $V_{ud}$ is the CKM matrix element, and $\Delta r_{\tau}(m_{u,d} \not= 0)$ is the chirality-violating contribution. The quantity $r_{\tau}$ stems from an OPE sum, therefore:
$r_{\tau} = \sum_{D \geq 0} r^{(D)}_{\tau}$. 
The here considered quantity $r^{(D=0)}_{\tau}$ is timelike, but it can be expressed theoretically, 
by using the Cauchy integral formula, by means of a spacelike quantity called (leading-twist and massless) Adler function $d(Q^2;D=0)$ \cite{Braaten,PichPra,Pivovarov:1991rh,LeDiberder:1992te}:
\be
r^{(D=0)}_{\tau, {\rm th}} = \frac{1}{2 \pi} \int_{-\pi}^{+ \pi}
d \phi \ (1 + e^{i \phi})^3 (1 - e^{i \phi}) \
d(Q^2=m_{\tau}^2 e^{i \phi};D=0) \ .
\label{rtaucont}
\ee
The Adler function $d(Q^2;D=0)$ is a derivative of the quark current correlator $\Pi$: $d(Q^2;D=0) = - 2 \pi^2 d \Pi(Q^2; D=0)/d \ln Q^2 - 1$, in the massless limit. Its perturbation expansion is known up to $\sim a^4$ \cite{d1,d2,d3}, cf.~Table \ref{tabAdl}
\bes
\label{dpt}
\bea
d(Q^2;D=0)_{\rm pt} &=&  d(Q^2,\mu^2;D=0)_{\rm pt}^{[4]} + {\cal O}(a^5)
,
\label{dpta}
\\
d(Q^2,\mu^2=\kappa Q^2; D=0)_{\rm pt}^{[4]} & = & a(\kappa Q^2) + \sum_{n=1}^{3} d_n(\kappa) a(\kappa Q^2)^{n+1},
\label{dptb}
\\
d(Q^2;D=0)_{\rm pt}^{[4]} & = & a(Q^2) + \sum_{n=1}^{3} d_n a(Q^2)^{n+1}.
\label{dptc}
\eea
\ees
\begin{table}
   \caption{The known perturbation coefficients $d_n$ of the $N_f=3$ massless Adler function $d(Q^2;D=0)$ in the $\MSbar$ and in the Lambert-MiniMOM (LMM) scheme. The transformation relations are given in Appendix \ref{app:tcp}.}
\label{tabAdl}  
\begin{ruledtabular}
\begin{tabular}{l|lll}
scheme &  $d_1$ & $d_2$  & $d_3$ 
  \\
  \hline
  $\MSbar$ & $1.63982$ & $6.37101$ & $49.0757$ 
  \\
       LMM & $1.63982$ & $1.54508$ & $8.01658$
\end{tabular}
\end{ruledtabular}
\end{table}
The truncated perturbation series (TPS) Eq.~(\ref{dptb}) is for a general renormalization scale (RScl) $\mu^2 = \kappa Q^2$ ($0 < \kappa \sim 1$), while Eq.~(\ref{dptc}) represents the notation for the special choice $\mu^2=Q^2$ ($\kappa=1$). The TPS (\ref{dptb}) has residual RScl-dependence (i.e., $\kappa$-dependence) due to truncation.
Since $\A(Q^2)$ is not a pQCD coupling, the evaluation of the truncated expansions (\ref{dptb}) [and thus of $r^{(D=0), {\rm [4]}}_{\tau, {\rm th}}$ Eq.~(\ref{rtaucont})] with $\A$-coupling should be performed with care, where the analogs of the pQCD powers $a(\kappa Q^2)^n$ are specific functions $\A_n(\kappa Q^2)$ [$\not= \A(\kappa Q^2)^n$]
\bes
\label{dan}
\bea
d(Q^2;D=0)_{\rm an} & = &  d(Q^2,\mu^2;D=0)_{\rm an}^{[4]} + {\cal O}(\A_5),
\label{dana}
\\
d(Q^2,\mu^2=\kappa Q^2 ;D=0)_{\rm an}^{[4]} & = & \A(\kappa Q^2) + d_1(\kappa) \A_{2}(\kappa Q^2) +  d_2(\kappa) \A_{3}(\kappa Q^2) +  d_3(\kappa) \A_{4}(\kappa Q^2),
\label{danb}
\\
 d(Q^2;D=0)_{\rm an}^{[4]} & = & \A(Q^2) + d_1 \A_{2}(Q^2) +  d_2 \A_{3}(Q^2) +  d_3 \A_{4}(Q^2).
\label{danc}
\eea
\ees
The power analogs $\A_n(Q^2)$ from $\A(Q^2)$ were constructed in general holomorphic theories from $\A(Q^2)$ in Ref.~\cite{CV12} for integer $n$ and in Ref.~\cite{GCAK} for general real $n$. In Appendix \ref{app:An} we present briefly the necessary formulas for obtaining $\A_n$ ($n=2,3,4$) from $\A$, relevant in the case of the truncated series (\ref{dan}). It turns out that this truncated series (\ref{danb}) can be resummed, in an efficient way, by an approach using a generalization  \cite{BGApQCD1,BGApQCD2,BGA,anOPE} of the diagonal Pad\'e approach ($[M/M]$, here $M=2$) \cite{GardiPA}. The generalization gives the result which is exactly independent of RScl (i.e., independent of $\kappa$) used in the original truncated series (\ref{danb}) from which the resummation is constructed, while the diagonal Pad\'e gives a result which does depend on the original RScl (is independent of RScl only at one-loop level precision, \cite{GardiPA}). In the case of truncated series (\ref{danb}), such resummed result can be written as
\be
d(Q^2;D=0)^{[4]}_{\rm res} = 
\tal_1 \; \A(\kappa_1 Q^2)+ (1-\tal_1) \; \A(\kappa_2 Q^2)  \ ,
\label{dBGan22}
\ee
where it can be shown that $d(Q^2;D=0)_{\rm res} - d(Q^2,\mu^2;D=0)^{[4]}_{\rm an} = {\cal O}(\A_5)$.
Here, $\tal_1$, $\kappa_1$, $\kappa_2$ are in general complex parameters, constructed directly from the  perturbation expansion coefficients $d_j(\kappa)$ ($j=1,2,3$) of the TPS (\ref{danb}), and turn out to be completely independent of the original RScl $\mu^2=\kappa Q^2$ used. In the four-loop $N_f=3$ Lambert-MiniMOM scheme, cf.~Eqs.~(\ref{c2c3}), the values of these parameters are $\tal_1=0.5 - 0.93404 i$, $\kappa_1=0.96904 - 0.43577 i$, $\kappa_2=0.96904 + 0.43577 i$.  Formally, the expressions (\ref{dBGan22}) and (\ref{danc}) differ from each other by $\sim \A_5$ ($\sim a^5$), as they should.\footnote{
  If the known TPS has $2 M$ terms, $d(Q^2;D=0)_{\rm pt}^{[2 M]}$, an analogous construction gives the resummed expression $d(Q^2;D=0)^{[2 M]}_{\rm res} = \sum_{j=1}^M \tal_j \A(\kappa_j Q^2)$ where $\sum_{j=1}^M \tal_j =1$.}
The application of the methods (\ref{danc}) and (\ref{dBGan22}) give very similar results in the considered scheme, leading to almost the same value for $r^{(D=0)}_{\tau, {\rm (th)}}$. We prefer to use the resummed expression (\ref{dBGan22}), both for practical reasons [it takes less time for computer evaluation than the truncated series (\ref{danc}) or (\ref{danb})] and for theoretical reasons which will be explained briefly in the next paragraph.

Namely, in Ref.~\cite{BGApQCD1} it was shown that the result (\ref{dBGan22}), and its extension $d_{\rm res}^{[2 M]}$ to the case of $2 M$ TPS terms, has no dependence on the RScl, the property clearly shared by the true (unknown) result. In Ref.~\cite{BGApQCD2} the approach was extended to truncated series with an uneven number of terms. The method originally did not work well, because it was used in pQCD (and in $\MSbar$ scheme) where the problem appeared with the evaluation of terms $\alpha_s(\kappa_j Q^2)$ with $|\kappa_j| \ll 1$, due to vicinity of Landau singularities in such expressions and the consequent impossibility of a reliable evaluation. In Refs.~\cite{BGA,anOPE} this method was brought back by applying it in QCD versions with holomorphic coupling, where it turned out to work remarkably well, basically due to the absence of Landau singularities (cuts and poles). Further, in Refs.~\cite{BGA,Techn} it was demonstrated that, when the QCD coupling is holomorphic (for $Q^2 \in \mathbb{C} \backslash (-\infty, -M_{\rm thr}^2]$) this approach gives a convergent sequence when the number of terms in the perturbation series increases. This intriguing property effectively eliminates the renormalon ambiguity problem in such (holomorphic) frameworks.\footnote{\label{ftresumm} For example, in the large-$\beta_0$ approximation, the entire (asymptotically divergent) series of the Adler function $d(Q^2)$ is known. If the truncated series are resummed by the mentioned approach $d_{\rm res}^{[2 M]}(Q^2)$, with any holomorphic QCD coupling, the resulting sequence converges \cite{BGA,Techn} as the number of terms increases, it converges for any spacelike $Q^2$, and it converges to the known large-$\beta_0$ result \cite{Neubert}. The latter is an integral containing in the integrand the coupling  $\A(t Q^2)$ ($0 \leq t < + \infty$). This integral is well defined precisely because the coupling is holomorphic (in $\MSbar$ pQCD it is ill-defined). The large-$\beta_0$ perturbation coefficients contain the main information about the renormalons of the observable $d(Q^2)$. Therefore, the mentioned resummation, with holomorphic coupling, effectively eliminates the renormalon ambiguity problem.}

For some other evaluations/resummations of the Adler function for complex $Q^2$, in pQCD approaches, we refer to Refs.~\cite{Capr,BenekeJamin}.

 \begin{table}
   \caption{The seven dimensionless parameters of the coupling $\A(Q^2)$: $s_j \equiv M_j^2/\Lambda^2$ ($j=0,1,2,3$); $f_j \equiv {\cal F}_j/\Lambda^2$ ($j=1,2,3$), for various representative cases: $\alpha_s(M_Z^2;\MSbar) = 0.1185$ with  $r^{(D=0)}_{\tau, {\rm th}}=0.201$ and $0.201 \pm 0.002$;  $\alpha_s(M_Z^2;\MSbar) = 0.1185 \pm 0.004$, and with $r^{(D=0)}_{\tau, {\rm th}}=0.201$ and $0.203$.
Included is the Lambert scale $\Lambda_L$ [cf.~Eq.~(\ref{LambdaL})], the location $Q^2_{\rm max}$ of the maximum of the coupling, and the value of the coupling at its maximum. The case of $\alpha_s(M_Z^2;\MSbar) = 0.1189$ with  $r^{(D=0)}_{\tau, {\rm th}}=0.201$ has somewhat different parameters [see the text around Eq.~(\ref{rhoAalt}) for details].}
\label{tabres}  
\begin{ruledtabular}
\begin{tabular}{lll|rrrrrrrrr}
${\overline \alpha}_s(M_Z^2)$ &  $s_0$ & $s_1$  & $s_2$ & $s_3$ & $f_1$ & $f_2$ & $f_3$  & $\Lambda_L$ [GeV] & $r^{(D=0)}_{\tau, {\rm th}}$ & $Q^2_{\rm max}$ $[{\rm GeV}^2]$ & $\pi \A(Q^2_{\rm max})$
  \\
  \hline
  $0.1185$ & $652.$ & $3.97$ & $18.495$ & $474.20$ & $-2.8603$ & $11.801$ & $5.2543$ & $0.11564$ & $0.2010$ & $0.1348$ & $0.9156$ 
  \\
  $0.1185$ & $692.$ & $2.93$ & $24.613$ & $504.52$ & $-1.5153$ & $10.724$ & $5.4432$ & $0.11564$ & $0.2030$ & $0.1349$ & $0.8600$ 
  \\
  $0.1185$ & $614.$ & $7.00$ & $10.661$ & $445.40$ & $-18.298$ & $26.978$ & $5.0727$ & $0.11564$ & $0.1990$ & $0.1345$ & $0.9816$ 
  \\
  \hline
  $0.1181$ & $740.$ & $2.40$ & $31.205$ & $540.88$ & $-1.0025$ & $10.526$ & $5.6674$ & $0.11360$ & $0.2010$ & $0.1343$ & $0.8003$ 
  \\
  $0.1181$ & $788.$ & $1.95$ & $37.805$ & $577.27$ & $-0.6980$ & $10.530$ & $5.8878$ & $0.11360$ & $0.2030$ & $0.1346$ & $0.7573$
  \\
  \hline
  $0.1189$ & $612$ & $6.21$ & $11.271$ & $443.90$ & $-11.791$ & $20.458$ & $5.0623$ & $0.11771$ & $0.2030$ & $0.1346$ & $0.9977$
  \\
  \hline
   \hline
 ${\overline \alpha}_s(M_Z^2)$ & $s_0$ & $s_1$ & $s_2$ & $f_1$ & $f_1^{(1)}$ &  $f_1^{(2)}$ & $f_2$ & $\Lambda_L$ [GeV] & $r^{(D=0)}_{\tau, {\rm th}}$ & $Q^2_{\rm max}$ $[{\rm GeV}^2]$ & $\pi \A(Q^2_{\rm max})$
  \\
  \hline 
 $0.1189$ & $577.1$ & $11.217$ & $417.29$ & $8.4175$ & $-20.192$ & $-475.92$ & $4.8891$ & $0.11771$ & $0.2010$ & $0.1348$ & $1.0567$
\end{tabular}
\end{ruledtabular}
\end{table}
Taking into account the seven conditions, we obtain the results presented in Table \ref{tabres}.
They are given for three different values $\alpha_s(M_Z^2; \MSbar)=0.1185$ and $0.1185 \pm 0.004$ (cf.~footnote \ref{PDGfootnote}) and for two different values of $r^{(D=0)}_{\tau, {\rm th}}=0.201, 0.203$. In the case of  $\alpha_s(M_Z^2; \MSbar)=0.1185$ we included three cases, $r^{(D=0)}_{\tau, {\rm th}}=0.201$ and $0.201 \pm 0.002$. We will consider henceforth as the central case the first line in Table \ref{tabres}, i.e., with $\alpha_s(M_Z^2; \MSbar)=0.1185$ and $r^{(D=0)}_{\tau, {\rm th}}=0.201$. The seven scale parameters are written in dimensionless form
\be
s_j \equiv \frac{M_j^2}{\Lambda_L^2}   \quad (j=0,1,2,3) \ ,
\qquad
f_k \equiv \frac{{\cal F}_k}{\Lambda_L^2}   \quad (k=1,2,3) \ .
\label{not}
\ee
In practice, the results of Table \ref{tabres} were obtained by first expressing the five parameters $s_j$ ($j=2,3$) and $f_k$ ($k=1,2,3$) in terms of the parameters $s_0$ and $s_1$, using the five conditions (\ref{cons}). Then, the two remaining parameters $s_0$ and $s_1$ are varied so that the maximum of the resulting $\A(Q^2)$ is reached at $Q^2_{\rm max} \approx 0.135 \ {\rm GeV}^2$ (as suggested by lattice results Fig.~\ref{FigAlatt}) and the resulting quantity $r^{(D=0)}_{\tau, {\rm th}}$, Eq.~(\ref{rtaucont}), acquires the value $0.201$ (or $0.203$, or $0.199$).

The values of $r_{\tau, {\rm th}}^{(D=0)}$ given in Table \ref{tabres} were obtained when applying the resummed expression (\ref{dBGan22}) to $d(Q^2;D=0)$ in Eq.~(\ref{rtaucont}). When applying, instead,  the TPS approach (\ref{danc}) [i.e., Eq.~(\ref{danb}) with $\kappa=1$] to $d(Q^2;D=0)$, the results differ insignificantly, by $0.0001$ or less, in all the cases of Table \ref{tabres}. The obtained truncated series for $r_{\tau, {\rm th}}^{(D=0)}$ has good convergence. For example, in the first case of Table \ref{tabres}, we have $r_{\tau,{\rm th}}(d^{[4]}_{\rm an})=0.158+0.054-0.010-0.001=0.201$. We recall that all the evaluations in the $\A$-coupling framework are performed in the  $N_f=3$ Lambert-MiniMOM renormalization scheme, cf.~Eq.~(\ref{c2c3}) and the usual $\Lambda_{\MSbar}$-scaling.

The last line in Table \ref{tabres}, for $\alpha_s(M_Z^2; \MSbar)=0.1189$ and $r_{\tau, {\rm th}}^{(D=0)}=0.201$, is the case with the highest value of $\pi \A(Q^2_{\rm max}) \approx 1.06$. In this case, in order to obtain simultaneously the value  $r_{\tau, {\rm th}}^{(D=0)}=0.201$ (suggested by $\tau$ decay physics) and $Q^2_{\rm max} \approx 0.135 \ {\rm GeV}^2$ (suggested by $\A_{\rm latt.}$), it turns out that two of the three delta functions, at low $\sigma$, appear practically at the same place, and this limiting case ($M_2^2 \to M_1^2$) can be equivalently described by a combination of $\delta(\sigma - M_1^2)$ and its first and second derivative. The same situation was encountered by us, already for lower values of $\alpha_s(M_Z^2; \MSbar)$, in the scheme which agrees with the Lambert-MiniMOM only to three loops, \cite{AQCDprev}. Specifically, in such a case we have
\be
\frac{1}{\pi} \rho_{\A}(\sigma) = \sum_{j=1}^2 {\cal F}_j \delta(\sigma - M_j^2) +{\cal F}_1^{(1)} \delta'(\sigma - M_1^2) +{\cal F}_1^{(2)} \delta^{\prime \prime}(\sigma - M_1^2) + \Theta(\sigma - M_0^2) \rho_a(\sigma).
\label{rhoAalt}
\ee
The new dimensionless parameters appearing for the last line in Table \ref{tabres} are $f_1^{(1)}={\cal F}_1^{(1)}/\Lambda_L^4$ and $f_1^{(2)}={\cal F}_1^{(2)}/\Lambda_L^6$. For details on such form of $\rho_{\A}(\sigma)$ and the corresponding $\A(Q^2)$, we refer to Ref.~\cite{AQCDprev}.

The obtained solutions of the coupling $\A(Q^2)$ are holomorphic in the complex $Q^2$-plane with the exception of the points $Q^2=-M_1^2, -M_2^2, -M_3^2$ and of the continuous cut $(-M_0^2,-\infty)$. We can regard that the cut starts at the point $Q^2=-M_1^2$, which is here approximated as the (closest to the origin) singularity point of $\A(Q^2)$ in the complex $Q^2$-plane. This means that in the dispersion integral (\ref{Adisp}) the squared threshold mass is $M_{\rm thr}^2=M_1^2$, i.e., the threshold mass is $M_{\rm thr}=M_1=\Lambda_L \sqrt{s_1}$. We note that this threshold mass, in all seven solutions in Table \ref{tabres}, lies in the interval $m_{\pi} < M_{\rm thr.} < 3 m_{\pi}$. This turns out to be close to the $e^+e^- \to$ hadrons production threshold $2 m_{\pi}$, cf.~Eq.~(\ref{DVm}) in Sec.~\ref{sec:pred} and the related Appendix \ref{app:mAQCD}. However, since our coupling $\A(Q^2)$ is considered to be a universal coupling, in that sense similar to the \textcolor{black}{underlying pQCD coupling $a(Q^2)$ or $\MSbar$ pQCD coupling $a(Q^2;\MSbar)$} (but constructed to describe better the regime $|Q^2| \sim 1 \ {\rm GeV}^2$),\footnote{\textcolor{black}{The term universal coupling is used here to mean a coupling that does not describe a specific observable as an effective charge (ECH), neither in the perturbative sense \cite{Grunberg} nor in the more general nonperturbative sense \cite{DBCK,MSS1,MSS2,MagrGl,mes2,DeRafael,MagrTau,Nest3a,Nest3b,NestBook}. Nonetheless, the coupling is in a specific MOM scheme called the MiniMOM (MM) \cite{MiniMOM}, in which the renormalized gluon and ghost dressing functions $Z_{\rm gl}(Q^2)$ and $Z_{\rm gh}(Q^2)$ are equal to one at a (large) renormalization scale $Q^2=\Lambda^2$, and the ghost-gluon vertex function is ${\widetilde Z}_1({\rm MM}) = {\widetilde Z}_1({\MSbar})=1$. In MOM schemes, such choices cannot be made simultaneously for all the dressing functions \cite{CelGon,DSE4glMM}.}}
  it is not a physical observable, it has renormalization scheme dependence, and it is not expected to have the cut coinciding with the cut of a specific physical observable. The coupling $\A(Q^2)$ has, however, the holomorphic properties in the complex $Q^2$-plane qualitatively similar to those of all spacelike QCD observables, in contrast to the (non)holomorphic properties of the usual pQCD couplings $a(Q^2)$ (Landau singularities). \textcolor{black}{We point out that the holomorphic behavior of the obtained $\A(Q^2)$ is the consequence of the fact that we obtained, from the mentioned seven conditions, solutions where all $M_j^2$ parameters turned out to have positive values, something that could not be predicted a priori.}

\textcolor{black}{The masses $M_j = \Lambda_{L} \sqrt{s_j}$, for the case of the first line of Table \ref{tabres}, are $0.230$ GeV, $0.497$ GeV, and $2.518$ GeV for $j=1,2,3$, respectively; and the pQCD-onset scale is $M_0 =2.953$ GeV. As mentioned, these masses do not have any direct physical meaning, as the coupling $\A(Q^2)$ is not an observable. They resulted from applying the coupling $\A(Q^2)$, with the $[2/3](Q^2)$ Pad\'e ansatz for $\Delta \A_{\rm IR}(Q^2)$ in Eq.~(\ref{ADA}), to (five) physically-motivated conditions at high and intermediate $Q^2$, and to (two) lattice-motivated conditions at low $Q^2$. Stated otherwise, a specific mathematical ansatz for the unknown part of the coupling $\A(Q^2)$, applied to the seven conditions, gave us these scales. One can regard the resulting terms in the coupling, ${\cal F}_j/(Q^2+M_j^2)$, as certain modified higher-twist terms. On the other hand, the terms resembling the effects from the gluon bremsstrahlung are not present in the coupling.}

Later in Sec.~\ref{sec:BSR} and Sec.~\ref{sec:pred} we will show that the coupling $\A$ in conjunction with OPE describes the physics in the regime $|Q^2| \sim 1 \ {\rm GeV}^2$ better than the usual ($\MSbar$) pQCD+OPE approach. In fact, this is the main purpose of the construction of such a coupling $\A$. The OPE method, unfortunately, cannot be used in the regime $|Q^2| < 1 \ {\rm GeV}^2$ because the OPE there becomes strongly divergent. \textcolor{black}{Further, the application of the $\A$-coupling framework to the (soft and collinear) gluon bremsstrahlung remains an outstanding problem to be addressed in a future; the $Q^2$-dependence in such processes is different from the simple inverse power terms.}

\begin{figure}[htb] 
\begin{minipage}[b]{.49\linewidth}
  \centering\includegraphics[width=85mm]{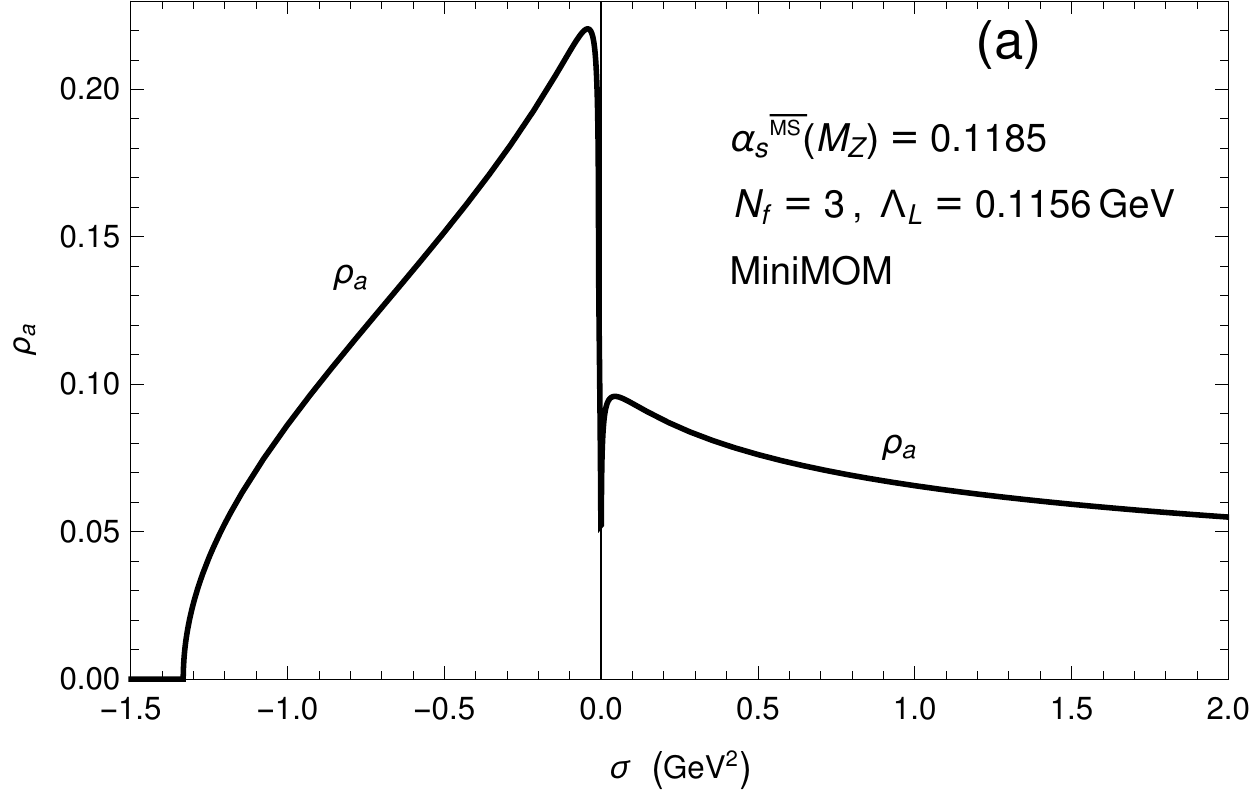}
  \end{minipage}
\begin{minipage}[b]{.49\linewidth}
  \centering\includegraphics[width=85mm]{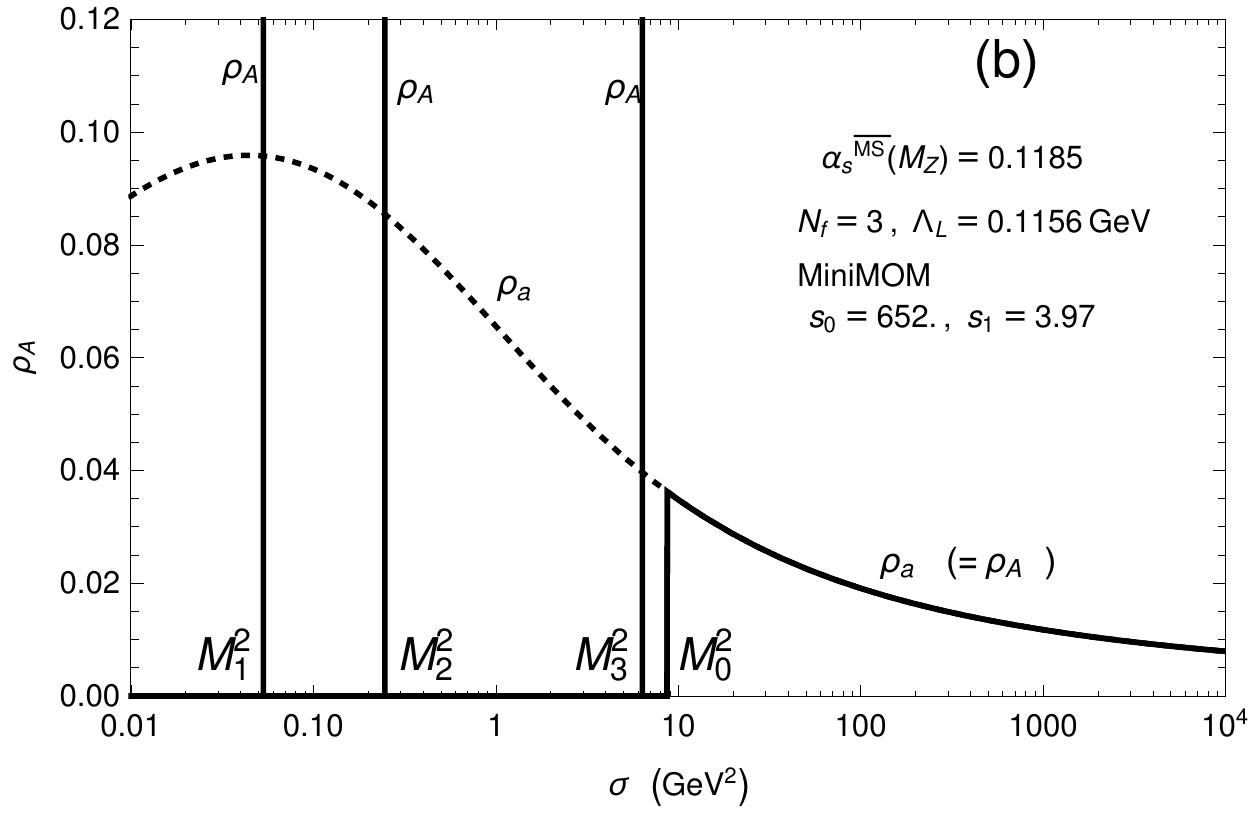}
\end{minipage}
\caption{\footnotesize  (a) The spectral (discontinuity) function $\rho_a(\sigma) = {\rm Im} \; a(Q^2=-\sigma - i \epsilon)$ for the underlying pQCD coupling in the four-loop Lambert-MiniMOM scheme, and $\sigma$ is on linear scale; (b) the spectral function $\rho_{\A}(\sigma) =  {\rm Im} \; \A(Q^2=-\sigma - i \epsilon)$ of the considered holomorphic coupling $\A(Q^2)$, where $\sigma > 0$ is on logarithmic scale. The $\delta$-spike at $M_1^2$ is in fact negative (cf.~Table \ref{tabres}), but is presented here in Fig.~(b) for simplicity as a positive spike.}
\label{Figrho}
\end{figure}
In Fig.~\ref{Figrho}(a), we present, for the case of $\alpha_s(M_Z^2; \MSbar)=0.1185$, the discontinuity function $\rho_a(\sigma) = {\rm Im} \; a(Q^2=-\sigma - i \epsilon)$ for the underlying pQCD coupling in the Lambert-MiniMOM scheme, cf.~Eqs.~(\ref{beta})-(\ref{c2c3}). We recall that this scheme agrees up to (and including) the four-loop level with the lattice MiniMOM scheme of Ref.~\cite{LattcoupNf0} (the MiniMOM scheme is at present known only to four-loop level \cite{MiniMOM}), except for the scale convention which in the Lambert-MiniMOM is the usual $\MSbar$-type, cf.~Eq.~(\ref{LambdaMMMSbar}). The branching point of this $a(Q^2)$ is at $Q^2_{\rm br} \approx 99.64 \times \Lambda_L^2 \approx 1.3326 \ {\rm GeV}^2$, i.e., $\sigma_{\rm min} = - Q^2_{\rm br} \approx -1.3326 \ {\rm GeV}^2$ is the minimal $\sigma$ for which   $\rho_a(\sigma)$ is nonzero. In Fig.~\ref{Figrho}(b), on the other hand, we present $\rho_{\A}(\sigma) =  {\rm Im} \; \A(Q^2=-\sigma - i \epsilon)$, Eq.~(\ref{rhoA}), i.e., where there is no Landau cut ($-Q^2_{\rm br} < \sigma < 0$) present any more, and in the low-$\sigma$  regime ($0 < \sigma < M_0$) we find the described parametrization of $\rho_{\A}(\sigma)$ by three delta functions.

\begin{figure}[htb] 
\begin{minipage}[b]{.49\linewidth}
  \centering\includegraphics[width=85mm]{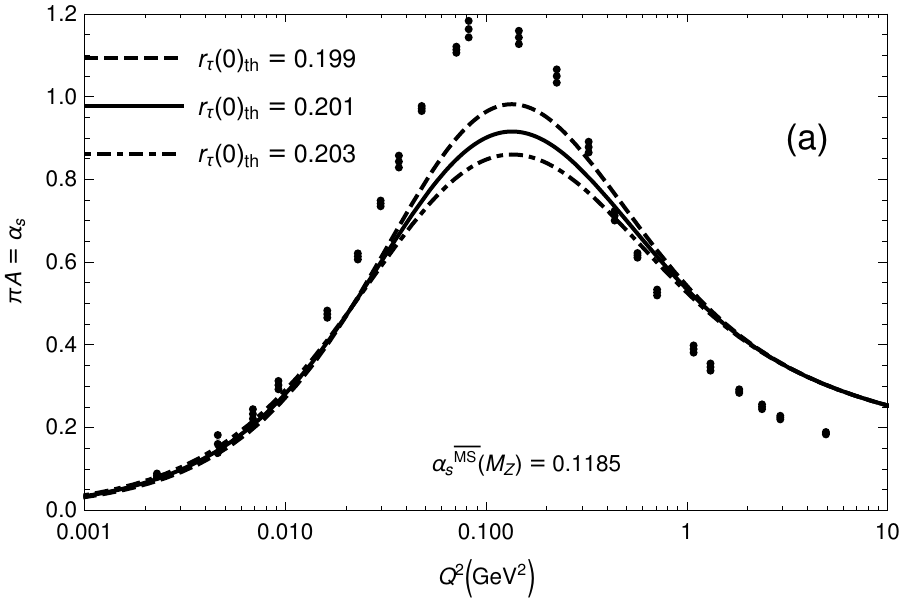}
\end{minipage}
\begin{minipage}[b]{.49\linewidth}
  \centering\includegraphics[width=85mm]{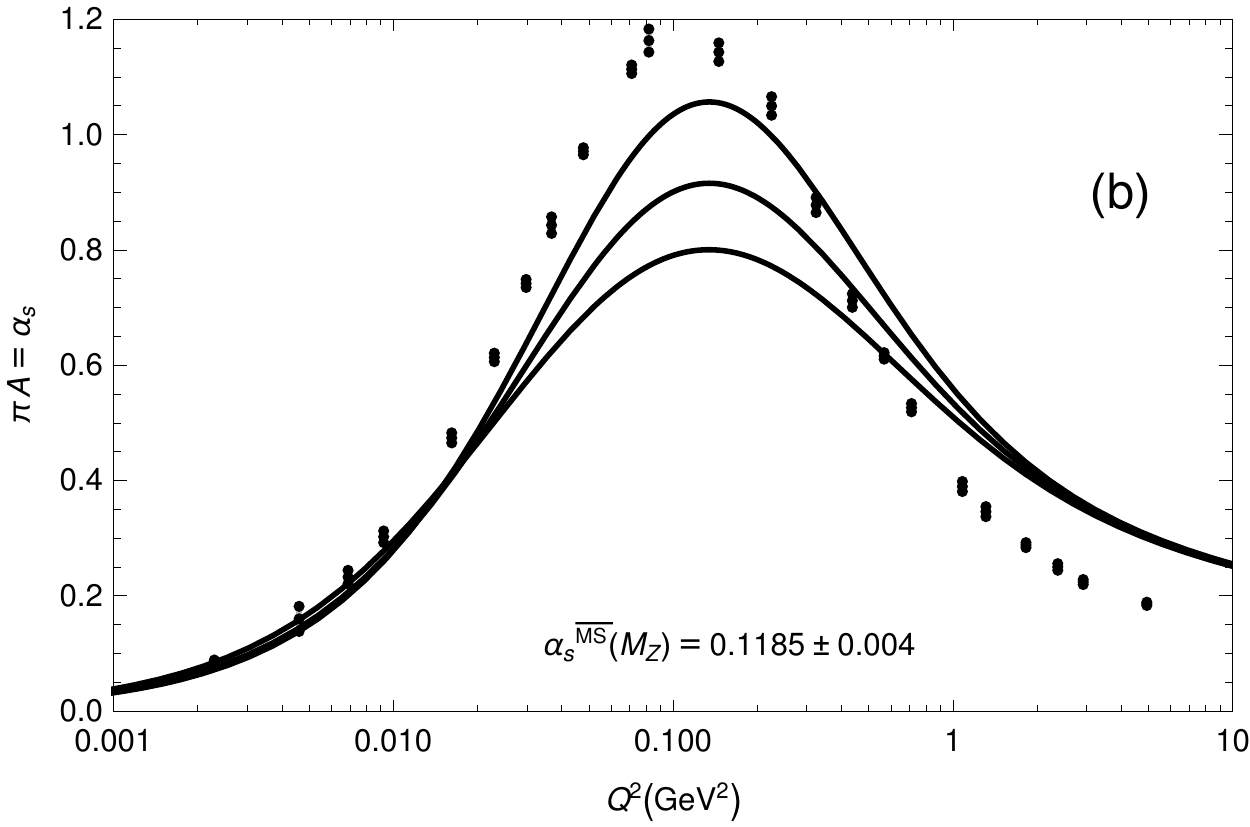}
\end{minipage}
\caption{\footnotesize  (a) The constructed $N_f=3$ coupling $\pi \A(Q^2)$ at low positive $Q^2$ for the case of $\alpha_s(M_Z^2; \MSbar)=0.1185$ with $r^{(D=0)}_{\tau, {\rm th}}=0.201$ and $0.201 \pm 0.002$. (b) The same, but now for the case of $r^{(D=0)}_{\tau, {\rm th}}=0.201$ with  $\alpha_s(M_Z^2; \MSbar)=0.1185$ and $0.1185 \pm 0.004$; the highest solid line corresponds to the value $\alpha_s(M_Z^2; \MSbar)=0.1189$, the lowest to $0.1181$. For comparison, the lattice $N_f=0$ coupling $\pi \A_{\rm latt.}(Q^2)$ of Ref.~\cite{LattcoupNf0} is included (as points), in both Figures, with $Q^2$ rescaled to the usual $\Lambda_{\MSbar}$ scale convention.}  
\label{FigComb}
\end{figure}
In Fig.~\ref{FigComb} we present the resulting coupling $\pi \A(Q^2)$ at low positive $Q^2$, for several mentioned cases. In general, we can see that increasing the value of $r^{(D=0)}_{\tau, {\rm th}}$, Eq.~(\ref{rtaucont}), tends to decrease the bump of the curve around its maximum, and the same effect occurs when $\alpha_s(M_Z^2;\MSbar)$ decreases.
We can further see that the running coupling $\pi \A(Q^2)$ in general agrees well with $\pi \A_{\rm latt.}(Q^2)$ at very low $Q^2$ ($Q \lesssim 0.01 \ {\rm GeV}^2$), and is lower than the lattice coupling near the maxima ($Q^2 \sim 0.1 \ {\rm GeV}^2$). We recall that we do not expect to have a good agreement between the theoretical and lattice coupling at $Q^2 \lesssim 0.1 \ {\rm GeV}^2$, but only a qualitative agreement, as argued earlier in Sec.~\ref{sec:latt}. At higher $Q^2$ ($Q^2 > 1 \ {\rm GeV}^2$), there is disagreement between $\pi \A(Q^2)$ and $\pi \A_{\rm latt.}(Q^2)$. One reason for this is that the theoretical curves have the number of active quark flavors $N_f=3$ while the lattice results \cite{LattcoupNf0} are for $N_f=0$. In fact, increasing $N_f$ in general decreases $\A_{\rm latt.}(Q^2)$, cf.~Fig.~5 of Ref.~\cite{LattcoupNf2}. However, the principal reason for the difference between the theoretical and lattice $\A(Q^2)$ for $Q^2 \sim 10^0$-$10^1 \ {\rm GeV}^2$ lies in the following: the lattice results here in Figs.~\ref{FigAlatt} and \ref{FigComb}, from Ref.~\cite{LattcoupNf0}, are close to the continuum limit only for the deep IR regime $Q^2 < 1 \ {\rm GeV}^2$, but for higher $Q^2$ these results have so called hypercubic lattice artifacts \cite{SternbeckComm}, because the lattice is coarse ($\beta = 5.7$, lattice spacing $a \approx 0.17$ fm). The authors of \cite{LattcoupNf0} concentrated on the deep IR regime, i.e., they had large lattice volume ($L \sim 10$ fm), but not small lattice spacing.

\begin{figure}[htb] 
\centering\includegraphics[width=110mm]{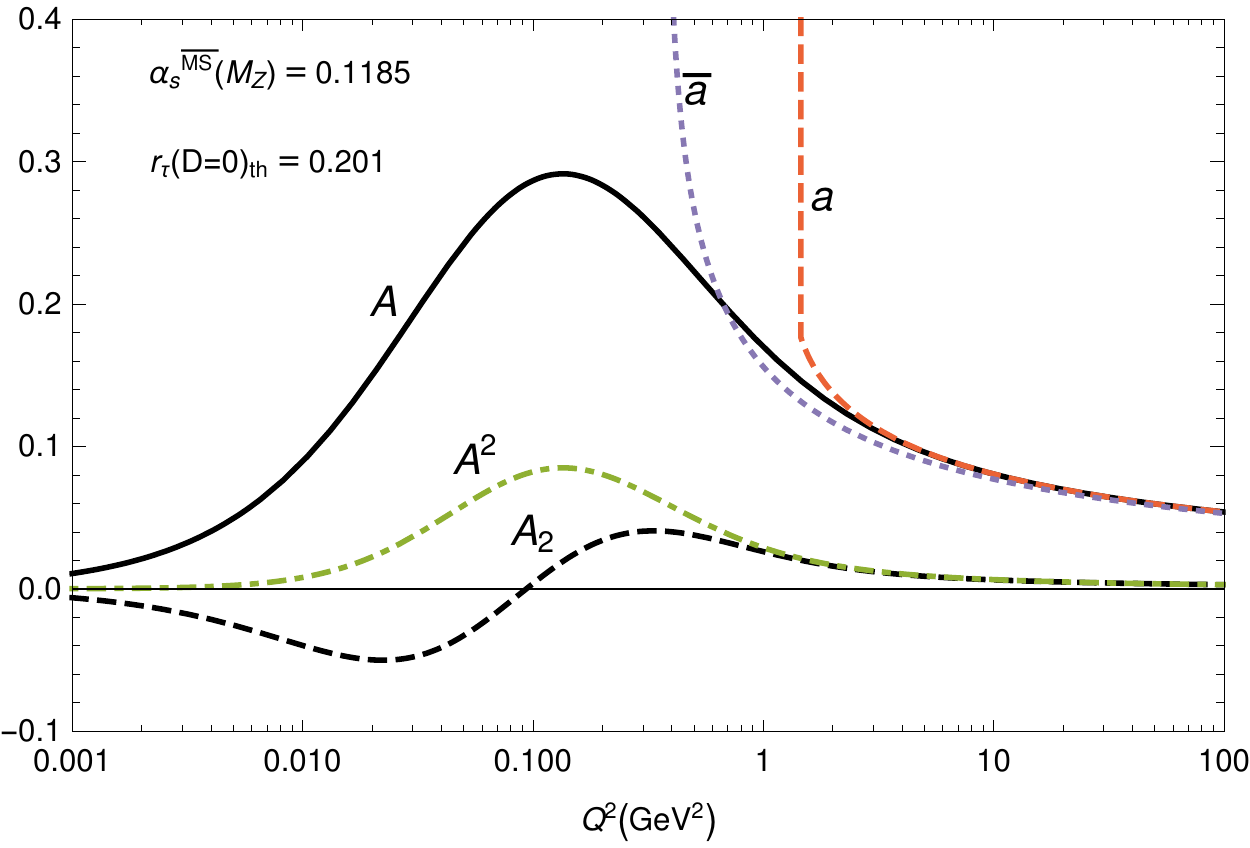}
\caption{\footnotesize  The considered holomorphic coupling $\A$ at positive $Q^2$ (solid curve) and its underlying pQCD coupling $a$ (light dashed curve). Included is $\A_2$ (dashed curve) which is the $\A$-analog of power $a^2$ [cf.~Eq.~(\ref{A2})], and the naive (i.e., unusable) power $\A^2$ (dot-dashed curve). Further, the usual $\MSbar$ coupling ${\overline a}$ (dotted curve) is included.}
\label{FigAa}
 \end{figure}

  Further, in Fig.~\ref{FigAa} the comparison of the coupling $\pi \A(Q^2)$ with its underlying pQCD coupling $\pi a(Q^2)$ is presented at positive $Q^2$, for the case $\alpha_s(M_Z^2; \MSbar)=0.1185$ and $r^{(D=0)}_{\tau, {\rm th}}=0.201$.  
The two couplings merge fast for $Q^2 > 4 \ {\rm GeV}^2$, in accordance with the condition (\ref{Aadiff2}) [i.e., Eqs.~(\ref{1u})]. For further comparison, the corresponding $\MSbar$ pQCD coupling $\pi {\overline a}(Q^2)$ is included in the Figure as well.

\section{Borel sum rules for semihadronic $\tau$ decays}
\label{sec:BSR}

For further application of the newly constructed coupling $\A(Q^2)$, we will now consider sum rules for Borel transforms in the semihadronic strangeless decays of $\tau$-lepton. These sum rules involve OPE (for inclusive quantities). Consequently we regard the considered coupling $\A(Q^2)$ as universal, in a sense similar to regarding the pQCD coupling (in any chosen scheme) as universal. The expectation values $\langle O_D \rangle$ of the local operators appearing in OPE are also considered to be universal. We are allowed to apply the OPE with $\A$-coupling in a way analogous to the OPE with pQCD $a$-coupling, because of the relation (\ref{Aadiff2}). The latter relation implies that we can include in OPE with $\A$-coupling unambiguously the terms of dimensionality $D < 10$.

\begin{figure}[htb] 
\begin{minipage}[b]{.49\linewidth}
  \centering\includegraphics[width=85mm]{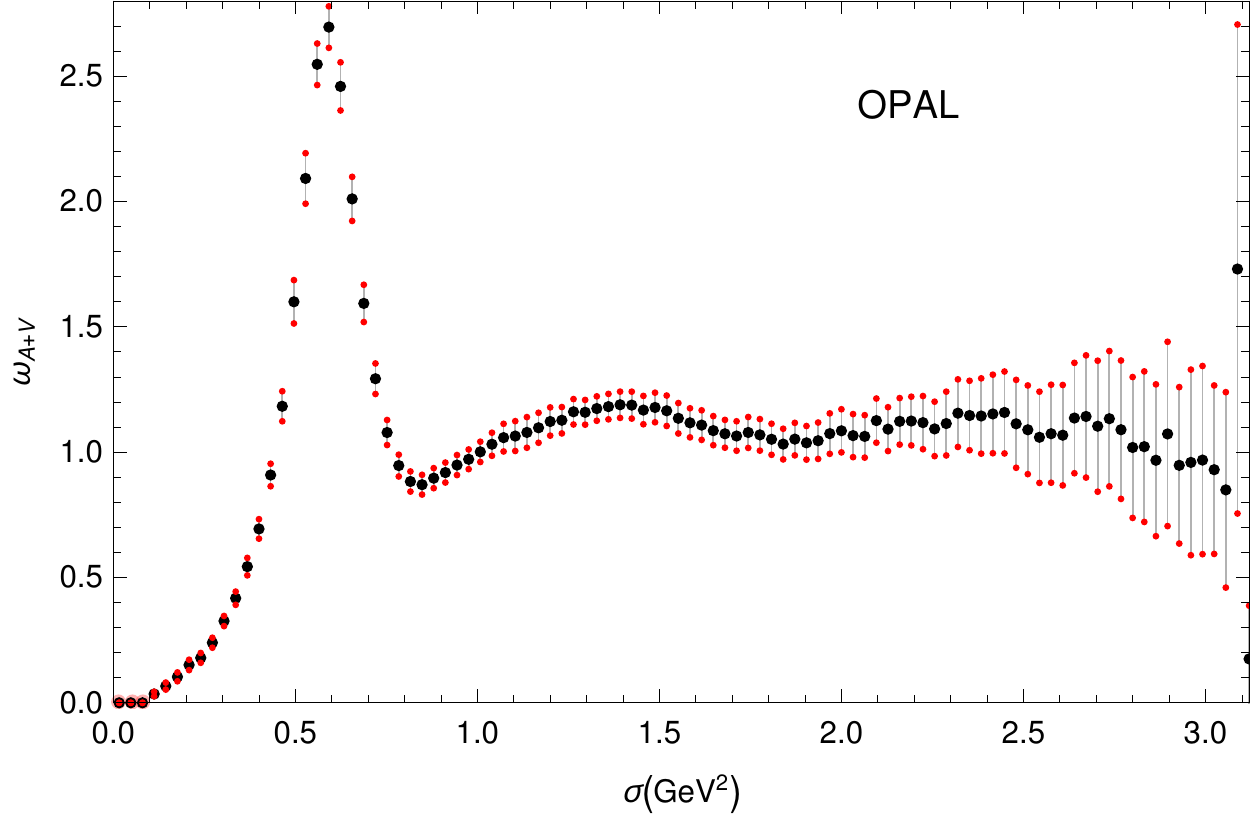}
  \end{minipage}
\begin{minipage}[b]{.49\linewidth}
  \centering\includegraphics[width=85mm]{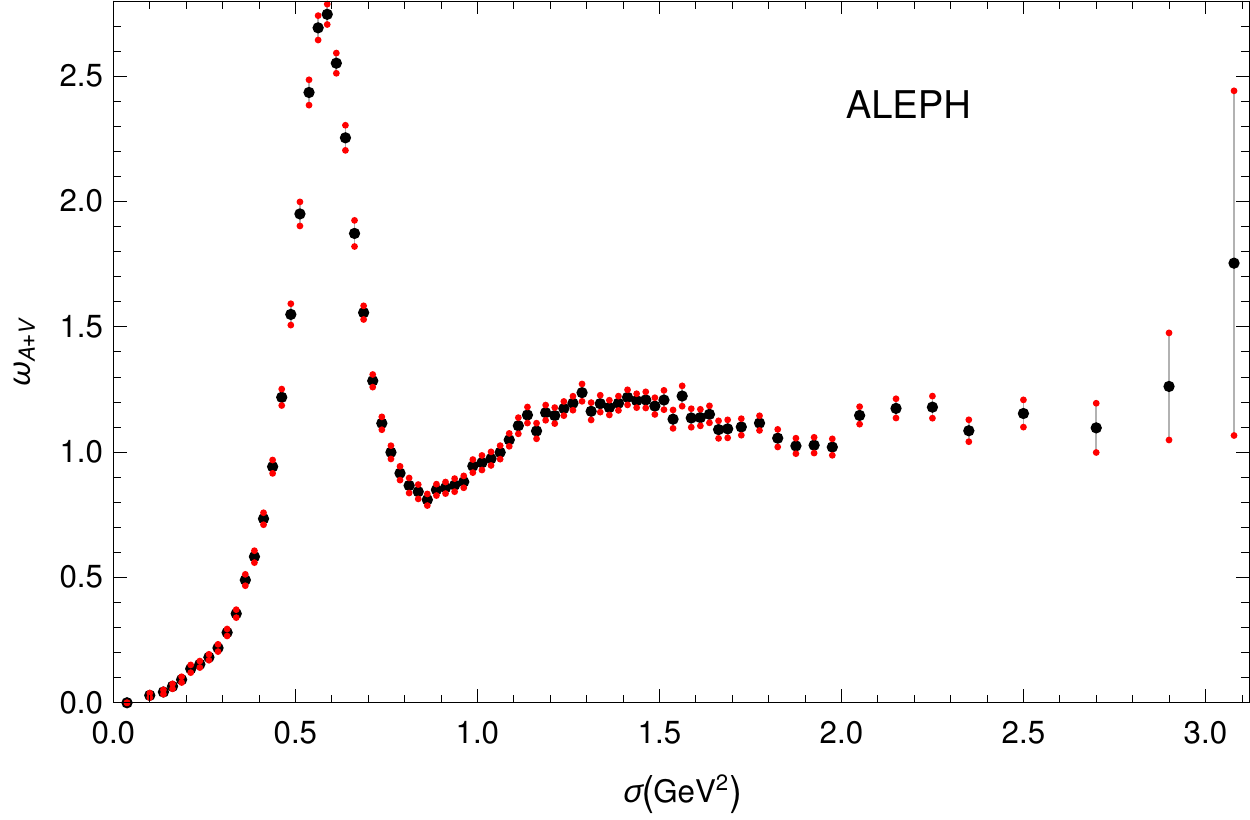}
\end{minipage}
\caption{\footnotesize  (a) The spectral function $\omega_{V+A}(\sigma) \equiv \omega(\sigma)$ es measured by OPAL Collaboration (left-hand figure) and by ALEPH Collaboration (right-hand figure). The pion peak contribution $2 \pi^2 f_{\pi}^2 \delta(\sigma - m_{\pi}^2)$ (where $f_{\pi}=0.1305$ GeV) must be added to this, accounting for the pion contribution.}
\label{FigOmega}
\end{figure}

The sum rules will be applied to the polarization (current correlation) function $\Pi(Q^2)$ of the strangeless vector (V) and axial (A) currents which play a central role in the semihadronic strangeless decays of the $\tau$ lepton
\be
\Pi_{J, \mu\nu}(q) =  i \int  d^4 x \; e^{i q \cdot x} 
\langle T J_{\mu}(x) J_{\nu}(0)^{\dagger} \rangle
=  (q_{\mu} q_{\nu} - g_{\mu \nu} q^2) \Pi_J^{(1)}(Q^2)
+ q_{\mu} q_{\nu} \Pi_J^{(0)}(Q^2) \ ,
\label{PiJ}
\ee
where $Q^2 \equiv -q^2$, $J=V,A$, and the quark currents are $J_{\mu} = {\overline u} \gamma_{\mu} d$ (when J=V), $J_{\mu} = {\overline u} \gamma_{\mu} \gamma_5 d$ (when J=A). We refer for more details to Refs.~\cite{Geshkenbein,Ioffe}. We will apply sum rules to the total semihadronic decay width (V+A); hence the polarization function is
\be
\Pi_{V+A}(Q^2) = \Pi^{(1)}_{V}(Q^2) +  \Pi^{(1)}_{A}(Q^2) + \Pi^{(0)}_{A}(Q^2) + \Pi^{(0)}_{V}(Q^2) \ ,
\label{Pi1}
\ee
where we will neglect the last term $\Pi^{(0)}_{V}(Q^2)$, because ${\rm Im} \Pi^{(0)}_{V}(-\sigma - i \epsilon) \propto (m_d-m_u)^2$. In the present analysis, the corrections ${\cal O}(m_{u,d}^2)$ and ${\cal O}(m_{u,d}^4)$ are considered numerically negligible and are not included, cf.~Refs.~\cite{Boitoetal} and \cite{Geshkenbein,Ioffe}. On the other hand, we will not neglect the numerically more important chirality-violating effects proportional to $(m_u+m_d) \langle {\bar q} q \rangle = - f_{\pi}^2 m_{\pi}^2/2$ [cf.~Eq.~(\ref{O4def})].
The correlator $\Pi_{V+A}(Q^2)$ is a spacelike physical quantity (it has RScl-dependence, which however disappears when applying derivative), and by the general principles of Quantum Field Theories it is a holomorphic (analytic) function in the complex $Q^2$-plane, for $Q^2 \in \mathbb{C} \backslash (-\infty, -M_{\rm thr}^2]$ where the hadron production threshold mass is $M_{\rm thr} \sim 0.1$ GeV. This quantity is then multiplied by any function $g(Q^2)$ analytic in the entire complex $Q^2$-plane, 
and the Cauchy integral formula can be applied to the integral of $g(Q^2) \Pi_{V+A}(Q^2)$ along the contour of Fig.~\ref{Figintpath}(b), where the radius $\sigma_{\rm max}$ of the circular path is in this case taken to be finite ($\sigma_{\rm max} \leq m_{\tau}^2$). This leads to the following sum rule:
  \be
\int_0^{\sigma_{\rm max}} d \sigma g(-\sigma) \omega_{\rm exp}(\sigma)  =
-i \pi  \oint_{|Q^2|=\sigma_{\rm max}} d Q^2 g(Q^2) \Pi_{V+A, {\rm th}}(Q^2)  \ ,
\label{sr1}
\ee
where $\omega(\sigma)$ is the spectral (discontinuity) function of $\Pi_{V+A}(Q^2)$ along the cut
\be
\omega(\sigma) \equiv 2 \pi \; {\rm Im} \ \Pi_{V+A}(Q^2=-\sigma - i \epsilon) \ ,
\label{om1}
\ee
The sum rule is then specified by the choice of $g(Q^2)$. The spectral functions $\omega_V(\sigma)$ and $\omega_A(\sigma)$ were measured in semihadronic strangeless $\tau$-lepton decays by the OPAL  \cite{OPAL,PerisPC1,PerisPC2} and ALEPH Collaboration \cite{ALEPH2,DDHMZ,ALEPHfin,ALEPHwww}. The resulting values for $\omega(\sigma) \equiv \omega_{V+A}(\sigma)$ are presented in Figs.~\ref{FigOmega}(a), (b) of the OPAL and ALEPH, correspondingly.
The integral on the right-hand side of the sum rule (\ref{sr1}) is to be evaluated theoretically. Namely, the correlator function can be evaluated by OPE
\be
\Pi_{V+A,{\rm th}}(Q^2) =  - \frac{1}{2 \pi^2} \ln(Q^2/\mu^2) + 
\Pi_{V+A,{\rm th}}(Q^2;D\!=\!0)
+ \sum_{n \geq 2} \frac{ \langle O_{2n} \rangle_{V+A}}{(Q^2)^n} \left(
1 + {\cal C}_n a(Q^2) \right) \ .
\label{OPE1}
\ee
Here, $\langle O_{2n} \rangle_{V+A}$ are the vacuum expectation values (condensates) of local operators with dimension $D=2 n \geq 4$ appearing in the OPE.
The terms ${\cal C}_n a(Q^2)$ in the Wilson coefficients of the contributions of dimension $D = 2n \geq 4$ turn out to be negligible. The term $D = 2n = 2$ is negligible because $\Pi_{V+A}(Q^2)$ has practically no $m_{u,d} \not= 0$ ($m_{\pi} \not= 0$) effects as mentioned earlier.\footnote{\textcolor{black}{The correlator (\ref{OPE1}) is a spacelike quantity, as is the related Adler function ${\cal D}(Q^2)$. Spacelike quantities do not have the OPE terms $\sim 1/(Q^2)^{1/2}$, $1/(Q^2)^{3/2}$, etc. For timelike physical quantities, on the other hand, such type of higher-twist terms do exist \cite{Renorm}. For one of the earlier works on higher-twist operators we refer to \cite{GrossWilczek}.}}  
In our approach, we will use for the analytic function $g(Q^2)$ the exponential function
\be
 g(Q^2) \equiv g_{M^2}(Q^2) = \frac{1}{M^2} \exp( Q^2/M^2) \ ,
 \label{gQ2}
 \ee
 where $M^2$ is a squared energy scale with complex values and ${\rm Re}(M^2) > 0$. For the choice (\ref{gQ2}), the sum rules are called Borel sum rules, cf.~Refs.~\cite{Geshkenbein,Ioffe}. Usually the right-hand side of Eq.~(\ref{sr1}) is performed by integration by parts. In the Borel case (\ref{gQ2}) this then leads to the following form of the sum rules:
 \bea
\frac{1}{M^2} \int_0^{\sigma_{\rm max}} \!\!\! d \sigma \exp( - \sigma/M^2) \omega_{\rm exp}(\sigma) &=&
- \frac{i}{2 \pi}  \int_{\phi=-\pi}^{\pi} \frac{d Q^2}{Q^2}
{\cal D}(Q^2) \left[ \exp( Q^2/M^2) -  \exp( -\sigma_{\rm max}/M^2) \right] {\big |}_{Q^2 = \sigma_{\rm max} \exp(i \phi)}.
\label{sr2}
\eea
The quantity ${\cal D}(Q^2)$ is the full massless Adler function
\bea
{\cal D}(Q^2) &\equiv&  - 2 \pi^2 \frac{d \Pi_{V+A,{\rm th}}(Q^2)}{d \ln Q^2} 
 =  1 + d(Q^2;D=0) + 2 \pi^2 \sum_{n \geq 2}
 \frac{ n \langle O_{2n} \rangle_{V+A}}{(Q^2)^n}  \ .
\label{Adlfull}
\eea
In this expression, the OPE expansion (\ref{OPE1}) was used and the negligible ${\cal C}_n$ terms were not included. We will use the real part of the Borel sum rule (\ref{sr2}), and this has then the following form:
\be
{\rm Re} B_{\rm exp}(M^2) =  {\rm Re} B_{\rm th}(M^2) \ ,
\label{sr3o0}
\ee
where
\bes
\label{sr3}
\bea
B_{\rm exp}(M^2) &\equiv& \int_0^{\sigma_{\rm max}} 
\frac{d \sigma}{M^2} \; \exp( - \sigma/M^2) \omega_{\rm exp}(\sigma)_{V+A} \ ,
\label{sr3a}
\\
B_{\rm th}(M^2) &\equiv&  \left( 1 - \exp(-\sigma_{\rm max}/M^2) \right)
+ B_{\rm th}(M^2;D\!=\!0) + 2 \pi^2 \sum_{n \geq 2}
 \frac{ \langle O_{2n} \rangle_{V+A}}{ (n-1)! \; (M^2)^n} \ ,
\label{sr3b}
\eea
\ees
where the leading-twist contributions ($D=0$) is
\bea
B_{\rm th}(M^2;D\!=\!0) &=&
\frac{1}{2 \pi}\int_{-\pi}^{\pi}
d \phi \; d(Q^2\!=\!\sigma_{\rm max} e^{i \phi};D=0) \left[ 
\exp \left( \frac{\sigma_{\rm max} e^{i \phi}}{M^2} \right) -
\exp \left( - \frac{\sigma_{\rm max}}{M^2} \right) \right] \ .
\label{BD0}
\eea
The total $D (\equiv 2 n) =2$ contribution in the OPE (\ref{sr3b}) is negligible, and we will include there the $D=4$ and $D=6$ terms.

The $D=4$ term is of particular interest because the condensate $\langle O_4 \rangle_{V+A}$ contains the gluon condensate $\langle (\alpha_s/\pi) G^a_{\mu \nu} G_a^{\mu \nu} \rangle \equiv  \langle a GG \rangle$. Namely, the $D=4$ term in OPE (\ref{sr3b}) has two parts, one from the gluon condensate, and the other from the main chirality-violating effects $m_{\pi} \not= 0$ (cf.~\cite{Braaten})
\bes
\label{O4def}
\bea
\langle O_4 \rangle_{V+A} & =& \frac{1}{6} \langle a GG \rangle + 2 (m_u + m_d) \langle {\bar q} q \rangle
\label{O4defa}
\\
 & =& \frac{1}{6} \langle a GG \rangle - f_{\pi}^2 m_{\pi}^2
= \frac{1}{6} \langle a GG \rangle - 3.31 \times 10^{-4} \ {\rm GeV}^4 \ ,
\label{O4defb}
\eea
\ees
where we denoted $\langle {\bar q} q \rangle \equiv \langle {\bar u} u \rangle = \langle {\bar d} d \rangle$, and neglected corrections of relative order ${\cal O}(a)$. In Eq.~(\ref{O4defb}) we used the PCAC relation \cite{PCAC} with the values $f_{\pi} = 0.1305$ GeV and $m_{\pi} =0.1396$ GeV \cite{PDG2016}. This means that the total gluon condensate is determined by the total $D=4$ condensate via the relation
\be
\langle a GG \rangle  =  6 \langle O_4 \rangle_{V+A} + 6 f_{\pi}^2 m_{\pi}^2
 \approx   6 \langle  O_4 \rangle_{V+A} + 0.00199 \ {\rm GeV}^4 \ .
\label{aGGdef}
\ee

\begin{figure}[htb] 
\begin{minipage}[b]{.49\linewidth}
  \centering\includegraphics[width=85mm]{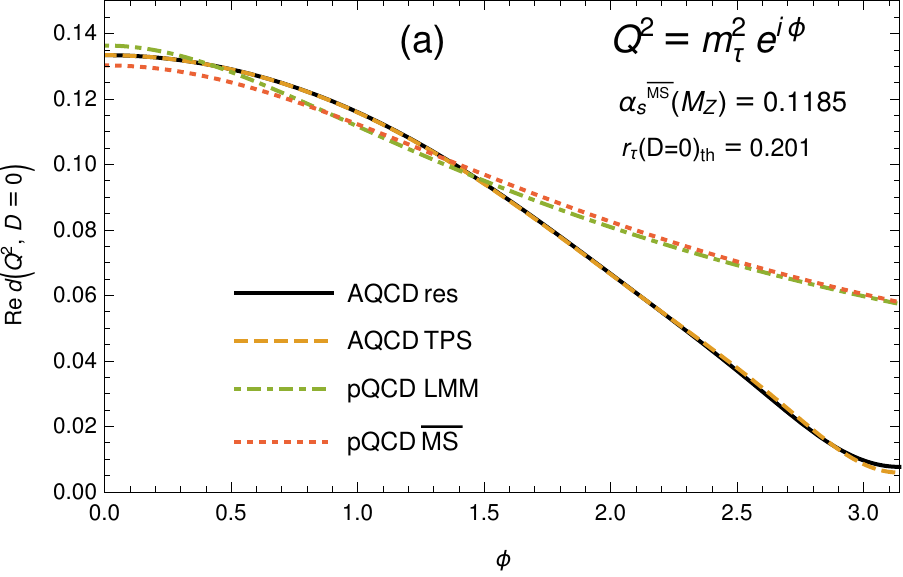}
  \end{minipage}
\begin{minipage}[b]{.49\linewidth}
  \centering\includegraphics[width=85mm]{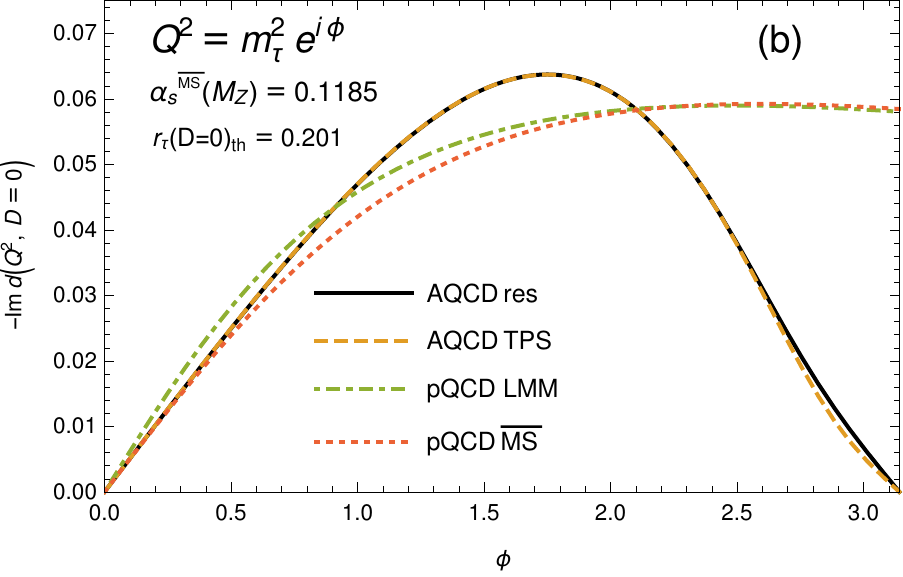}
\end{minipage}
\caption{\footnotesize  (a) The real part of the Adler function $d(Q^2=m_{\tau}^2 e^{i \phi}; D=0)$ as a function of $\phi$, calculated in the $\A$-coupling framework ($\A$QCD), either in the resummed form (\ref{dBGan22}) or in the TPS form (\ref{danc}). Included is also the (TPS) evaluated form for pQCD, in the Lambert-MiniMOM (LMM) scheme, and in $\MSbar$ scheme. The $\A$QCD resummed and $\A$QCD TPS curves practically overlap. (b) The same, but for the negative of the imaginary part; again, the $\A$QCD resummed and $\A$QCD TPS curves practically overlap. }
\label{FigD0}
\end{figure}

The crucial part of the evaluation of the OPE (\ref{sr3b}) is the $D=0$ (leading-twist) massless Adler function $d(Q^2;D=0)$ appearing in Eqs.~(\ref{Adlfull}) and (\ref{BD0}). This quantity will be evaluated with our considered holomorphic coupling $\A(Q^2)$, using either the truncated (TPS) form (\ref{danc}), or the generalization of the diagonal Pad\'e approach, i.e., the resummed form (\ref{dBGan22}). In Fig.~\ref{FigD0} we present the Adler function $d(Q^2=m_{\tau}^2 e^{i \phi}; D=0)$ for these two approaches.
In practice, the presented results will be for the resummed approach (\ref{dBGan22}), although the TPS approach  (\ref{danc}) gives similar results. Namely, the considered values of ${\rm Re} B_{\rm th}(M^2)$ differ in the two approaches by less than $4 \times 10^{-3}$ in the case of $\sigma_{\rm max} = m_{\tau}^2$ (relevant for OPAL), and by less than $3 \times 10^{-3}$ in the case of $\sigma_{\rm max} = 2.80 \ {\rm GeV}^2$ (relevant for ALEPH).

Contour integrals of the part  $d(Q^2;D=0)$ of the Adler function, but with polynomial in $Q^2$ functions $W_i(Q^2/\sigma_{\rm max})$ instead of the exponential function (\ref{gQ2}), were studied within pQCD in Refs.~\cite{BenekeJamin}. But we choose here the exponential function, i.e., the Borel sum rule approach. This approach has two very attractive aspects \cite{Ioffe}:
\begin{enumerate}
\item
  At low Borel scales $M^2$ the Borel transform $B(M^2)$ probes the low-$\sigma$ (IR) regime. On the other hand, the high-$\sigma$ (UV) contributions have larger experimental uncertainties $\delta \omega(\sigma)$ and are suppressed in the Borel transform (\ref{sr3a}).
\item
  When $M^2=|M^2| \exp(i \pi/6)$, it is straightforward to see that the $D=6$ term in ${\rm Re} B_{\rm th}(M^2)$ is zero (and thus only the $D=4$ higher-twist term survives). Analogously, when  $M^2=|M^2| \exp(i \pi/4)$, the corresponding $D=4$ term is zero (and thus only the $D=6$ higher-twist term survives). This helps us extract more easily the values of the condensates $\langle O_{4} \rangle_{V+A}$ and $\langle O_{6} \rangle_{V+A}$ for $M^2=|M^2| \exp(i \pi/6)$, $|M^2| \exp(i \pi/4)$, respectively.
\end{enumerate}
Hence, the value of the gluon condensate $\langle a GG \rangle = 6 \langle O_4 \rangle_{V+A} + 0.00199 \ {\rm GeV}^4$ will be determined by comparing ${\rm Re}B_{\rm th}(M^2)$ with ${\rm Re}B_{\rm exp}(M^2)$ along the ray $M^2=|M^2| \exp(i \pi/6)$ in the complex $M^2$-plane, through fitting in an interval for $|M^2|$. Analogously, the value of the condensate  $\langle O_6 \rangle_{V+A}$ will be determined by choosing the ray  $M^2=|M^2| \exp(i \pi/4)$. After determining the values of both condensates, the theoretical Borel sum rule can be applied along the positive semiaxis $M^2 > 0$, where both condensates affect the result, and comparison with the experimental values there will give us a verification of the quality of the obtained description. In practice, the quantity ${\rm Re} B_{\rm th}(M^2)$ in the case of $\A$-coupling approach will be evaluated according to Eqs.~(\ref{sr3b})-(\ref{BD0}) with $d(Q^2;D=0)$ evalated in the resummed form (\ref{dBGan22}); the evaluation with $d(Q^2;D=0)$ in the TPS form (\ref{danc}) gives similar results.

\subsection{Borel sum rules with OPAL data}
\label{subs:OPAL}

\begin{figure}[htb] 
\begin{minipage}[b]{.49\linewidth}
  \centering\includegraphics[width=85mm]{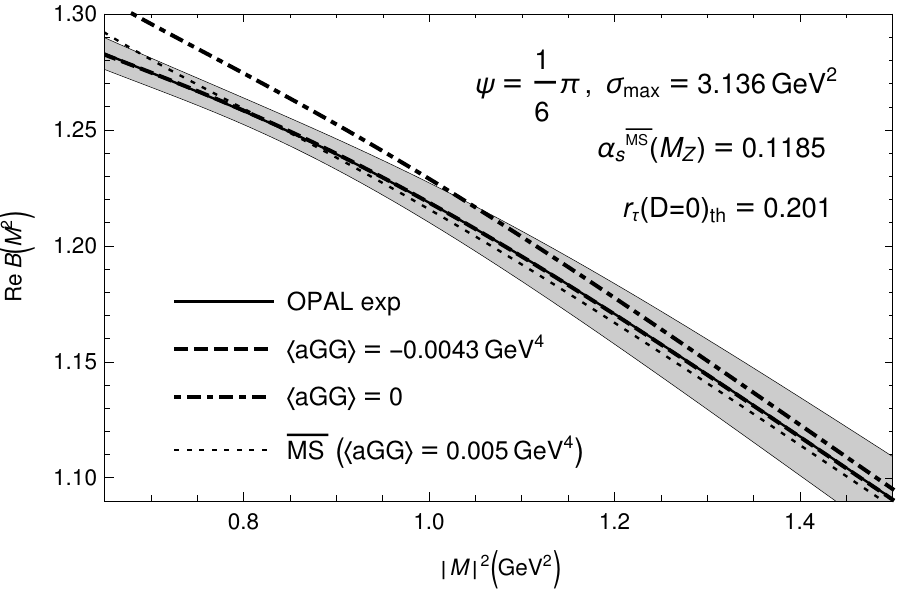}
  \end{minipage}
\begin{minipage}[b]{.49\linewidth}
  \centering\includegraphics[width=85mm]{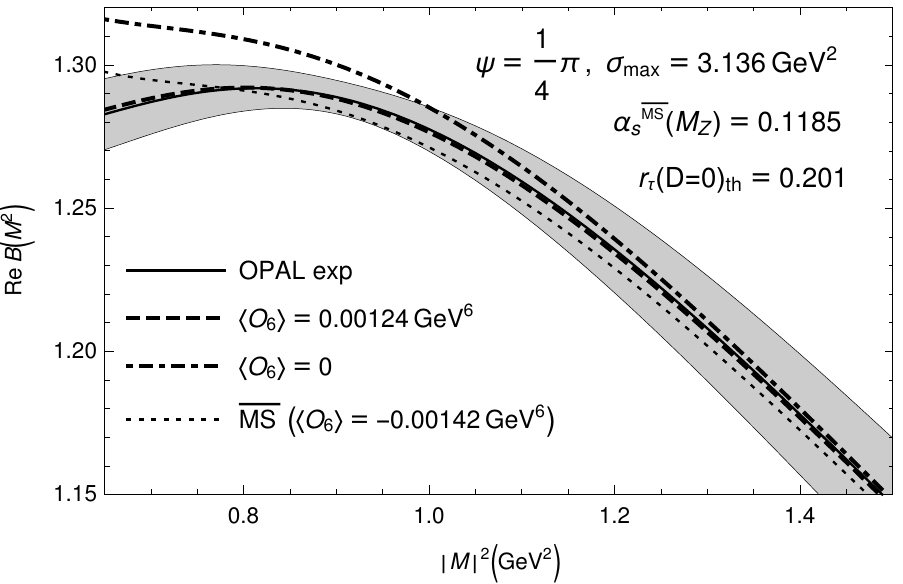}
\end{minipage}
\vspace{-0.2cm}
\caption{\footnotesize Borel transforms ${\rm Re} B(M^2)$ along the rays $M^2 = |M^2| \exp(i \Psi)$ with $\Psi=\pi/6$ (left-hand side) and $\Psi = \pi/4$ (right-hand side), as a function of $|M^2|$. The grey band represents the experimental results. In both Figures the central experimental curve (grey line) is almost identical with the theoretical curve (dashed line) of the holomorphic $\A$-coupling approach ($\A$QCD), with the fitted value $\langle a G G \rangle = - 0.0068 \ {\rm GeV}^4$ (left-hand side) and $\langle O_{6} \rangle_{V+A} = +0.0013 \ {\rm GeV}^6$ (right-hand side). The corresponding curve with zero values of condensates is included (dot-dashed)for comparison. Further, the fitted theoretical curve of $\MSbar$ pQCD approach is included (dotted).}
\label{FigPi64}
\end{figure}

We present in Figs.~\ref{FigPi64} the theoretical and OPAL experimental Borel transforms ${\rm Re}B(M^2)$, for the maximal value of the sum rule upper bound $\sigma_{\rm max}=3.136 \ {\rm GeV}^2$,\footnote{The spectral function data of OPAL \cite{PerisPC2} have 98 bins of width $\Delta \sigma =0.032 \ {\rm GeV}^2$, and reach the maximal value $\sigma_{\rm max}=3.136 \ {\rm GeV}^2$, which is somewhat lower than $m_{\tau}^2 \approx 3.157 \ {\rm GeV}^2$.}
  along the rays ${\rm Arg}(M^2) \equiv \Psi =\pi/6$ and $\pi/4$, respectively, for the case of the considered $\A(Q^2)$, with the choice of the parameters as given in the first line of Table \ref{tabres}, i.e., $\alpha_s(M_Z^2;\MSbar)=0.1185$, $s_0=652$, $s_1=3.97$ [$\Rightarrow r^{(D=0)}_{\tau,{\rm th}}=0.201$].
The resulting values of the obtained condensates $\langle a GG \rangle$ and $\langle O_6 \rangle_{V+A}$ are also given, obtained by the least-square fitting of the theoretical curve with the central experimental curve. The latter curve, and the experimental bands, are for the data obtained by the OPAL Collaboration  \cite{OPAL,PerisPC1,PerisPC2}. The experimental bands are obtained by taking into account the full covariance matrix of the OPAL spectral function data. The $\chi^2$ values are calculated by dividing the considered $|M^2|$ interval $[0.65,1.50] \ {\rm GeV^2}$ in $n=85$ equidistant intervals, and dividing the corresponding sum of 86 squared deviations by 85. We refer for details to Appendix \ref{app:fit}, especially Eqs.~(\ref{chi2a}) and (\ref{chi2aexp}) there. The described procedure then gives us the condensate values and the corresponding $\chi^2$
\bes
\label{cenres}
\bea
\langle O_4 \rangle_{V+A} & = & (-0.00104 \pm 0.00028)  \ {\rm GeV}^4
\nonumber\\
\Rightarrow \; \langle a GG \rangle & = & (-0.0043 \pm 0.0017) \ {\rm GeV}^4 \quad  (\chi^2=4.6 \times 10^{-8}; \chi^2_{\rm exp}=1.4 \times 10^{-4}),
\label{aGG}
\\
\langle O_6 \rangle_{V+A} & = & (+0.00124 \pm 0.00033) \ {\rm GeV}^6 \quad (\chi^2=1.2 \times 10^{-6}; \chi^2_{\rm exp}=2.0 \times 10^{-4}).
\label{O6}
\eea
\ees
The experimental uncertainties in the values of the condensates in Eqs.~(\ref{cenres}), due to the experimental (OPAL) bands in Figs.~\ref{FigPi64}, were estimated in the following way: the quantity $\chi^2_{\rm cov}(\Psi)$ of Eq.~(\ref{chi2c}) in Appendix \ref{app:fit} was evaluated, involving the $2 \times 2$ covariance matrix $U(\Psi)$ of the Borel transforms ${\rm Re} B(|M^2| e^{i \Psi})$ with $\Psi=\pi/6, \pi/4$ respectively, and with $n=1$ (i.e., $|M^2_0|=0.65 \ {\rm GeV}^2$ and $|M^2_1|=1.5 \ {\rm GeV}^2$ points). The minimum of that quantity was searched and obtained at a specific value of the corresponding condensate $\langle O \rangle$  ($O = a GG$, $(O_6)_{V+A}$), this value being close to the central values in Eqs.~(\ref{cenres}). Then the corresponding condensate value was varied around that obtained value, in such a way that $\chi^2_{\rm cov}(\Psi) = \chi^2_{\rm cov}(\Psi)_{\rm min} + 1$.  This corresponds to the condensate value variations $\delta \langle a GG \rangle = \pm 0.0017 \ {\rm GeV}^4$ and $\delta \langle O_6 \rangle_{V+A} = \pm 0.00033 \ {\rm GeV}^6$, for $\Psi = \pi/6$ and $\pi/4$, respectively. These variations are then given in Eqs. ~(\ref{cenres}). For further details and comments on these aspects, we refer to Appendix \ref{app:fit}.
In Figs.~\ref{FigPi64} we also included the $\MSbar$ pQCD coupling approach, with the same value $\alpha_s(M_Z^2;\MSbar)=0.1185$. This gave the following analogous results:
\bes
\label{cenresMS}
\bea
\langle a GG \rangle_{\MSbar} & = & (+0.0050 \pm 0.0017) \ {\rm GeV}^4 \quad  (\chi^2_{\MSbar}=1.4 \times 10^{-5}),
\label{aGGMS}
\\
\langle O_6 \rangle_{V+A,\MSbar} & = & (-0.00142 \pm 0.00033) \ {\rm GeV}^6 \quad (\chi^2_{\MSbar}=3.8 \times 10^{-5}).
\label{O6MS}
\eea
\ees
For some OPE analyses of OPAL data with $\MSbar$ pQCD, see Refs.~\cite{OPAL,PerisPC1,BGJMP}. 
Comparing various $\chi^2$ values in Eqs.~(\ref{cenres})-(\ref{cenresMS}), we can see that the quality of the fit with the considered $\A(Q^2)$ coupling is considerably better than in the $\MSbar$ pQCD approach. Further, in Eqs.~(\ref{cenres}) we included also the values of $\chi^2_{\rm exp}$, i.e., the corresponding variation between the central experimental values and the upper (or lower) experimental bound values, cf.~Eq.~(\ref{chi2aexp}). These values are considerably higher than the $\chi^2$ values in both the $\A$-coupling and $\MSbar$ pQCD approach.

\begin{figure}[htb] 
  \centering\includegraphics[width=100mm]{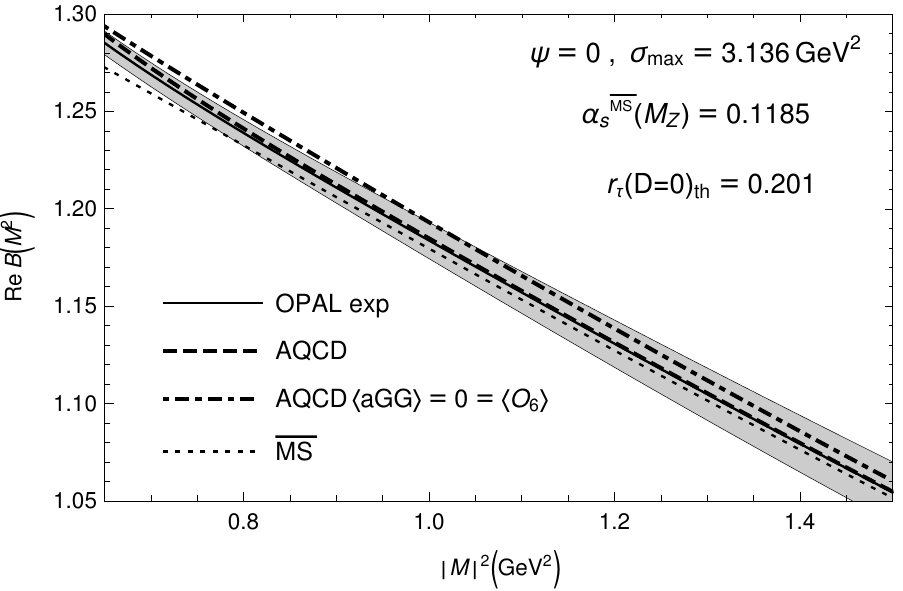}
\vspace{-0.3cm} 
 \caption{\footnotesize  Analogous to Figs.~\ref{FigPi64}, but now the Borel transforms $B(M^2)$ are for real $M^2 > 0$, and the values of the condensates are those determined in Figs.~\ref{FigPi64}. Again, the theoretical curve with $\A$-coupling ($\A$QCD) almost agrees with the central experimental curve.}
\label{FigPsi0}
 \end{figure}

In Fig.~\ref{FigPsi0}, the curves for ${\rm Arg} M^2=0$ are presented, with the corresponding central values of the condensates that were obtained in Figs.~\ref{FigPi64}.
We note that the values of both condensates affect ${\rm Re} B(M^2)$ for $M^2 = |M^2| > 0$.  The obtained $\chi^2$ values in the case $\Psi \equiv {\rm Arg} M^2=0$, and with the mentioned obtained condensate central values, are: 
\be
\Psi=0, \; {\rm OPAL:} \quad \chi^2=2.8 \times 10^{-6}, \quad \chi^2_{\MSbar}=2.8 \times 10^{-5}, \quad \chi^2_{\rm exp}=1.2 \times 10^{-4}.
\label{xi2Psi0}
\ee
The obtained $\chi^2$ is again very small ($\sim 10^{-6}$), as reflected in Fig.~\ref{FigPsi0}, which represents a good cross-check of consistency of the considered approach with $\A(Q^2)$ coupling.

  Another verification of consistency is to check whether the theoretical value  $r^{(D=0)}_{\tau, {\rm th}}=0.201$ of the quantity (\ref{rtaucont}) is consistent with the corresponding experimental value. If we apply the same type of sum rule approach, but now with the weight function $g(Q^2) = 2 (1 + Q^2/m_{\tau}^2)^2 (1 - 2  Q^2/m_{\tau}^2)$, then the right-hand side of the sum rule (\ref{sr1}), i.e., the theoretical value, is the expression $r_{\tau, {\rm th}}^{(D=0)}$ Eq.~(\ref{rtaucont}), and the left-hand side (the experimental value) is
\bes
\label{rtauexp}
\bea
r_{\tau}(\Delta S=0; m_{u,d}=0)_{\rm exp} &=& 2 \int_{0}^{m_{\tau}^2} \frac{d \sigma}{m_{\tau}^2} \left( 1 - \frac{\sigma}{m_{\tau}^2} \right)^2 \left( 1 + 2  \frac{\sigma}{m_{\tau}^2} \right) \omega_{\rm exp}(\sigma) - 1  \approx 0.198 \pm 0.006 \ ,
\label{rtauexpa}
\\
\Rightarrow \; r^{(D=0)}_{\tau,{\rm exp}} &=& 2 \int_{0}^{m_{\tau}^2} \frac{d \sigma}{m_{\tau}^2} \left( 1 - \frac{\sigma}{m_{\tau}^2} \right)^2 \left( 1 + 2  \frac{\sigma}{m_{\tau}^2} \right) \omega_{\rm exp}(\sigma) - 1 + 12 \pi^2 \frac{\langle O_6 \rangle_{V+A}}{m_{\tau}^6}
  \label{rtauexpb}
\\
  & \approx & (0.198 \pm 0.006) + 0.005 = 0.203 \pm 0.006 \ .
  \label{rtauexpc}
\eea
\ees  
We note that there is no $D=4$ condensate contribution to $r_{\tau}$, in view to our approximation of constant Wilson coefficients in the OPE of ${\cal D}(Q^2)$: $(1 + {\cal C}_n a) \mapsto 1$, cf.~Eqs.~(\ref{OPE1}) and (\ref{Adlfull}). Therefore, in Eq.~(\ref{rtauexpb}) for the $D=0$ part, there is only subtraction of the $D=6$ contribution. As always, the pion peak  term $2 \pi^2 f_{\pi}^2 \delta(\sigma - m_{\pi}^2)$ (where $f_{\pi}=0.1305$ GeV) is included in $\omega_{\rm exp}(\sigma)$; this accounts for the pion contribution, but nonetheless does not include in $r_{\tau,{\rm exp}}$ the main part of the mass effects $m_{\pi} \not=0$, i.e., $\Delta r_{\tau}(m_{u,d} \not=0)$; this is then consistent with the nonpresence of the mass effects $m_{u,d} \not=0$ in the theoretical expression (\ref{rtaucont}) for $r_{\tau}$ and $r_{\tau}^{(D=0)}$.
\footnote{The leading chirality-violating effects in $r_{\tau}$ can be shown to be
\be
\Delta r_{\tau}(\Delta S=0, m_{u,d} \not=0)
= 2 \int_0^{m_{\tau}^2} (d \sigma/m_{\tau}^2) (1 - \sigma/m_{\tau}^2)^2
\omega^{(0)}(\sigma), \ee
where $\omega^{(0)}(\sigma) = 2 \pi {\rm Im} \Pi^{(0)}_{V+A}(-\sigma - i \epsilon) \approx 2 \pi {\rm Im} \Pi^{(0)}_{A}(-\sigma - i \epsilon) \approx 2 \pi^2 f_{\pi}^2 ( \delta(\sigma - m_{\pi}^2) - \delta(\sigma) )$, which gives $\Delta r_{\tau}(\Delta S=0, m_{u,d} \not=0) \approx - 8 \pi^2 f_{\pi}^2 m_{\pi}^2/m_{\tau}^4 \approx -0.003$; cf.~\cite{Braaten,GCTL}.}

  Further, $(1.198 \pm 0.006)$ is the value of the above integral over $\sigma$, and $0.005$ represents the subtraction of the $D=6$ term with $\langle O_6 \rangle_{V+A} =  +0.0013 \ {\rm GeV}^6$, cf.~Eq.~(\ref{O6}). The obtained experimental value $0.203 \pm 0.006$ is consistent with the theoretical result $0.201$ that we started with. We recall that this latter value was obtained by evaluating the Adler function $d(Q^2;D=0)$ in Eq.~(\ref{rtaucont}) for $r_{\tau,{\rm th}}^{(D=0)}$ by the resummed approach (\ref{dBGan22}), $r_{\tau,{\rm th}}^{(D=0)}(d^{[4]}_{\rm res}) = 0.201$. The TPS approach (\ref{danc}) [i.e., Eq.~(\ref{danb}) with $\kappa=1$] for $d(Q^2,D=0)$ gave via Eq.~(\ref{rtaucont}) practically the same value $r_{\tau,{\rm th}}^{(D=0)}(d^{[4]}_{\rm an})=0.158 + 0.054 - 0.010 - 0.001 = 0.201$.

In the $\MSbar$ pQCD case, this type of consistency is lost: in this case $\langle O_6 \rangle_{V+A} =  -0.0014 \ {\rm GeV}^6$ and thus $r^{(D=0)}_{\tau,{\rm exp}, \MSbar} = (0.198 \pm 0.006) - 0.005 = 0.193 \pm 0.006$, which is different by about two standard deviations from the theoretical value (\ref{rtaucont}) in the $\MSbar$ approach, $r^{(D=0)}_{\tau, {\rm th}}(d^{[4]}_{{\rm pt},\MSbar})=0.182$. We point out that the Adler function $d(Q^2;D=0)$ in $\MSbar$ pQCD approach is evaluated as TPS, leading via Eq.~(\ref{rtaucont}) to: $r_{\tau, {\rm th}}(d^{[4]}_{{\rm pt},\MSbar})^{(D=0)}({\rm TPS}) = 0.138 + 0.026 + 0.010 + 0.007 = 0.182$. 

In Eq.~(\ref{rtauexpa}), $\sigma_{\rm max}=m_{\tau}^2 \approx 3.157 \ {\rm GeV}^2$. However, OPAL data for $\omega_{\rm exp}(\sigma)$ have $\sigma_{\rm max} = 3.136 \ {\rm GeV}^2$, somewhat lower. Nonetheless, the interval between these two values contributes to the integral (\ref{rtauexpa}) only $\sim 10^{-6}$, which is entirely negligible. In the Borel sum rules, this effect was not negligible and we had to evaluate the Borel transforms with $\sigma_{\rm max} = 3.136 \ {\rm GeV}^2$.

It may seem, at first sight, that the quantity $r_{\tau, \rm th}^{(D=0)}$, Eq.~(\ref{rtaucont}), is the theoretical prediction for the quantity whose experimental value is given in Eqs.~(\ref{rtauexpb})-(\ref{rtauexpc}). This is really so in the $\MSbar$ pQCD+OPE approach, where the only adjustable parameter is $\alpha_s(M_Z^2;\MSbar)$, and the condensate values (including $\langle O_6 \rangle_{V+A}$) are obtained by the described sum rule approach.
On the other hand, in the considered $\A$QCD+OPE approach, the value of  $r_{\tau, \rm th}^{(D=0)}$ is not a prediction, but an adjustable input parameter which then represents one of the seven conditions fixing the seven parameters of the $\A$-coupling, cf.~Sec.~\ref{sec:constr}. In fact, the (adjusted) value of  $r_{\tau, \rm th}^{(D=0)}$ was the only input parameter for the construction of $\A$ coming from the $\tau$-decay physics, or equivalently, from a QCD correlation function.

\subsection{Borel sum rules with ALEPH data}
\label{subs:ALEPH}

We performed the same type of analysis also with ALEPH data \cite{ALEPH2,DDHMZ,ALEPHfin,ALEPHwww}. To extract $\omega_{\rm exp}(\sigma)$ (V+A channel) from ALEPH data \cite{ALEPHwww}, cf.~the right-hand Fig.~\ref{FigOmega}, we applied the procedure as described in Ref.~\cite{BGMOS} (Sec.~III there), using the updated values \cite{BGMOS} for the parameters $B_e=0.17827$, $V_{ud}=0.97425$, $S_{\rm EW}=1.0201$, $m_{\pi}=139.57018$ MeV, $f_{\pi} = \sqrt{2} \times 92.21$ MeV. We further applied a rescaling factor $0.9987$ to the extracted spectral functions, as explained in Ref.~\cite{BGMOS}, principally due to the updated value of $f_{\pi}$ obtained from $\pi_{\mu 2}$ decays. For the $\tau$ lepton mass we used (throughout) the updated value $m_{\tau}=1.77686$ GeV \cite{PDG2016}. Further, it turned out that $\chi^2$ is quite large if we took into account the largest ALEPH bins (with $\sigma > 2.80 \ {\rm GeV}^2$). In Fig.~\ref{FigOmega}, the right-hand figure, we can see that the uncertainties for $\omega_{\rm exp}(\sigma)$ in such bins are quite large, and that these bins are quite wide.\footnote{This last aspect is not shared by the OPAL data, where all the bins are narrow, and thus the large uncertainties in the last few bins do not affect much the obtained values $B_{\rm exp}(M^2)$ and $r^{(D=0)}_{\tau}$.} Therefore, we decided to eliminate these wide bins with large uncertainties, and considered only the 77 ALEPH bins, reaching $\sigma_{\rm max}=2.80 \ {\rm GeV}^2$. This choice then favorably affects the sum rules (\ref{sr2})  and (\ref{rtauexpa}), significantly decreasing the experimental uncertainties of these quantities.

\begin{figure}[htb] 
\begin{minipage}[b]{.49\linewidth}
  \centering\includegraphics[width=85mm]{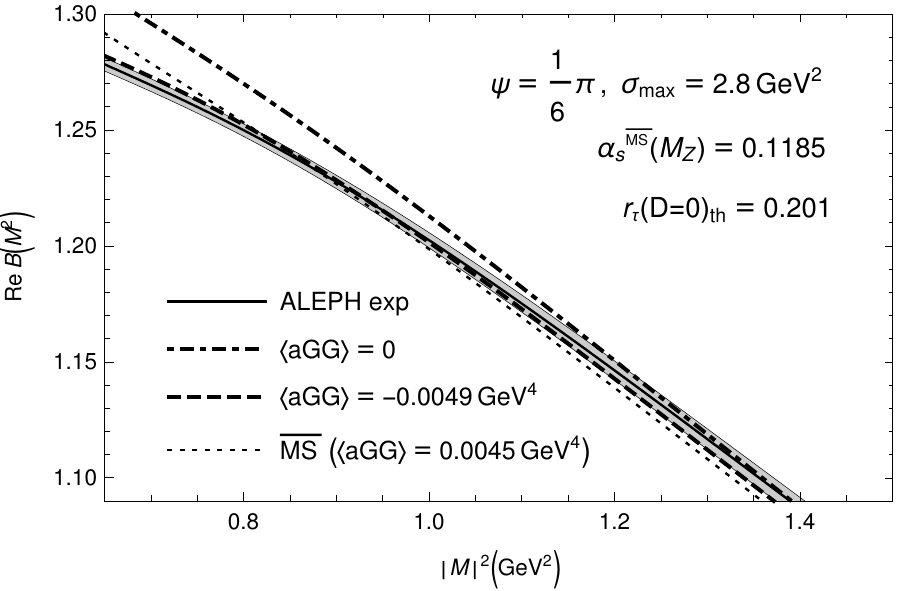}
  \end{minipage}
\begin{minipage}[b]{.49\linewidth}
  \centering\includegraphics[width=85mm]{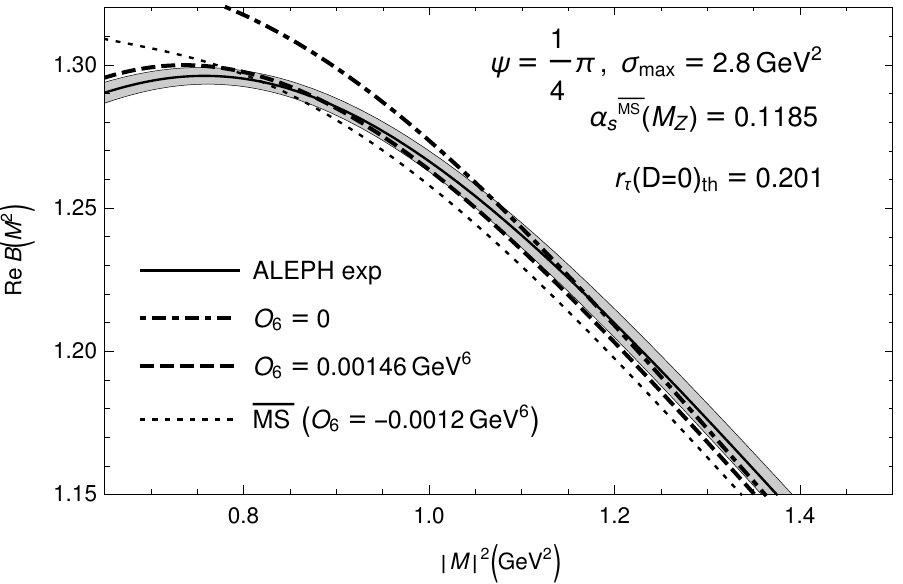}
\end{minipage}
\vspace{-0.2cm}
\caption{\footnotesize As Figs.~\ref{FigPi64}, but using the ALEPH data as described in the text.}
\label{FigPi64AL}
\end{figure}
\begin{figure}[htb] 
  \centering\includegraphics[width=100mm]{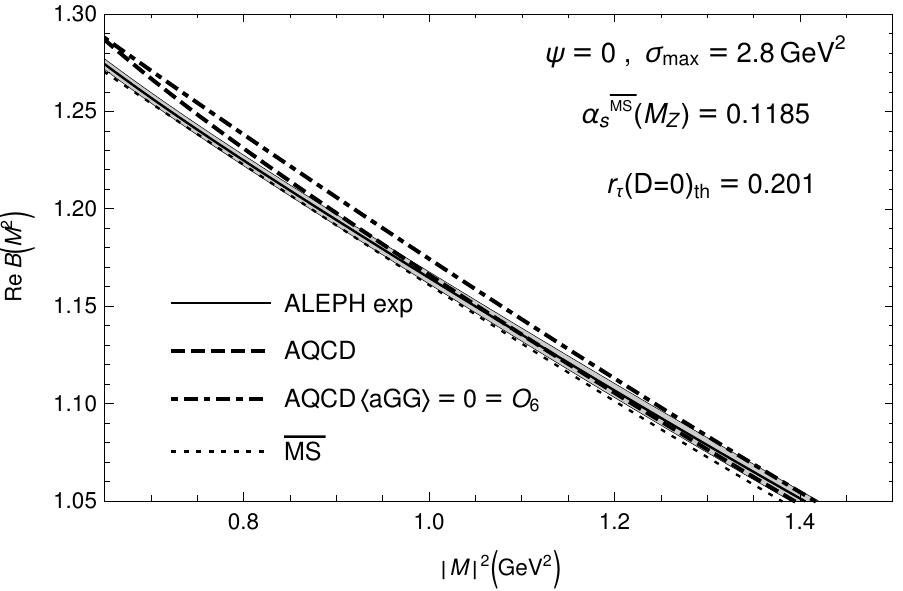}
\vspace{-0.3cm}
  \caption{\footnotesize  As Fig.~\ref{FigPsi0}, but using the ALEPH data and the corresponding condensate values obtained in Figs.~\ref{FigPi64AL}, as described in the text.}
\label{FigPsi0AL}
 \end{figure}
In analogy with the OPAL case, Figs.~\ref{FigPi64}-\ref{FigPsi0}, we present the analysis with the aforedescribed procedure with the ALEPH data in Fig.~\ref{FigPi64AL}-\ref{FigPsi0AL}. The theoretical $\A(Q^2)$ is the same as in the OPAL case: $s_0=652.$, $s_1=3.97$, $\alpha_s(M_Z^2;\MSbar)=0.1185$ [$\Rightarrow r^{(D=0)}_{\tau, {\rm th}}(\sigma_{\rm max}=m^2_{\tau})=0.201$]. In this case, the fitting procedure for the condensates in the Borel transforms gives
\bes
\label{cenresAL}
\bea
\langle a GG \rangle ({\rm ALEPH})& = & (-0.0049 \pm 0.0007) \ {\rm GeV}^4 \quad  (\chi^2=1.3 \times 10^{-5}; \chi^2_{\rm exp}=1.4 \times 10^{-5}),
\label{aGGAL}
\\
\langle O_6 \rangle_{V+A}({\rm ALEPH}) & = & (+0.00146 \pm 0.00014) \ {\rm GeV}^6 \quad (\chi^2=3.5 \times 10^{-5}; \chi^2_{\rm exp}=2.0 \times 10^{-5}),
\label{O6AL}
\eea
\ees
whereas the $\MSbar$ pQCD coupling approach, with the same value $\alpha_s(M_Z^2;\MSbar)=0.1185$, would give
\bes
\label{cenresMSAL}
\bea
\langle a GG \rangle_{\MSbar} ({\rm ALEPH}) & = & (+0.0045 \pm 0.0007) \ {\rm GeV}^4 \quad  (\chi^2_{\MSbar}=5.2 \times 10^{-5}),
\label{aGGMSAL}
\\
\langle O_6 \rangle_{V+A,\MSbar} ({\rm ALEPH}) & = & (-0.00120 \pm 0.00014) \ {\rm GeV}^6 \quad (\chi^2_{\MSbar}=1.2 \times 10^{-4}).
\label{O6MSAL}
\eea
\ees
For some other recent OPE analyses of ALEPH data with $\MSbar$ pQCD, see Refs.~\cite{ALEPHfin,BGMOS,DV,Pichrev1,Pichrev2}.   
The experimental uncertainties in the values of the condensates in Eqs.~(\ref{cenresAL})-(\ref{cenresMSAL}) were estimated in the same way as in the case of OPAL data, Eqs.~(\ref{cenres})-(\ref{cenresMS}). For details on the estimates of these uncertainties, we refer to Appendix \ref{app:fit}. The resulting $\chi^2$ values in the case $\Psi \equiv {\rm Arg} M^2=0$, with ALEPH data and the above condensate central values, are: 
\be
\Psi=0, \; {\rm ALEPH}: \quad \chi^2=2.3 \times 10^{-5}, \quad \chi^2_{\MSbar}=2.3 \times 10^{-5}, \quad \chi^2_{\rm exp}=1.2 \times 10^{-5}.
\label{xi2Psi0AL}
\ee
In comparison with the results obtained from the OPAL data, Eqs.~(\ref{cenres})-(\ref{xi2Psi0}), we see that in the ALEPH case the extracted values of the condensates are very similar. However, the experimental $\chi^2_{\rm exp}$ are in ALEPH case much smaller, by about one order of magnitude (this to a large extent because we chose $\sigma_{\rm max}=2.80 \ {\rm GeV}^2 < m_{\tau}^2$ in the ALEPH case). The other difference is that the fit quality in the ALEPH case ($\chi^2$, $\chi^2_{\MSbar}$) is also worse than in the OPAL by about one order of magnitude. As a consequence, the theoretical $\chi^2$ (i.e., with $\A$-coupling approach) and the experimental $\chi^2_{\rm exp}$ are comparable in the ALEPH case, while in the OPAL case $\chi^2$ is much smaller than $\chi^2_{\rm exp}$ (by about two orders of magnitude). In both cases (OPAL and ALEPH), the $\A$-coupling approach is consistently better than the $\MSbar$ pQCD approach, i.e., $\chi^2 < \chi^2_{\MSbar}$. 

An additional cross-check as in Eq.~(\ref{rtauexp}) can be performed also in the present (ALEPH) case. However, since we used $\sigma_{\rm max}=2.80 \ {\rm GeV}^2$ for the ALEPH experimental input, the quantities to compare are those of Eq.~(\ref{rtaucont}) but with radius of contour integration $\sigma_{\rm max}=2.80 \ {\rm GeV}^2$, and Eq.~(\ref{rtauexp}) with ALEPH value of $\omega_{\rm exp}(\sigma)$ and upper bound of integration $\sigma_{\rm max}=2.80 \ {\rm GeV}^2$.
The corresponding results are
\bea
r^{(D=0)}_{\tau, {\rm th}}(\sigma_{\rm max}=2.80 \ {\rm GeV}^2 ) & \equiv & \frac{1}{2 \pi} \int_{-\pi}^{+ \pi}
d \phi \ (1 + e^{i \phi})^3 (1 - e^{i \phi}) \
d(Q^2=\sigma_{\rm max} e^{i \phi};D=0) = 0.216.
\label{rtaucontAL}
\eea
\bes
\label{rtauexpAL}
\bea
  r^{(D=0)}_{\tau, {\rm exp}}(\sigma_{\rm max}=2.80 \ {\rm GeV}^2) &=& 2 \int_{0}^{\sigma_{\rm max}} \frac{d \sigma}{\sigma_{\rm max}} \left( 1 - \frac{\sigma}{\sigma_{\rm max}} \right)^2 \left( 1 + 2  \frac{\sigma}{\sigma_{\rm max}} \right) \omega_{\rm exp}(\sigma) - 1 + 12 \pi^2 \frac{\langle O_6 \rangle_{V+A}}{\sigma_{\rm max}^3}
  \label{rtauexpALa}
  \\
  & \approx & (0.206 \pm 0.003) + 0.008 = 0.214 \pm 0.003 \ .
  \label{rtauexpALb}
\eea
\ees
In Eq.~(\ref{rtauexpALb}), the contribution $0.008$ stems from the subtraction of the $D=6$ contribution with $\langle O_6 \rangle_{V+A} = 0.0015 \ {\rm GeV}^6$ [cf.~Eq.~(\ref{O6AL})].
Comparing the results in Eqs.~(\ref{rtaucontAL}) and (\ref{rtauexpALb}), we can see that they are consistent with each other. 

In $\MSbar$ case, for $\sigma_{\rm max}= 2.80 \ {\rm GeV}^2$, the corresponding results are $r^{(D=0)}_{\tau, {\rm th}, \MSbar}(\sigma_{\rm max}=2.80 \ {\rm GeV}^2 ) =0.191$ and $r^{(D=0)}_{\tau, {\rm exp}, \MSbar}(\sigma_{\rm max}=2.80 \ {\rm GeV}^2)=(0.206 \pm 0.003) -0.006 = 0.200 \pm 0.003$, which are not consistent with each other.

\subsection{Varying parameters of the coupling and of sum rules}
\label{subs:vary}

In the analyses of the sum rules so far, we have used the input parameters of the first line of Table \ref{tabres}, i.e., when $\alpha_s(M_Z^2;\MSbar)=0.1185$ and $r^{(D=0)}_{\tau,{\rm th}}=0.201$. Below in Tables \ref{tabSROPAL} and  \ref{tabSRALEPH} we present the results of the Borel sum rules for all {\bf seven} choices of the input parameters of the $\A$-coupling (cf.~Table \ref{tabres}), and for the $\MSbar$ pQCD approach, for the OPAL and ALEPH data, respectively. We use the notations of the previous two Subsections.
\begin{table}
  \caption{The central values of the $D=4, 6$ condensates and of the corresponding $\chi^2$ ($\Psi=\pi/6$, $\pi/4$), for the seven cases of $\A$QCD of Table \ref{tabres}, and for the corresponding cases of $\MSbar$ pQCD. Included are the resulting values of $\chi^2$ for $\Psi=0$, as well as $r^{(D=0)}_{\tau, {\rm exp}}$, as extracted from OPAL data, with $\sigma_{\rm max}=m_{\tau}^2$. For comparison of the $\chi^2$-values, the OPAL values for $\Psi=\pi/6$, $\pi/4$, $0$ are: $\chi^2_{\rm exp}=1.4 \times 10^{-4}$, $2.0 \times 10^{-4}$ and $1.2 \times 10^{-4}$, respectively. The experimental uncertainties due to the (OPAL) experimental bands are always the same, $10^3 \times \delta \langle a GG \rangle = \pm 1.71 \ {\rm GeV}^4$ and $10^3 \times \delta \langle O_6 \rangle_{V+A} = \pm 0.33 \ {\rm GeV}^6$.}
\label{tabSROPAL}  
\begin{ruledtabular}
\begin{tabular}{lll|llll}
  method & ${\overline \alpha}_s(M_Z^2)$ &  $r^{(D=0)}_{\tau, {\rm th}}$ & $10^3 \times \langle a GG \rangle [{\rm GeV}^4]$ $(\chi^2)$ & $10^3 \times \langle O_6 \rangle_{V+A} [{\rm GeV}^6]$ $(\chi^2)$ & $\Psi=0: \chi^2$ &   $r^{(D=0)}_{\tau, {\rm exp}}$
  \\
  \hline
  $\A$QCD & $0.1185$ & $0.201$ & $-4.26 \; (4.6 \times 10^{-8})$ & $+1.24 \; (1.2 \times 10^{-6})$ & $2.8 \times 10^{-6} $ & $0.203 \pm 0.006$
  \\
 $\A$QCD & $0.1185$ & $0.203$ & $-3.82 \; (3.5 \times 10^{-7})$ & $+1.39 \; (3.6 \times 10^{-8})$ & $5.6 \times 10^{-6}$ & $0.203 \pm 0.006$ 
  \\
 $\A$QCD & $0.1185$ & $0.199$ & $-4.56 \; (6.1 \times 10^{-7})$ & $+1.08 \; (4.1 \times 10^{-6})$ & $1.1 \times 10^{-6}$ & $0.202 \pm 0.006$ 
 \\
  $\MSbar$pQCD & $0.1185$ & $0.182$ & $+5.00 \; (1.4 \times 10^{-5})$ & $-1.42 \; (3.8 \times 10^{-5})$ & $2.8 \times 10^{-5}$ & $0.193 \pm 0.006$
  \\
  \hline
  $\A$QCD & $0.1181$ & $0.201$ & $-1.03 \; (4.4 \times 10^{-8})$ & $+1.12 \; (8.5 \times 10^{-7})$ & $3.6 \times 10^{-6}$ & $0.202 \pm 0.006$
  \\
 $\A$QCD & $0.1181$ & $0.203$ & $-0.31 \; (5.7 \times 10^{-7})$ & $+1.22 \; (9.2 \times 10^{-8})$ & $6.5 \times 10^{-6}$ & $0.203 \pm 0.006$ 
  \\
  $\MSbar$pQCD & $0.1181$ & $0.179$ & $+6.94 \; (2.1 \times 10^{-5})$ & $-1.65 \; (5.6 \times 10^{-5})$ & $3.7 \times 10^{-5}$ & $0.192 \pm 0.006$
 \\
  \hline
 $\A$QCD & $0.1189$ & $0.201$ & $-7.24 \; (3.2 \times 10^{-8})$ & $+1.32 \; (1.2 \times 10^{-6})$ & $2.0 \times 10^{-6} $ & $0.203 \pm 0.006$
  \\
 $\A$QCD & $0.1189$ & $0.203$ & $-6.98 \; (2.9 \times 10^{-7})$ & $+1.48 \; (5.7 \times 10^{-8})$ & $4.6 \times 10^{-6}$ & $0.204 \pm 0.006$ 
  \\
  $\MSbar$pQCD & $0.1189$ & $0.184$ & $+3.01 \; (7.8 \times 10^{-6})$ & $-1.18 \; (2.3 \times 10^{-5})$ & $2.0 \times 10^{-5}$ & $0.194 \pm 0.006$
 \end{tabular}
\end{ruledtabular}
\end{table}
\begin{table}
  \caption{As Table \ref{tabSROPAL}, but for ALEPH data and with $\sigma_{\rm max}=2.80 \ {\rm GeV}^2$. The last two columns compare the experimental and theoretical values of $r^{(D=0)}_{\tau}(\sigma_{\rm max})$. For comparison of the $\chi^2$-values, the ALEPH values for $\Psi=\pi/6$, $\pi/4$, $0$ (with $\sigma_{\rm max}=2.80 \ {\rm GeV}^2$) are: $\chi^2_{\rm exp}=1.4 \times 10^{-5}$, $2.0 \times 10^{-5}$ and $1.2 \times 10^{-5}$, respectively.}
\label{tabSRALEPH}  
\begin{ruledtabular}
\begin{tabular}{lll|lllll}
  method & ${\overline \alpha}_s(M_Z^2)$ &  $r^{(D=0)}_{\tau, {\rm th}}$ & $10^3 \times \langle a GG \rangle [{\rm GeV}^4]$ $(\chi^2)$ & $10^3 \times \langle O_6 \rangle_{V+A} [{\rm GeV}^6]$ $(\chi^2)$ & $\Psi=0: \chi^2$ &   $r^{(D=0)}_{\tau}(\sigma_{\rm max})_{\rm exp}$ &  $r^{(D=0)}_{\tau}(\sigma_{\rm max})_{\rm th}$
  \\
  \hline
  $\A$QCD & $0.1185$ & $0.201$ & $-4.89 \; (1.3 \times 10^{-5})$ & $+1.46 \; (3.5 \times 10^{-5})$ & $2.2 \times 10^{-5}$ & $0.214 \pm 0.003$ & $0.216$
  \\
 $\A$QCD & $0.1185$ & $0.203$ & $-4.45 \; (8.7 \times 10^{-6})$ & $+1.61 \; (2.5 \times 10^{-5})$ & $2.6 \times 10^{-5}$ & $0.215 \pm 0.003$ & $0.218$
  \\
  $\A$QCD & $0.1185$ & $0.199$ & $-5.19 \; (1.8 \times 10^{-5})$ & $+1.30 \; (4.7 \times 10^{-5})$ & $1.9 \times 10^{-5}$ & $0.214 \pm 0.003$ & $0.214$
\\
  $\MSbar$pQCD & $0.1185$ & $0.182$ & $+4.45 \; (5.2 \times 10^{-5})$ & $-1.20 \; (1.2 \times 10^{-4})$ & $2.4 \times 10^{-5}$ & $0.200 \pm 0.003$ & $0.191$
  \\
 \hline
  $\A$QCD & $0.1181$ & $0.201$ & $-1.68 \; (1.2 \times 10^{-5})$ & $+1.34 \; (3.3 \times 10^{-5})$ & $2.4 \times 10^{-5}$ & $0.214 \pm 0.003$ & $0.216$
  \\
 $\A$QCD & $0.1181$ & $0.203$ & $-0.96 \; (7.7 \times 10^{-6})$ & $+1.44 \; (2.2 \times 10^{-5})$ & $2.7 \times 10^{-5}$ & $0.214 \pm 0.003$ & $0.218$
  \\
  $\MSbar$pQCD & $0.1181$ & $0.179$ & $+6.39 \; (6.6 \times 10^{-5})$ & $-1.43 \; (1.5 \times 10^{-4})$ & $3.2 \times 10^{-5}$ & $0.199 \pm 0.003$ & $0.188$
 \\
 \hline
  $\A$QCD & $0.1189$ & $0.201$ & $-7.86 \; (1.3 \times 10^{-5})$ & $+1.54 \; (3.5 \times 10^{-5})$ & $2.0 \times 10^{-5}$ & $0.215 \pm 0.003$ & $0.216$
 \\
 $\A$QCD & $0.1189$ & $0.203$ & $-7.61 \; (9.1 \times 10^{-6})$ & $+1.71 \; (2.6 \times 10^{-5})$ & $2.4 \times 10^{-5}$ & $0.21x \pm 0.003$ & $0.218$
 \\
  $\MSbar$pQCD & $0.1189$ & $0.184$ & $+2.47 \; (3.9 \times 10^{-5})$ & $-0.97 \; (9.1 \times 10^{-5})$ & $1.7 \times 10^{-5}$ & $0.201 \pm 0.003$ & $0.193$
\end{tabular}
\end{ruledtabular}
\end{table}  
The obtained values of the condensates from OPAL with the $\A$QCD+OPE approach, Table \ref{tabSROPAL}, can be summarized in the following way:
\bes
\label{resOPAL}
\bea
\langle O_4 \rangle_{V+A}({\rm OPAL}) & = & -0.00104^{-0.00050}_{+0.00054}(\delta \alpha_s)^{+0.00007}_{-0.00005}(\delta r_{\tau}) \pm 0.00028({\rm exp})  \ [{\rm GeV}^4]
\nonumber\\
\Rightarrow \; \langle a GG \rangle({\rm OPAL}) & = & -0.0043^{-0.0029}_{+0.0033}(\delta \alpha_s)^{+0.0005}_{-0.0003}(\delta r_{\tau}) \pm 0.0017 ({\rm exp}) \ [{\rm GeV}^4] ,
\label{aGGOPAL}
\\
\langle O_6 \rangle_{V+A}({\rm OPAL}) & = & +0.00124^{+0.00008}_{-0.00012}(\delta \alpha_s)^{+0.00015}_{-0.00016}(\delta r_{\tau}) \pm 0.00033 ({\rm exp}) \ [{\rm GeV}^6].
\label{O6OPAL}
\eea
\ees
Here, the central value refers to the central case $\alpha_s(M_Z^2;\MSbar)=0.1185$ and $r_{\tau,{\rm th}}^{(D=0)}=0.201$ (the first line of Table \ref{tabres}); the first variation corresponds to the values of $\alpha_s(M_Z^2;\MSbar)=0.1185 \pm 0.0004$ [and $r_{\tau,{\rm th}}^{(D=0)}=0.201$]; the second variation corresponds to the values $r_{\tau,{\rm th}}^{(D=0)}=0.201 \pm 0.002$ [and $\alpha_s(M_Z^2;\MSbar)=0.1185$]; the third uncertainty is due to the experimental (OPAL) bands in Figs.~\ref{FigPi64}. For example, the central value in the case of  $\alpha_s(M_Z^2;\MSbar)=0.1181$ and $r_{\tau,{\rm th}}^{(D=0)}=0.201$ is $\langle a GG \rangle=-0.0010 \ {\rm GeV}^4$; in the case of  $\alpha_s(M_Z^2;\MSbar)=0.1185$ and $r_{\tau,{\rm th}}^{(D=0)}=0.203$ is $\langle a GG \rangle=-0.0038 \ {\rm GeV}^4$.

The corresponding values for the $\MSbar$ pQCD+OPE approach are (cf.~Table \ref{tabSROPAL})
\bes
\label{resOPALMS}
\bea
\langle O_4 \rangle_{V+A}({\rm OPAL}; \MSbar) & = & +0.00050 \mp 0.00033(\delta \alpha_s) \pm 0.00028({\rm exp})  \ [{\rm GeV}^4]
\nonumber\\
\Rightarrow \; \langle a GG \rangle({\rm OPAL}; \MSbar) & = & +0.0050^{-0.0020}_{+0.0019}(\delta \alpha_s) \pm 0.0017 ({\rm exp}) \ [{\rm GeV}^4] ,
\label{aGGOPALMS}
\\
\langle O_6 \rangle_{V+A}({\rm OPAL}; \MSbar) & = & -0.00142^{+0.00024}_{-0.00023}(\delta \alpha_s) \pm 0.00033 ({\rm exp}) \ [{\rm GeV}^6].
\label{O6OPALMS}
\eea
\ees
In this case, there is no dependence on the value of $r_{\tau,{\rm th}}^{(D=0)}$ because this value is in the $\MSbar$ pQCD+OPE approach not an input parameter but a result of the approach once the value of $\alpha_s(M_Z^2;\MSbar)$ has been fixed. In this case, there is a clear tension between the values of $r_{\tau,{\rm th}}^{(D=0)}$ and $r_{\tau,{\rm exp}}^{(D=0)}$ (the difference being about two standard deviations), in contrast to the $\A$QCD+OPE approach, cf.~Table \ref{tabSROPAL} columns 3 and 7.

The corresponding results with the ALEPH data (and $\sigma_{\rm max}=2.80 \ {\rm GeV}^2$) are
\bes
\label{resALEPH}
\bea
\langle O_4 \rangle_{V+A}({\rm ALEPH}) & = & -0.00115^{-0.00049}_{+0.00054}(\delta \alpha_s)^{+0.00008}_{-0.00005}(\delta r_{\tau}) \pm 0.00012({\rm exp})  \ [{\rm GeV}^4]
\nonumber\\
\Rightarrow \; \langle a GG \rangle({\rm ALEPH}) & = & -0.0049^{-0.0030}_{+0.0032}(\delta \alpha_s)^{+0.0004}_{-0.0003}(\delta r_{\tau}) \pm 0.0007 ({\rm exp}) \ [{\rm GeV}^4] ,
\label{aGGALEPH}
\\
\langle O_6 \rangle_{V+A}({\rm ALEPH}) & = & +0.00146^{+0.00008}_{-0.00012}(\delta \alpha_s)^{+0.00015}_{-0.00016}(\delta r_{\tau}) \pm 0.00014 ({\rm exp}) \ [{\rm GeV}^6].
\label{O6ALEPH}
\eea
\ees
\bes
\label{resALEPHMS}
\bea
\langle O_4 \rangle_{V+A}({\rm ALEPH}; \MSbar) & = & +0.00041^{-0.00033}_{+0.00032}(\delta \alpha_s) \pm 0.00012({\rm exp})  \ [{\rm GeV}^4]
\nonumber\\
\Rightarrow \; \langle a GG \rangle({\rm ALEPH}; \MSbar) & = & +0.0045^{-0.0020}_{+0.0019}(\delta \alpha_s) \pm 0.0007 ({\rm exp}) \ [{\rm GeV}^4] ,
\label{aGGALEPHMS}
\\
\langle O_6 \rangle_{V+A}({\rm ALEPH}; \MSbar) & = & -0.00120 \pm 0.00023(\delta \alpha_s) \pm 0.00014 ({\rm exp}) \ [{\rm GeV}^6].
\label{O6ALEPHMS}
\eea
\ees

Inspection of Tables \ref{tabSROPAL} and \ref{tabSRALEPH} shows that, in the $\A$QCD+OPE approach, the values of $\chi^2$ for the Borel sum rule predictions at $\Psi=0$ in the case of OPAL data are by one order of magnitude smaller (better) than those of the $\MSbar$ pQCD+OPE approach, and comparable with each other in the case of ALEPH data. The quality of fits (in the $\Psi=\pi/6$ and $\pi/4$ Borel sum rules) is considerably better in the $\A$QCD+OPE case, for OPAL as well as for ALEPH data. Further, the quality of fits (when $\Psi=\pi/6$, $\pi/4$) and of predictions (when $\psi=0$) is comparably good in all seven cases of $\A$QCD, either for OPAL data, or for ALEPH data.

From Tables \ref{tabSROPAL} and \ref{tabSRALEPH} and Eqs.~(\ref{resOPAL})-(\ref{resALEPHMS}) we can also see that the value of the gluon condensate $\langle a GG \rangle$ in the $\A$-coupling framework depends strongly on the value of ${\overline \alpha}_s(M_Z^2)$; these results indicate that for low values  ${\overline \alpha}_s(M_Z^2) \lesssim 0.1180$ we get $\langle a GG \rangle \gtrsim 0$. We recall that according to Particle Data Group 2016  \cite{PDG2016}, the world average value is ${\overline \alpha}_s(M_Z^2)=0.1181 \pm 0.0011$. In the case of OPAL data with $\A$QCD+OPE, the best results (the smallest $\chi^2$ at $\Psi=\pi/6$ and $\Psi=0$) are obtained for $r^{(D=0)}_{\tau, {\rm th}} = 0.201$ (for all choices of $\alpha_s$), and in the case of ALEPH data the quality is comparable in all cases of $\A$QCD+OPE.

\section{Some predictions using $\A$QCD, and discussions}
\label{sec:pred}

In the previous Section we extracted the values of the condensates $\langle O_n \rangle_{V+A}$ ($n=4,6$) from OPAL and ALEPH data, by using the Borel sum rules along specific rays $\Psi \equiv {\rm arg}(M^2) =\pi/6$ and $\pi/4$, respectively. Strictly speaking, these are not predictions but rather extracted values of some OPE parameters for the considered $\A$QCD framework. On the other hand, the quality of the Borel transforms for $\Psi=0$ (with the obtained values of the two condensates), as compared to the corresponding experimental bands, is the quality of predictions of the considered $\A$QCD+OPE approach, cf.~Figs.~\ref{FigPsi0}, \ref{FigPsi0AL}. While this quality is equally good in the $\A$QCD and $\MSbar$ pQCD approaches for ALEPH data [cf.~Fig~\ref{FigPsi0AL} and Table \ref{tabSRALEPH} (6th column)], it is considerably better in the $\A$QCD than the $\MSbar$ pQCD approach for OPAL data [cf.~Fig~\ref{FigPsi0} and Table \ref{tabSROPAL} (6th column)]. Concerning the quality of the fits for rays $\Psi=\pi/6$ and $\pi/4$, it is considerably better in the $\A$QCD than the $\MSbar$ pQCD approach, for both OPAL and ALEPH data, cf.~Figs.~\ref{FigPi64}, \ref{FigPi64AL}, and Tables \ref{tabSROPAL} and \ref{tabSRALEPH} ($\chi^2$ in the 4th and 5th columns).

We notice that in the Borel sum rules we used $\sigma_{\rm max}=3.136 \ {\rm GeV}^2$ in the OPAL case, and $\sigma_{\rm max}=2.80 \ {\rm GeV}^2$ in the ALEPH case. One may ask what is the quality of the fits for Borel sum rules when we decrease the value of $\sigma_{\rm max}$ while keeping the obtained original values of the condensates. We wish to point out that the quality of such fits represents predictions of the considered approach, because the condensate values were extracted with different sum rules, i.e., with a significantly higher  $\sigma_{\rm max}$ ($3.136$ and $2.80 \ {\rm GeV}^2$, for OPAL and ALEPH, respectively). In Table \ref{tabsigmax} we present the resulting quality parameters $\chi^2$, for the $\A$-coupling and $\MSbar$ pQCD approach for OPAL Borel sum rules, in the case of  $\alpha_s(M_Z^2; \MSbar)=0.1185$ and $r^{(D=0)}_{\tau, {\rm th}}=0.201$ (i.e., the first line of Table \ref{tabres}), for five different values of $\sigma_{\rm max}$, keeping the values of the condensates, $\langle a GG \rangle =-0.00426 \ {\rm GeV}^4$ and $\langle O_6 \rangle_{V+A} = + 0.00124 \ {\rm GeV}^6$ as obtained from the fit with the maximal possible value $\sigma_{\rm max} = 3.136 \ {\rm GeV}^2$, cf.~the first line of Table \ref{tabSROPAL}. For the $\MSbar$ pQCD approach, we use the corresponding condensate values $\langle a GG \rangle_{\MSbar} =+0.00500 \ {\rm GeV}^4$ and  $\langle O_6 \rangle_{V+A, \MSbar} = - 0.00142 \ {\rm GeV}^6$ obtained analogously, cf.~the fourth line of Table \ref{tabSROPAL}.
\begin{table}
 \caption{Values of $\chi^2$ for Borel sum rules with OPAL data, for different number of bins $N_{\rm bins}$ (and thus different $\sigma_{\rm max}$), and for different angles $\Psi$ ($\equiv {\rm Arg}(M^2)$)$= \pi/6, \pi/4, 0$. The $\chi^2$ values are for the $\A$-coupling approach, $\MSbar$ pQCD approach, and (OPAL) experimental values. The input values are $\alpha_s(M_Z^2; \MSbar)=0.1185$ (relevant for $\A$QCD and $\MSbar$ pQCD) and $r^{(D=0)}_{\tau, {\rm th}}=0.201$ (relevant for $\A$QCD). The values of condensates are $\langle a GG \rangle =-0.00426 \ {\rm GeV}^4$ and $\langle O_6 \rangle_{V+A} = + 0.00124 \ {\rm GeV}^6$ for $\A$QCD, and $\langle a GG \rangle_{\MSbar} =+0.00500 \ {\rm GeV}^4$ and  $\langle O_6 \rangle_{V+A, \MSbar} = - 0.00142 \ {\rm GeV}^6$ for $\MSbar$ pQCD approach, as obtained by fitting the Borel sum rules at the highest value $\sigma_{\rm max} = 3.136 \ {\rm GeV}^2$, cf. the first and the fourth lines of Table \ref{tabSROPAL}.}
    \label{tabsigmax}
  \begin{ruledtabular}
\begin{tabular}{ll|lll|lll|lll}
$N_{\rm bins}$ & $\sigma_{\rm max} [{\rm GeV}^2]$ &  $\chi^2_{\Psi=\pi/6}$ & $\chi^2_{\Psi=\pi/6, \MSbar}$ & $\chi^2_{\Psi=\pi/6, {\rm exp}}$ &  $\chi^2_{\Psi=\pi/4}$ & $\chi^2_{\Psi=\pi/4, \MSbar}$ & $\chi^2_{\Psi=\pi/4, {\rm exp}}$ &  $\chi^2_{\Psi=0}$ & $\chi^2_{\Psi=0, \MSbar}$ & $\chi^2_{\Psi=0, {\rm exp}}$ 
\\
\hline
$98$ & $3.136$ & $4.6 \cdot 10^{-8}$ & $1.4 \cdot 10^{-5}$ & $1.4 \cdot 10^{-4}$ & $1.2 \cdot 10^{-6}$ & $3.8 \cdot 10^{-5}$ & $2.0 \cdot 10^{-4}$ & $2.8 \cdot 10^{-6}$ & $2.8 \cdot 10^{-5}$ & $1.2 \cdot 10^{-4}$
\\
$70$ & $2.240$ & $8.0 \cdot 10^{-7}$ & $2.9 \cdot 10^{-5}$ & $4.5 \cdot 10^{-5}$ & $3.6 \cdot 10^{-6}$ & $6.4 \cdot 10^{-5}$ & $6.1 \cdot 10^{-5}$ & $2.5 \cdot 10^{-6}$ & $4.6 \cdot 10^{-5}$ & $3.7 \cdot 10^{-5}$
\\
$50$ & $1.600$ & $2.0 \cdot 10^{-6}$ & $6.7 \cdot 10^{-5}$ & $4.5 \cdot 10^{-5}$ & $5.6 \cdot 10^{-6}$ & $1.3 \cdot 10^{-4}$ & $5.4 \cdot 10^{-5}$ & $2.3 \cdot 10^{-6}$ & $9.0 \cdot 10^{-5}$ & $3.9 \cdot 10^{-5}$
\\
$30$ & $0.960$ & $2.0 \cdot 10^{-5}$ & $8.0 \cdot 10^{-5}$ & $3.7 \cdot 10^{-5}$ & $3.2 \cdot 10^{-5}$ & $1.4 \cdot 10^{-4}$ & $4.2 \cdot 10^{-5}$ & $2.3 \cdot 10^{-5}$ & $1.2 \cdot 10^{-4}$ & $3.4 \cdot 10^{-5}$
\\
$26$ & $0.832$ & $3.1 \cdot 10^{-5}$ & $4.2 \cdot 10^{-4}$ & $3.0 \cdot 10^{-5}$ & $3.3 \cdot 10^{-5}$ & $6.0 \cdot 10^{-4}$ & $3.3 \cdot 10^{-5}$ & $1.8 \cdot 10^{-5}$ & $4.7 \cdot 10^{-4}$ & $2.8 \cdot 10^{-5}$
\end{tabular}
\end{ruledtabular}
\end{table}  
We recall that, in practice, all these $\chi^2$ values were calculated as sums of the squares of the corresponding deviations from the central experimental (OPAL) values, for 86 equidistant points in the considered $|M^2|$ intervals $0.65 \ {\rm GeV}^2 \leq |M^2| \leq 1.50 \ {\rm GeV}^2$, and dividing by $n=85$, cf.~Eq.~(\ref{chi2a}) in Appendix \ref{app:fit}.

\begin{figure}[htb] 
\begin{minipage}[b]{.49\linewidth}
  \centering\includegraphics[width=85mm]{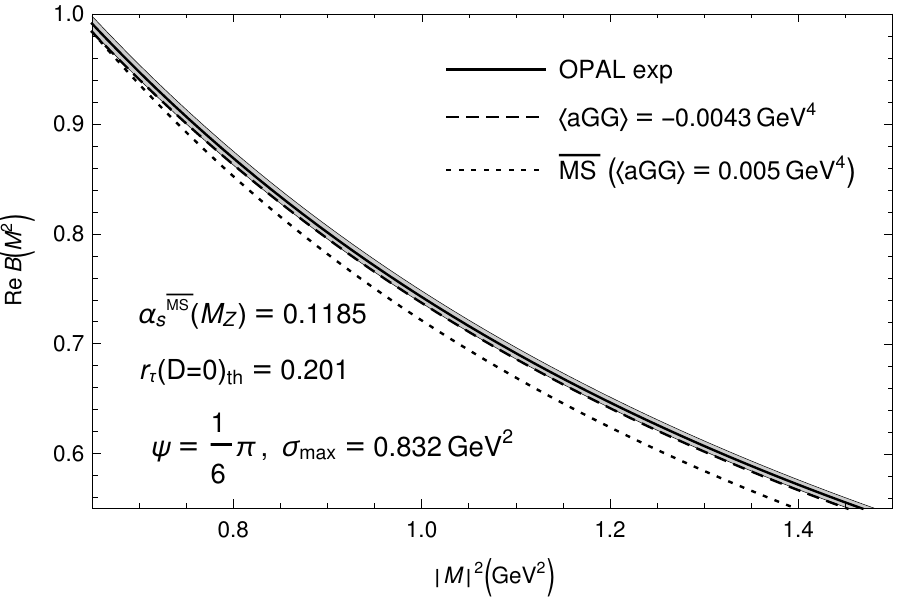}
  \end{minipage}
\begin{minipage}[b]{.49\linewidth}
  \centering\includegraphics[width=85mm]{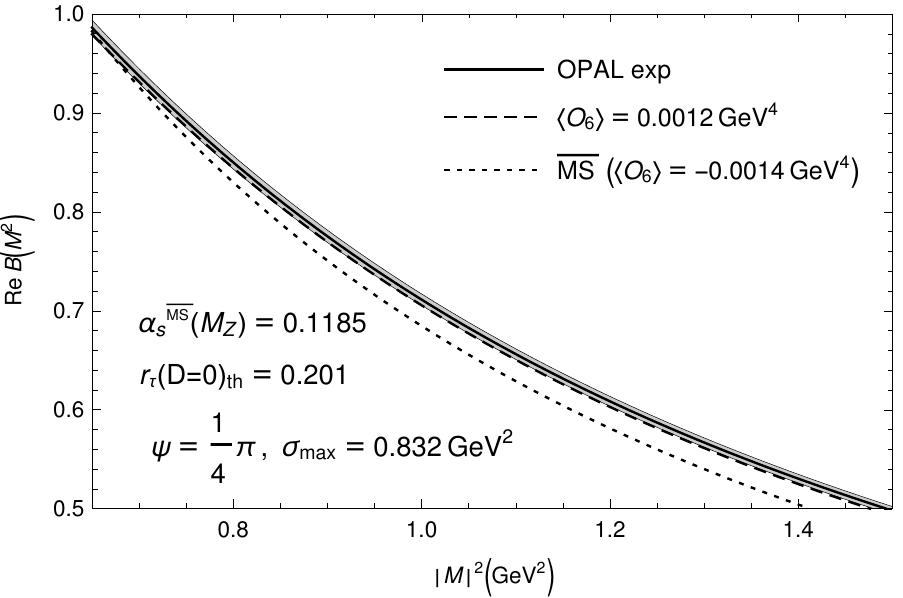}
\end{minipage}
\vspace{-0.2cm}
\caption{\footnotesize As Figs.~\ref{FigPi64}, with the same condensate values as there, but now for $\sigma_{\rm max}=0.832 \ {\rm GeV}^2$. The $\A$QCD curve (dashed) is at the lower edge of the experimental band.}
\label{FigPi64o26}
\end{figure}

\begin{figure}[htb] 
  \centering\includegraphics[width=100mm]{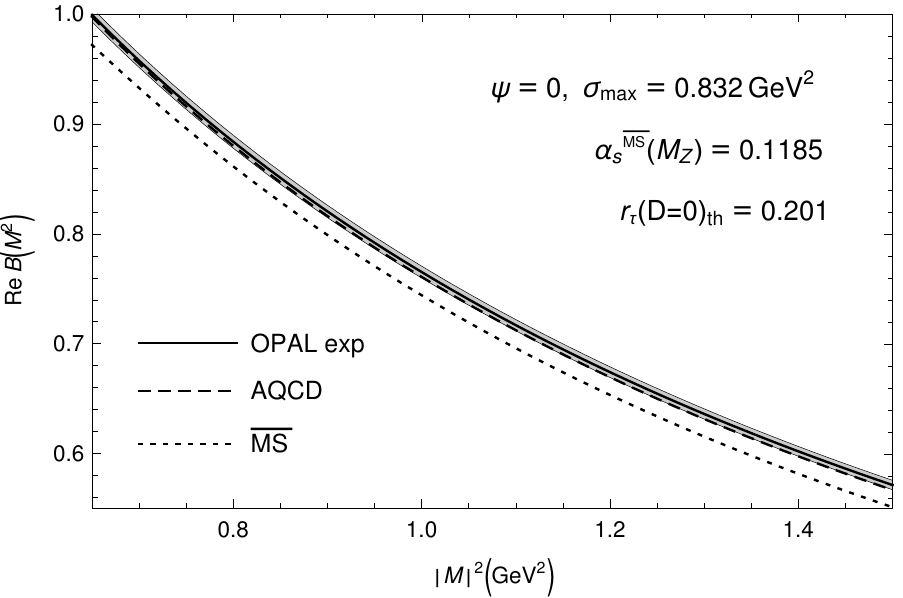}
\vspace{-0.3cm}
  \caption{\footnotesize  Analogous to Figs.~\ref{FigPsi0}, with the same condensate values as there, but now for $\sigma_{\rm max}=0.832 \ {\rm GeV}^2$. The $\A$QCD curve (dashed) is at the lower edge of the experimental band.}
\label{FigPsi0o26}
 \end{figure}

Inspection of Table \ref{tabsigmax} shows that the fit with the $\A$-coupling approach ($\A$QCD+OPE) keeps a remarkable level of quality when the upper bound of the considered energy interval of experimental data is decreased. The quality starts deteriorating only when $\sigma_{\rm max} < 1 \ {\rm GeV}^2$. In the $\MSbar$ pQCD approach the quality deteriorates continuously and significantly when $\sigma_{\rm max}$ decreases; the latter behavior is a manifestation of the quark-hadron duality violation \cite{BGMOS,DV,Pichrev1,Pichrev2} in the $\MSbar$ pQCD approach. For visualization, we present in Figs.~\ref{FigPi64o26} and \ref{FigPsi0o26} the curves of the Borel transforms for $\Psi=\pi/6$, $\pi/4$ and $\Psi=0$. They are as Figs.~\ref{FigPi64} and \ref{FigPsi0} but now for the significantly lower value $\sigma_{\rm max}=0.832 \ {\rm GeV}^2$.
We cannot decrease $\sigma_{\rm max}$ in the Borel sum rules below that value, because then the experimental spectral function $\omega(\sigma)$ becomes dominated by the $\rho$-resonance, cf.~Fig.~\ref{FigOmega}.
In this context, we point out that, since $\A$QCD practically agrees with its underlying pQCD version at high momenta and energies, in this regime it accounts for the quark-hadron duality equally well as pQCD. On the other hand, Table \ref{tabsigmax} and Figs.~\ref{FigPi64o26}-\ref{FigPsi0o26} indicate that it accounts for the quark-hadron duality better than pQCD even at lower momenta and energies, $|Q^2| \approx 1 \ {\rm GeV}^2$.
  
We present here predictions of the considered approach for yet another physical quantity, the V-channel Adler function ${\cal D}_V(Q^2)$ which is closely related with $R(\sigma)$, the production ratio for $e^+ e^- \to$ hadrons at the center-of-mass squared energy $\sigma$. Namely, the V-channel Adler function is
\bea
{\cal D}_V(Q^2) &\equiv&  - 4 \pi^2 \frac{d \Pi_V(Q^2)}{d \ln Q^2} 
 =  1 + d(Q^2;D=0) + 2 \pi^2 \sum_{n \geq 2}
 \frac{ n 2 \langle O_{2n} \rangle_V}{(Q^2)^n}  \ ,
\label{DV}
\eea
where we use the conventional notation $\langle O_{2 n} \rangle_{V+A} =\langle O_{2 n} \rangle_{V}+\langle O_{2 n} \rangle_{A}$, and the $D=0$ massless term  $d(Q^2;D=0)$ is the same as in the V+A channel case considered earlier (that case was related with the semihadronic $\tau$ decay physics). The theoretical connection with  the production ratio for $e^+ e^- \to$ hadrons, $R(\sigma)$, is the following:
\be
{\cal D}_V(Q^2) = Q^2 \int_0^{\infty} \frac{R(\sigma)}{(\sigma + Q^2)^2} d \sigma.
\label{DVR}
\ee
We note that ${\cal D}_V$ is normalized so that ${\cal D}_V \to 1$ when $|Q^2| \to \infty$. The experimental value of this function is then obtained in the following way:
\be
{\cal D}_{V, {\rm exp}}(Q^2) = Q^2 \int_{m^2}^{\sigma_0} \frac{R_{\rm data}(\sigma)}{(\sigma + Q^2)^2} d \sigma + Q^2 \int_{\sigma_0}^{\infty} \frac{R_{\rm pert}(\sigma)}{(\sigma + Q^2)^2} d \sigma,
\label{DVexp}
\ee
where $m^2 = (2 m_{\pi})^2$ ($m_{\pi}=0.13957$ GeV) is the kinematical production threshold, and $\sigma_0$ is a sufficiently high squared energy where pQCD approach is good. The quantity ${\cal D}_{V, {\rm exp}}(Q^2)$ in this convention and for $Q^2>0$ was obtained and presented in Refs.~\cite{Nest3a,NestBook}.\footnote{Cf.~Refs.~\cite{Eidel} when a different normalization convention is used.}

\begin{figure}[htb] 
  \centering\includegraphics[width=120mm]{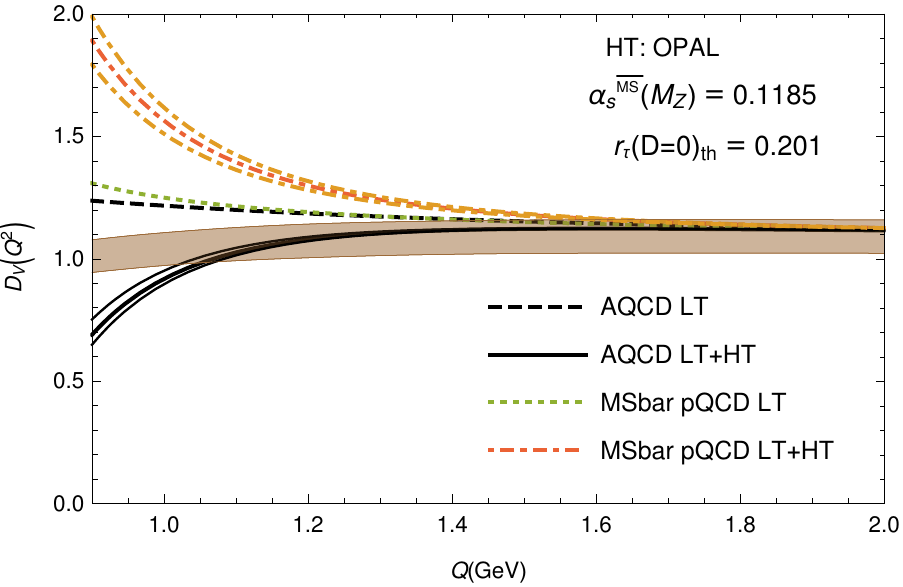}
  \caption{\footnotesize The V-channel Adler function at $Q^2>0$ ($Q \equiv \sqrt{Q^2}$): the grey band (online: brown band) are the experimental values, \cite{NestBook} (Fig. 1.7 there), cf.~also Refs.~\cite{Nest3a}. The solid lines are the theoretical curves for $\alpha_s(M_Z^2)=0.1181$ (upper), $0.1185$ (middle), $0.1189$ (lower curve) in the $\A$QCD+OPE approach, and the dash-dotted lines are in the $\MSbar$ pQCD+OPE approach. The dashed line is the leading twist (LT) contribution in $\A$QCD, and the dotted line in $\MSbar$ pQCD, for $\alpha_s(M_Z^2)=0.1185$. The $D=4$ and $D=6$ terms (higher-twist) are with the corresponding values of the condensates as explained in the text. $N_f=3$ is used throughout.}
  \label{FigDVOPE}
\end{figure}

Concerning the theoretical expression for ${\cal D}_V$ Eq.~(\ref{DV}), it turns out that we can obtain, or estimate, the values of the V-channel condensates appearing there from the values of the V+A channel condensates obtained in the previous Section. In the approximations applied in the Borel sum rules of the previous Section, we have for the $D=4$ condensates \cite{Braaten,PichPra}
\be
2 \langle O_4 \rangle_V =  2 \langle O_4 \rangle_A  = \langle O_4 \rangle_{V+A} \ .
\label{O4V}
\ee
Furthermore, using the factorization hypothesis \cite{Ioffe}, we obtain the relation
\be
\langle O_6 \rangle_V \approx - \frac{7}{4} \langle O_6 \rangle_{V+A},
\label{O6V}
\ee
where we refer for details to Refs.~\cite{Ioffe,AQCDprev}. The factorization hypothesis is expected to have a relative error of about $1/N_c^2 \approx 10\%$. Using the relations (\ref{O4V})-(\ref{O6V}) and the values of the V+A condensates obtained in the previous Section, we can evaluate the theoretical value of the V-channel Adler function ${\cal D}_V(Q^2)$ Eq.~(\ref{DV}) and compare it with the experimental value (\ref{DVexp}). In Fig.~\ref{FigDVOPE} we present this comparison in an interval of positive $Q^2$. 
The V+A channel condensate values, taken from the previous Section, were those extracted from OPAL data, cf.~Eqs.~(\ref{resOPAL})-(\ref{resOPALMS}). The results are presented for the cases of $\alpha_s(M_Z^2)=0.1185$ and $\alpha_s(M_Z^2)=0.1185 \pm 0.004$. In the $\A$QCD+OPE approach, the cases with $r^{(D=0)}_{\tau, {\rm th}}=0.201$ were taken. The $D=0$ part of the Adler function was calculated by the resummation method (\ref{dBGan22}), but the truncated version (\ref{danc}) gives practically the same results.

We can theoretically evaluate the V-channel Adler function also in a different way, namely by incorporating the additional low-energy nonperturbative contributions not as ($D=4, 6$) OPE terms but by accounting for the kinematical threshold $\sigma \geq m^2$ in $R(\sigma)$, Eq.~(\ref{DVexp}), where $m= 2 m_{\pi^-}$ \cite{Nest3a,Nest3b,NestBook}. We refer to those references and to Appendix \ref{app:mAQCD} for details. Namely, due to that kinematical threshold, the function ${\cal D}(Q^2)$ obtains the following ``massive'' form in $\A$QCD (or in fact, in any analytic QCD):
\bea
{\cal D}_V(Q^2)_{m \A QCD} & = & {\cal D}^{(0)}(Q^2)_m + \frac{Q^2}{(Q^2+m^2)} \frac{1}{\pi} \int_{m^2}^{+\infty} d \sigma \left(1 - \frac{m^2}{\sigma} \right) \frac{\rho_d(\sigma)}{(\sigma + Q^2)} ,
\label{DVm}
\eea
where $\rho_d(\sigma) \equiv {\rm Im} \; d(-\sigma - i \epsilon; D=0)$ is the spectral function of the $D=0$ Adler function $d(Q^2; D=0)$ of Eqs.~(\ref{danc}) and (\ref{dBGan22}) for $\A$QCD. We used the truncated version (\ref{danc}), but the resummed version (\ref{dBGan22}) gives practically the same results. The term ${\cal D}^{(0)}(Q^2)_m$ is the leading order term with the massive kinematic threshold (mLO), given in Eq.~(\ref{D0m}) in Appendix \ref{app:mAQCD}.

\begin{figure}[htb] 
  \centering\includegraphics[width=120mm]{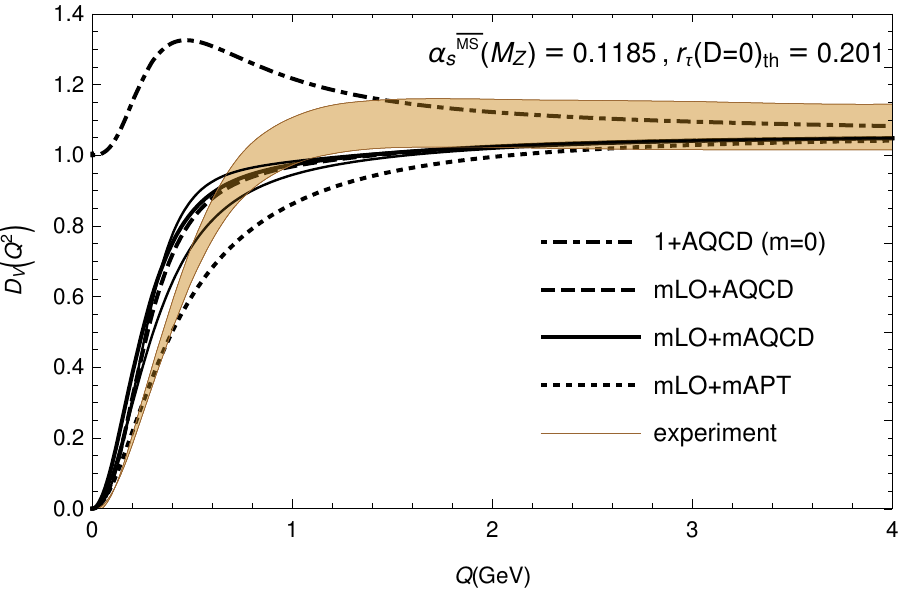}
  \caption{\footnotesize The V-channel Adler function at $Q^2>0$ ($Q \equiv \sqrt{Q^2}$): the grey band (online: brown band) are the experimental values as in Fig.~\ref{FigDVOPE}. The three solid lines are the theoretical curves for $\alpha_s(M_Z^2)=0.1189$ (upper), $0.1185$ (middle), $0.1181$ (lower curve) in the massive $\A$QCD approach (mLO+m$\A$QCD).  The dash-dotted line is the massless limit ($m^2 \mapsto 0$), for $\alpha_s(M_Z^2)=0.1185$. The dashed line is for the massive leading order term ${\cal D}^{(0)}_V(Q^2)_m$ and massless $\A$QCD term (mLO+$\A$QCD), for $\alpha_s(M_Z^2)=0.1185$. The dotted line (mLO+mAPT) is the case where $\rho_d(\sigma)$ in Eq.~(\ref{DVm}) is the pQCD spectral function, as explained in the text. $N_f=3$ is used throughout for $\rho_d(\sigma)$.}
 \label{FigDVm}
\end{figure}

We present the results of this approach in Fig.~\ref{FigDVm}.
In all cases, the $\A$ couplings used are such that $r^{(D=0)}_{\tau,{\rm th}}=0.201$. The additional dotted curve (mLO+mAPT, i.e., massive APT, or DPT, \cite{NestBook}) is for the case when the spectral function $\rho_d(\sigma)$ is that of the pQCD expression Eq.~(\ref{dptc}), $\rho_d(\sigma)=\rho_d^{\rm (pt)}(\sigma) = {\rm Im} \; d(-\sigma-i \epsilon;D=0)^{[4]}_{\rm pt}$, in the same 4-loop Lambert-MiniMOM scheme and for  $\alpha_s(M_Z^2)=0.1185$. We can see that the three solid curves (mLO+m$\A$QCD), for massive $\A$QCD Eq.~(\ref{DVm})  with $\alpha_s(M_Z^2)=0.1185, 0.1185 \pm 0.004$, are either within the experimental band or follow it reasonably closely. When massive kinematic threshold condition is entirely lifted ($m^2 \mapsto 0$), the theoretical curve becomes the dot-dashed curve (1+$\A$QCD), indicating that the massive kinematic threshold is an important ingredient at very low $Q^2 < 1 \ {\rm GeV}^2$.  

We wish to point out that the kinematic threshold $\sigma = m^2$ with $m= 2 m_{\pi}$, in the massive approach Eq.~(\ref{DVm}), is an additional external input, its value being specific for the specific process ($e^+e^- \to$ hadrons), i.e., the threshold parameter $m^2$ is process dependent. It does not affect, in principle, the coupling $\A(Q^2)$ which is considered universal. On the other hand, the approach $\A$QCD+OPE for the same quantity ${\cal D}_V(Q^2)$, Eq.~(\ref{DV}), does not require any process dependent parameter. However, this approach has a more limited application regime ($|Q^2| \gtrsim 1 \ {\rm GeV}^2$), as can be seen also by comparing Figs.~\ref{FigDVOPE} and \ref{FigDVm}. This limitation comes from the divergent nature of the OPE series (\ref{DV}) at $|Q^2| < 1 \ {\rm GeV}^2$. Nonetheless, as seen in Fig.~\ref{FigDVOPE}, the $\A$QCD+OPE approach starts failing at about $Q^2 \approx 1 \ {\rm GeV}^2$, while the $\MSbar$ pQCD+OPE starts failing already at $Q^2 \approx 2.5 \ {\rm GeV}^2$ ($Q \approx 1.6$ GeV). We point out that the incorporation of the lattice-motivated behavior for $\A(Q^2)$ at $|Q^2| \lesssim 0.1 \ {\rm GeV}^2$ affects significantly the behavior of $\A(Q^2)$ in the entire complex $Q^2$-plane, including in the regime of the utmost interest at present, $|Q^2| \sim 1 \ {\rm GeV}^2$.

One may raise the following question: what would happen with the applicability of the $\A$QCD+OPE approach to QCD inclusive quantities if the OPE were improved in the sense of having a convergent series even for $|Q^2| < 1 \ {\rm GeV}^2$?\footnote{For example, if the terms $\langle O_{2n} \rangle/(Q^2)^n$ were replaced by terms $\langle O_{2n} \rangle/(Q^2+M_n^2)^n$; and not only the condensate values, but also the effective masses $M_n$ were to be calculated by a well defined procedure. Such a procedure is not known at present. The usual ($\MSbar$) pQCD+OPE approach would not work even in such a case, because the pQCD coupling $a(Q^2)$ would hit Landau singularities for such $Q^2 < 1 {\rm GeV}^2$.} This is an open question. The question of description of resonances or hadronic bounds states with $\A$QCD is another open problem in the regime $|Q^2| \sim 0.1 \ {\rm GeV}^2$. We believe that the application of the presented coupling $\A$ in such problems, using the methods of Bethe-Salpeter (BS) and DS equations, is worth investigating. 
 Resonances may be treated via poles in the integral transforms of solutions to inhomogeneous Bethe-Salpeter equations \cite{Lipatov:1996ts}. Such equations may be solved explicitly when the coupling does not run \cite{Allendes:2012mr}. However when the gauge coupling is running, for example in the analytical way $\A(Q^2),$  to solve these equations is not simple.
At the present stage, we regard that the $\A$-coupling approach has limitations on applicability to QCD phenomena at $|Q^2| < 1 \ {\rm GeV}^2$ ($|Q^2| \sim 0.1 \ {\rm GeV}^2$).
Nonetheless, in models with nonperturbative process-dependent parameters the $\A$ coupling can also be regarded as an improvement in the entire range of the applicability of the model, cf.~Fig.~\ref{FigDVm}.

At the present stage, we regard as the main benefit of the constructed universal coupling $\A$ that it is suitable for extension of the known calculations of physical quantities by the pQCD(+OPE) methods down to squared momenta of $|Q^2| \sim 1 \ {\rm GeV}^2$, cf.~Fig.~\ref{FigDVOPE}. This advantage is based on the way the $\A(Q^2)$ coupling was constructed. Namely, the known pQCD coupling behavior in the UV regime, and the lattice-motivated coupling behavior in the deep IR regime, were connected into a single universal coupling $\A(Q^2)$ which is applicable, in principle, in the entire complex $Q^2$-plane. Further, the resulting coupling $\A(Q^2)$ has no Landau singularities, i.e., its holomorphic properties are qualitatively the same as those of the spacelike QCD observables. As a result, evaluations of truncated perturbation series for spacelike QCD observables ${\cal D}(Q^2)$ can be resummed into a sequence of partial sums which converges as the number of terms in the perturbation theory increases \cite{BGA,Techn} [in this context, we refer back to the discussion in the two paragraphs after Eq.~(\ref{dBGan22}), and footnote \ref{ftresumm} there]. The unknown behavior of the spectral function $\rho_{\A}(\sigma) \equiv {\rm Im} \A(-\sigma - i \epsilon)$ of the coupling in the low-$\sigma$ regime was approximated as a linear combination of delta functions, motivated by the large-$N_c$ QCD limit. The remaining (seventh) parameter was adjusted so that one crucial measured quantity of the physics at $|Q^2| \sim 1 \ {\rm GeV}^2$, namely the semihadronic $\tau$-lepton decay ratio $r_{\tau}(\Delta S =0)$, was reproduced. It remains open to interpretation whether this coupling $\A$ represents a model, or a direct construction based on incorporation of a host of physically-motivated conditions.

\section{Comparison with the work \cite{AQCDprev} and other works}
\label{sec:diff}

In comparison with our previous, shorter, work \cite{AQCDprev}, the present paper reproduces qualitatively (not quantitatively) the same results, but contains several differences, extensions and improvements.
\begin{itemize}
\item
In the present work, in the $\A$-coupling ($\A$QCD) approach, we worked in the renormalization scheme which agrees to four loops with the MiniMOM (MM) lattice scheme \cite{MiniMOM,BoucaudMM,CheRet} (plus the rescaling from $\Lambda_{\rm MM}$ to $\Lambda_{\MSbar}$), while in \cite{AQCDprev} we worked in a scheme that agrees only to three loops with the MiniMOM scheme. We recall that, at present, the MiniMOM (MM) lattice scheme is known to four loops. Further, the perturbative change of the renormalization scheme is presented with more details, cf.~Sec.~\ref{sec:pQCD}. 
\item
  As a consequence, the resulting $\A(Q^2)$ coupling, which fulfills various conditions at high ($|Q^2| > \Lambda_L^2$), low ($|Q^2| \to 0$) and intermediate ($|Q^2| \sim m_{\tau}^2$) squared momenta, was constructed in the present work in a dispersive way by using for the spectral function $\rho_{\A}(\sigma) \equiv {\rm Im} \A(Q^2 = -\sigma - i \varepsilon)$ at low $\sigma < M_0^2$ simply the sum of three delta functions at three different locations, cf.~Eqs.~(\ref{rhoA})-(\ref{AQ2}), which is motivated by attractive properties of Pad\'e approximants. In our previous work, the mentioned conditions were fulfilled when two of those three delta functions appeared at practically the same location ($\sigma=M_1^2$), i.e., $\rho_{\A}(\sigma)$ had equivalently the form Eq.~(\ref{rhoAalt}). In the present work we had to use this version only in the case of $\alpha_s(M_Z^2;\MSbar)=0.1189$ and $r^{(D=0)}_{\tau, {\rm th}}=0.201$ (i.e., when large $\alpha_s$ and low $r_{\tau}^{(D=0)}$).
\item
  In the previous work we fixed the $\A(Q^2)$ coupling to only one value of  $r_{\tau,{\rm th}}^{(D=0)}=0.201$ [by using the resummation method (\ref{dBGan22}) in the integral (\ref{rtaucont})], while $\alpha_s(M_Z^2;\MSbar)$ had three values as here ($0.1181$, $0.1185$, $0.1189$). In the present work, we present results for more (seven) possibilities, namely for $r_{\tau,{\rm th}}^{(D=0)}=0.201$ and $0.203$, and in the case of $\alpha_s(M_Z^2;\MSbar)=0.1185$ also $r_{\tau,{\rm th}}^{(D=0)}=0.199$, cf.~Table \ref{tabres}.
\item
The Borel sum rule analysis, and the comparison of $r_{\tau,{\rm th}}^{(D=0)}$ with $r_{\tau,{\rm exp}}^{(D=0)}$ is now performed for all seven mentioned choices of input parameters, for OPAL and ALEPH data (cf.~Tables \ref{tabSROPAL} and \ref{tabSRALEPH}), while in \cite{AQCDprev} the analysis was performed only for three choices of parameter $\alpha_s(M_Z^2;\MSbar)$ and only for OPAL data.
\end{itemize}

In Ref.~\cite{AQCDprev} we obtained the following values of the condensates:
\bes
\label{resprev}
\bea
\langle O_4 \rangle_{V+A} &=& -0.00157^{-0.00066}_{+0.00070} (\delta \alpha_s)  
\pm 0.00030({\rm exp}) \; [{\rm GeV}^4] 
\nonumber\\
\Rightarrow \;
\langle a GG \rangle  &=& -0.0074^{-0.0040}_{+0.0042}(\delta \alpha_s)  \pm 0.0018({\rm exp}) \; [{\rm GeV}^4]  \ ,
\label{aGGprev}
\\
\langle O_{6} \rangle_{V+A} & = &  +0.00136 \pm 0.00022(\delta \alpha_s) \pm 0.00042({\rm exp}) \; [{\rm GeV}^6] \ ,
\label{O6prev}
\eea
\ees
where the uncertainties from the experimental band were obtained by an ``educated guess'' approach. On the other hand, in the present work the obtained values of the $D=4$ and $D=6$ condensates, Eqs.~(\ref{resOPAL}) and (\ref{resALEPH}) and Tables \ref{tabSROPAL} and \ref{tabSRALEPH}, are now in general somewhat closer to zero, but the quality indicators $\chi^2$ are similarly good. In all cases we obtain negative values of the gluon condensate $\langle a GG \rangle$. 

On the other hand, the values of the gluon condensate obtained in the literature vary significantly.
\begin{table}
\caption{The gluon condensate values from the literature.}
\label{tabgl}  
\begin{ruledtabular}
\begin{tabular}{r l}
  $\langle a GG \rangle$ $[{\rm GeV}^4]$ &  work and method
  \\
  \hline
  $0.012$ & \cite{Shifman:1978bx},  charmonimum sum rules
  \\
  $0.037 \pm 0.015$ & \cite{Doming14}, Finite Energy Sum Rules (FESRs) in $e^+e^-$
  \\
  $0.022 \pm 0.004$ & \cite{O4sr1}, QCD-moment sum rules
  \\
  $0.024 \pm 0.006$ & \cite{O4sr1}, QCD-exponential moment sum rules
  \\
  $0.007 \pm 0.005$ & \cite{anOPE}, $\tau$ V+A sum rules with holomorphic coupling, $\A(0)>0$
  \\
  $0.012 \pm 0.005$ & \cite{anpQCD2},  $\tau$ V+A sum rules with hol.~pQCD (not $\MSbar$) $a(Q^2)$ with $a(0)>0$
  \\
  $0.077 \pm 0.087$ & \cite{BaBaPi1}, stochastic pQCD approach for $SU(3)$ plaquette
  \\
  $0.006 \pm 0.012$ & \cite{Geshkenbein}, three-loop Borel sum rules in V+A-channel $\tau$ decay
    \\
  $0.005 \pm 0.004$ &  \cite{Ioffe}, charmonium sum rules
  \\
 $0.005 \pm 0.004$ &  \cite{DHSS}, FESRs from $\tau$ decay data
 \\
  $-0.005 \pm 0.003$ &  \cite{ALEPHfin}, V-channel $\tau$ decay multiparameter fit with FESRs (CIPT)
  \\
  $-0.034 \pm 0.004$ & \cite{ALEPHfin}, A-channel $\tau$ decay multiparameter fit with FESRs (CIPT)
  \\
  $-0.020 \pm 0.003$ & \cite{ALEPHfin}, V+A-channel $\tau$ decay multiparameter fit with FESRs (CIPT)
  \\
 $-0.018^{+0.006}_{-0.005}$ & \cite{Pichrev1}, V+A-channel $\tau$ decay multiparameter fit with FESRs (CIPT)
\\
$-0.007^{+0.007}_{-0.016}$ & \cite{Pichrev1}, V+A-channel $\tau$ decay multiparameter fit with FESRs (FOPT)
\\
  $-0.007 \pm 0.005$ & \cite{AQCDprev}, $\tau$ V+A sum rules with holomorphic coupling, $\A(0)=0$
  \\
  $-0.004 \pm 0.004$ & this work, from OPAL data
  \\
  $-0.005 \pm 0.003$ & this work, from ALEPH data
\end{tabular}
\end{ruledtabular}
\end{table}
A positive value $\langle a GG \rangle \approx 0.012  \ {\rm GeV}^4$ was obtained in the original work on the sum rules \cite{Shifman:1978bx}. Positive values are obtained also by the finite energy sum rules (FESRs), e.g. $(0.037 \pm 0.015) \ {\rm GeV}^4$ in Ref.~\cite{Doming14} (from $e^+e^-$ annihilation in the $c$-quark region), and by the QCD-moment and QCD-exponential moment sum rules for heavy quarkonia \cite{O4sr1} which give $\langle a G G \rangle \approx (0.022 \pm 0.004) \ {\rm GeV}^4$ and $(0.024 \pm 0.006) \ {\rm GeV}^4$, respectively. The values $0.005$-$0.010 \ {\rm GeV}^4$ were obtained \cite{anOPE,anpQCD2} using a model with holomorphic coupling $\A(Q^2)$ and $0< \A(0) < \infty$ \cite{2danQCD} and in pQCD in a scheme with holomorphic coupling $a(Q^2)$ \cite{anpQCD2} with $0< a(0) < \infty$. Within a numerical stochastic pQCD approach, calculation of the plaquette in $SU(3)$ to high orders \cite{BaBaPi1} gives the values $0.077 \pm 0.087 \ {\rm GeV}^4$ \cite{BaBaPi2}. The uncertainty in the latter result originates from the $D=4$ renormalon in pQCD, which was revealed by calculation to such a high order to be able to see the onset of the ($D=4$) renormalon.
Further, the value $(0.006 \pm 0.012) \ {\rm GeV}^4$, compatible with zero, was obtained in Ref.~\cite{Geshkenbein} from Borel sum rules (in $\MSbar$ at three loops) for $\tau$ decay data in the $V+A$ channel of $\tau$ decay for ALEPH data; the value $(0.005 \pm 0.004) \ {\rm GeV}^4$ was obtained in Ref.~\cite{Ioffe} where the charmonium sum rules were used; and the same value was obtained in Ref.~\cite{DHSS} by FESRs for the $\tau$ decay data of ALEPH. The latter small values are compatible with our results for the $\MSbar$ pQCD(+OPE) case, cf.~Eqs.~(\ref{aGGOPALMS}) and (\ref{aGGALEPHMS}). On the other hand,  consistently negative values $(-0.005 \pm 0.003) \ {\rm GeV}^4$  (from $V$ channel),  $(-0.034 \pm 0.004) \ {\rm GeV}^4$ (from $A$ channel), and $(-0.020 \pm 0.003) \ {\rm GeV}^4$ (from $V+A$ channel) were obtained in \cite{ALEPHfin} (Table 4 there), from an updated multiparameter fit of ALEPH data on the $\tau$ decay were FESRs were used and $\alpha_s(m_{\tau}^2; \MSbar)$ was determined, too. Various extensions and improvements of the latter analysis were performed in Ref.~\cite{Pichrev1}, where the negative values $(-0.018^{+0.006}_{-0.005}) \ {\rm GeV}^4$ and $(-0.007^{+0.007}_{-0.016}) \ {\rm GeV}^4$ were obtained for the $V+A$ channel in the case of the CIPT and of the fixed order perturbation theory (FOPT) approach, respectively (cf.~Table VI there). In Table \ref{tabgl} we summarize all these values, for comparison. At the end of the Table we also included the results of our previous work \cite{AQCDprev}, and of the present work, Eqs.~(\ref{aGGOPAL}) and (\ref{aGGALEPH}), where the uncertainties were added in quadrature.

All these values of $\langle a GG \rangle$ in the literature were obtained by using the $\MSbar$ pQCD coupling (which is singular in the IR regime), and in Refs.~\cite{anOPE,anpQCD2} couplings were used which are finite positive in the deep IR regime. In the present work (and in our previous work \cite{AQCDprev}), on the other hand, a holomorphic coupling with $\A(0)=0$ was used. Further, we recall that the $\MSbar$ pQCD approach gives in general values of the condensates which vary significantly when the number of terms in the leading-twist truncated series increases \cite{KatParSid}. At the moment, it is not clear whether the value of the gluon condensate is positive or negative. It is not even clear how to interprete such condensates. For instance, in Ref.~\cite{BrodShr} it is argued that the QCD condensates are associated  with the internal dynamics of hadrons and with color confinement, and are not associated with the vacuum and, as a consequnece, they give zero contribution to the cosmological constant.

\section{Summary}
\label{sec:summ}

In this work we presented a construction of a QCD running coupling $\A(Q^2)$ which is applicable \textcolor{black}{in the regimes of high and intermediate $Q^2$}. It simultaneously fulfills several physically motivated restrictions \textcolor{black}{in the two mentioned regimes. In addition, it fulfills two restrictions in the regime of low $Q^2 < 1 \ {\rm GeV}^2$, motivated by the lattice evaluations of a specifically defined $\A_{\rm latt.}(Q^2)$ in that regime in the Landau gauge and MiniMOM scheme. We applied $\A(Q^2)$ to} the well-measured $\tau$-lepton semihadronic decay physics, and to a quantity related to the $e^+e^- \to$ hadrons production.

The starting idea is to have a QCD coupling which is universal in the same sense as \textcolor{black}{(the $\MSbar$ or MiniMOM)} pQCD coupling is universal, i.e., the coupling can be applied to the evaluation of QCD physical quantities at high ($|Q^2| > 1 \ {\rm GeV}^2$) and intermediate momenta ($|Q^2| \sim 1 \ {\rm GeV}^2$), via (resummed) perturbation expansion series for the leading-twist contribution, and additional contributions of the higher-twist terms which, for inclusive quantities, involve universal consensate values (i.e., OPE expansion). However, in contrast to the pQCD+OPE approach involving the pQCD coupling $a(Q^2)$, in our $\A$QCD+OPE approach the running coupling $\A(Q^2)$ has no Landau singularities, making therefore this approach more consistent theoretically, and more stable numerically at low $|Q^2| \sim 1 \ {\rm GeV}^2$. Stated otherwise, the obtained coupling $\A(Q^2)$, unlike the pQCD coupling $a(Q^2)$, as a byproduct of its construction shares with the QCD spacelike physical quantities (such as current correlators, \textcolor{black}{DIS differential cross sections}, etc.) the holomorphic (analytic) behavior in the complex $Q^2$-plane as required by the general properties of Quantum Field Theories. However, the way how the coupling $\A(Q^2)$ \textcolor{black}{turned out to be} holomorphic depended to a significant degree on the input information at low $|Q^2| \lesssim 1 \ {\rm GeV}^2$. In this work, we incorporated this information by requiring: (I) $\A(Q^2) \sim Q^2$ at $Q^2 \to 0$ ($|Q^2| < 1 \ {\rm GeV}^2$) and $\A(Q^2)$ at positive $Q^2$ has the maximum at $Q^2 \approx 0.13 \ {\rm GeV}^2$, both properties suggested by lattice results; (II) in addition, requiring that the measured $\tau$-semihadronic decay (V+A) ratio $r_{\tau}^{(D=0)} \approx 0.20$ be reproduced ($|Q^2| \sim 1 \ {\rm GeV}^2$); (III) as well as requiring that at high momenta  ($|Q^2| > 1 \ {\rm GeV}^2$) the coupling practically merge with the underlying pQCD coupling  $a(Q^2)$ (in the same lattice MiniMOM scheme): $\A(Q^2) - a(Q^2) \sim (\Lambda_{\rm QCD}^2/Q^2)^5$. The coupling was constructed by the dispersive approach, where the spectral function $\rho_{\A}(\sigma) \equiv {\rm Im} \A(Q^2=-\sigma - i \varepsilon)$ of the coupling was parametrized in the unknown infrared regime ($\sqrt{\sigma} \lesssim 1 \ {\rm GeV}^2$) by three delta functions whose parameters were then fixed by the aforementioned conditions. The construction was performed in the four-loop Lambert-MiniMOM renormalization scheme, i.e., in the scheme which coincides with the four-loop lattice MiniMOM (MM) renormalization scheme in which the lattice results are available, but with the rescaling to the usual scale $\Lambda_{\MSbar}$ (from the $\Lambda_{\rm MM}$ scale).

In order to cope with the present experimental uncertainties of $\alpha_s$ and $r_{\tau}$, we provided the construction of the $\A$QCD coupling $\A(Q^2)$ for seven choices of input parameters $\alpha_s(M_Z^2;\MSbar)$ and $r_{\tau, {\rm th}}^{(D=0)}$: $\alpha_s(M_Z^2;\MSbar)=0.1185$, $0.1181$ and $0.1189$ [$\Rightarrow \alpha_s(m_{\tau}^2; \MSbar)=0.3188$, $0.3156$ and $0.3221$], and for the values $r_{\tau, {\rm th}}^{(D=0)}=0.201$ and $0.203$; and in the case $\alpha_s(M_Z^2;\MSbar)=0.1185$ also the value $r_{\tau, {\rm th}}^{(D=0)}=0.199$. Two of the chosen $\alpha_s(M_Z^2; \MSbar)$ values ($0.1185$, $0.1181$) are the world average values of the PDG2014 and PDG2016, respectively \cite{PDG2014,PDG2016}.

Subsequent analysis of the Borel sum rules with OPAL and ALEPH data then allowed us to determine the values of the corresponding dimension $D=4$ and $D=6$ condensates: $\langle a GG \rangle$ (gluon condensate) and $\langle O_6  \rangle_{V+A}$. In all seven cases of input parameters, the fitting to the OPAL and ALEPH data is significaltly better than with the usual $\MSbar$ pQCD(+OPE) approach, and is quite stable even when the upper bound $\sigma_{\rm max}$ in the sum rules is decreased down to $\sigma_{\rm max} \approx 1 \ {\rm GeV}^2$. 
The extracted value of the gluon condensate $\langle a GG \rangle$ is significantly dependent on the value of $\alpha_s(M_Z^2; \MSbar)$. Application of the obtained $\A$QCD+OPE approach to the V-channel Adler function ${\cal D}_V(Q^2)$ for positive $Q^2$, closely related with the production ratio of $e^+e^- \to$ hadrons, gives results which agree with experimental data down to the value $Q^2 \approx 1 \ {\rm GeV}^2$, in contrast to the usual ($\MSbar$) pQCD+OPE approach which gives results which agree with experimental data only down to $Q^2 \approx 2.5 \ {\rm GeV}^2$. Good results for ${\cal D}_V(Q^2)$ were obtained also by applying the mass-modified dispersive approach \cite{Nest3a,Nest3b,NestBook} with $\A$ coupling.

The program packages, written in Mathematica \cite{Math}, which calculate $\A(Q^2)$ and the (holomorphic) analogs $\A_{n}(Q^2)$ of the pQCD powers $a(Q^2)^{n}$ are freely available \cite{prgs}, for the seven mentioned cases of the input parameters $\alpha_s(M_Z^2;\MSbar)$ and  $r_{\tau, {\rm th}}^{(D=0)}$.

\begin{acknowledgments}
C.A.~acknowledges the support by FONDECYT (Chile) Postdoctoral Grant No.~3170116. G.C.~acknowledges the support by FONDECYT (Chile) Grant No.~1130599. The work of I.K.~was supported in part by FONDECYT (Chile) Grants Nos.~1040368, 1050512 and 1121030, by DIUBB (Chile) Grant Nos.~125009,  GI 153209/C  and GI 152606/VC.

 We are grateful to S.~Peris and A.~Sternbeck for valuable discussions and clarifications on the subject. We thank A.~L.~Kataev for a clarification on NSVZ $\beta$-function. C.A.~thanks the IFAE group at Universitat Aut{\`o}noma de Barcelona for warm hospitality during part of this work.
\end{acknowledgments}

\appendix

\section{Transformation of perturbation coefficients under the change of scheme}
\label{app:tcp}

When the renormalization scheme coefficients $c_j \equiv \beta_j/\beta_0$ ($j \geq 2$) change from the $\MSbar$ to another scheme (${\overline c}_j \mapsto c_j$), the perturbation coefficients $d_j$ of any scheme invariant quantity $d(Q^2)$, Eq.~(\ref{dptc}), change accordingly (cf.~\cite{Stevenson})
\bes
\label{tcp}
\bea
d_1 &=& {\overline d}_1, \quad d_2 = {\overline d}_2 - (c_2 - {\overline c}_2),
\label{tcpd1d2}
\\
d_3 &=& {\overline d}_3 - 2 {\overline d}_1 (c_2 - {\overline c}_2) - \frac{1}{2}(c_3 - {\overline c}_3),
\label{tcpd3}
\eea
\ees
Here, the coefficients in $\MSbar$ scheme have bar, and those in another scheme (e.g., in Lambert MiniMOM) are without bar.

\section{Analogs $\A_n(Q^2)$ of powers $a(Q^2)^n$}
\label{app:An}

The analytic version $(a^n)_{\rm an} = \A_n$ of the analogs of higher powers $a^n$ of the (underlying) pQCD coupling, for integer $n$, was constructed in the general case of holomophic QCD in \cite{CV12}. We recapitulate it briefly here. The construction goes via a detour by considering first, instead of the powers $a^n$, the logarithmic derivatives
\be
{\ta}_{n+1}(Q^2)
\equiv \frac{(-1)^{n}}{\beta_0^{n} n!}
\frac{ \partial^n a(Q^2)}{\partial (\ln Q^2)^n} \ , 
\qquad (n=1,2,\ldots) \ .
\label{tan}
\ee
According to RGE, we have ${\ta}_{n+1}(Q^2) = a(Q^2)^{n+1} + {\cal O}(a^{n+2})$. Specifically, we have
\bea
\ta_{2} &=& a^2 + c_1 a^3 + c_2 a^4 + \cdots \ ,
\label{ta2}
\\
\ta_{3} &=& a^3 + \frac{5}{2} c_1 a^4  + \cdots \ ,
\qquad
\ta_{4} = a^4 +   \cdots \ ,  
\qquad {\rm etc.} \ .
\label{ta3ta4}
\eea 
Inverting these relations gives
\bea
a^2 & = & \ta_{2}
- c_1 \ta_{3} 
+ \left( \frac{5}{2} c_1^2 - c_2 \right) {\tilde a}_{4} + \cdots \ ,
\label{a2}
\\
a^3 & = & \ta_{3} - \frac{5}{2} c_1 \ta_{4} + \cdots \ ,
\qquad
 a^4  =  \ta_{4}  +  \cdots \ ,
\qquad {\rm etc.}
\label{a3a4}
\eea
The linearity of ``analytization'' implies that in holomorphic QCD the correponding analogs of logarithmic derivatives are constructed in the very same way, therefore we define
\be
\tA_{n+1}(Q^2)
\equiv \frac{(-1)^n}{\beta_0^n n!}
\frac{ \partial^n \A(Q^2)}{\partial (\ln Q^2)^n} \ .
\qquad (n=1,2,\ldots) \ .
\label{tAn}
\ee
Further, the linearity of the relations (\ref{a3a4}) implies that the analogs $\A_2$, $\A_3$, $\A_4$ of the powers $a^n$ are obtained in the same way
\bea
\A_2 & \equiv & \left( a^2 \right)_{\rm an} 
= \tA_2 - c_1 \tA_3
+ \left( \frac{5}{2} c_1^2 - c_2 \right) \tA_{4} + \cdots \ ,
\label{A2}
\\
\A_3 & \equiv & \left( a^3 \right)_{\rm an} 
=  \tA_3 - \frac{5}{2} c_1 \tA_4 + \cdots \ ,
\quad
 \A_4 \equiv \left( a^4 \right)_{\rm an} 
=  \tA_{4}  +  \cdots \ .
\qquad {\rm etc.}
\label{A3A4}
\eea
In the case of the truncated series $d^{[4]}$, Eqs.~(\ref{danb})-(\ref{danc}), we truncated the above relations at $\tA_{4}$ (including $\tA_{4}$).

The analogs $\A_{\nu}$ for the powers $a^{\nu}$, when $\nu$ is real (and not necessarily integer) were constructed for general holomorphic couplings $\A$ in \cite{GCAK}. They are based on the following generalization of the logarithmic derivatives (\ref{tAn}) to general noninteger $\nu$:
\be
\tA_{\nu+1}(Q^2) = \frac{1}{\pi} \frac{(-1)}{\beta_0^{\nu} \Gamma(\nu+1)}
\int_{0}^{\infty} \ \frac{d \sigma}{\sigma} \rho_{\A}(\sigma)  
{\rm Li}_{-\nu}\left( - \frac{\sigma}{Q^2} \right) \quad (-1 < \nu) \ ,
\label{disptAn3}
\ee
where ${\rm Li}_{-\nu}$ is the polylogarithm function. This formula can be extended in the index $\nu$ to: $-2 < \nu$, cf.~Eq.~(22) of \cite{GCAK}. The expression $\A_{\nu}$, i.e., the analog of the power $a^{\nu}$, is then obtained in the following way

\be
\A_{\nu} \equiv {\tA}_{\nu} + \sum_{m \geq 1}
\tk_m(\nu) {\tA}_{\nu + m} \quad (\nu > -2) \ ,
\label{AnutAnu}
\ee
where the coefficients $\tk_m$ are given, for general $\nu$, in App.~A of \cite{GCAK}; they involve the coefficients of the RGE $\beta$ function, and derivatives of Gamma functions.

If the expressions (\ref{A2})-(\ref{A3A4}) are truncated at ${\tA}_4$, it is straightforward to show that the truncated series (\ref{danc}) can be expressed in terms of ${\tA}_n$'s
\bes
\label{dandtan}
\bea
 d(Q^2;D=0)_{\rm an}^{[4]} & = & \A(Q^2) + d_1 \A_{2}(Q^2) +  d_2 \A_{3}(Q^2) +  d_3 \A_{4}(Q^2)
 \label{dandtana}
\\
 & = & 
 \A(Q^2) + {\td}_1 \tA_{2}(Q^2) +  \td_2 \tA_{3}(Q^2) +  \td_3 \tA_{4}(Q^2),
 \label{dandtanb}
 \eea
 \ees
 where the modified expansion coefficients ${\td}_n$ are simple combinations of the original expansion coefficients
 \bes
\label{tdn}
\ba
\td_1 & = & d_1 \ , \qquad  \td_2 =  d_2 - c_1 d_1 \ ,
\label{td1td2}
\\
\td_3 & = & d_3 - \frac{5}{2} c_1 d_2 + 
\left(\frac{5}{2} c_1^2 - c_2 \right) d_1 \ .
\label{td3}
\ea
\ees
It is evident that the form (\ref{dandtanb}) is more time-efficient in numerical evaluations than the form (\ref{dandtana}).

\section{Fit procedures for Borel transforms}
\label{app:fit}

In this Appendix we explain how the covariance matrix of the Borel transforms ${\rm Re} B(M^2)$ are obtained, and thus the variance (half width) of the experimental bands in Figs.~\ref{FigPi64}-\ref{FigPsi0o26} in Sec.~\ref{sec:BSR}. Further, we explain how various $\chi^2$ values were evaluated in Sec.~\ref{sec:BSR}.

The covariance matrix $U_{ij} \equiv U(\sigma_i,\sigma_j)$ for the experimental $\omega_{V+A}(\sigma)$ spectral functions (of OPAL and ALEPH Collaboration) can be probabilistically defined by means of the following expectation values:
\bes
\label{covM}
\bea
U_{ij} \equiv U(\sigma_i,\sigma_j) & \equiv & \langle \Delta \omega_{V+A}(\sigma_i) \Delta \omega_{V+A}(\sigma_j) \rangle
\label{covMa}
\\
& = & U^{(VV)}_{ij} + U^{(AA)}_{ij} + U^{VA}_{ij} + U^{AV}_{ij} ,
\label{covMb}
\eea
\ees
where in the first line, Eq.~(\ref{covMa}), we denoted
\be
\Delta \omega(\sigma_i)  = \omega(\sigma_i) - \langle \omega(\sigma_i) \rangle,
\label{DOm}
\ee
with $\langle \omega(\sigma_i) \rangle = \omega_{\rm exp}(\sigma_i)$ representing the experimental average (central) value for the measurements of $\omega(\sigma_i)$ in the i'th $\sigma$-bin whose length is $(\Delta \sigma)_i$ and central point $\sigma_i$. In the second line, Eq.~(\ref{covMb}), we denoted the corresponding covariance matrices for combinations of different (V and A) channels. These matrices are available, or extractable, from the corresponding OPAL \cite{OPAL,PerisPC1,PerisPC2} and ALEPH data \cite{ALEPHfin}.

The covariance matrix $U_B$ for the Borel transforms ${\rm Re} B(M^2)$ can then obtained from the ``sum over bins'' evaluation of ${\rm Re} B(M^2)$ [cf.~Eq.~(\ref{sr3a})]
\be
   {\rm Re} B(M^2) = \sum_{j=1}^N (\Delta \sigma)_j f(\sigma_j;M^2) \omega(\sigma_j)_{V+A}
\label{ReBexp}
\ee
where
\be
f(\sigma_j;M^2) \equiv {\rm Re} \left( \frac{\exp(-\sigma_j/M^2)}{M^2} \right) .
\label{fsj}
\ee
The Borel transform covariance matrix $(U_B(\Psi))_{\alpha \beta} \equiv U_B(M^2_{\alpha},M^2_{\beta})$, where $M^2_{\alpha} = |M^2_{\alpha}| \exp(i \Psi)$ ($\Psi$ fixed), can then be expressed in terms of the aforementioned covariance matrix $U$, Eq.~(\ref{covM}), in the following way:
\bes
\label{covMB}
\bea
(U_B(\Psi))_{\alpha \beta} \equiv U_B(M^2_{\alpha},M^2_{\beta}) & = &
\langle \Delta {\rm Re} B(M^2_{\alpha}) \Delta {\rm Re} B(M^2_{\beta}) \rangle
\label{covMBa}
\\
& = & 
\sum_{j=1}^N \sum_{k=1}^N (\Delta \sigma)_j (\Delta \sigma)_k f(\sigma_j; M^2_{\alpha}) f(\sigma_k; M^2_{\beta}) U_{jk}.
\label{covMBb}
\eea
\ees
Here we denoted
\be
\Delta {\rm Re} B(M^2) \equiv  {\rm Re} B(M^2) - \langle  {\rm Re} B(M^2) \rangle,
\label{DReB}
\ee
where $\langle {\rm Re} B(M^2) \rangle$ is the central experimental value of the Borel transform [at a scale $M^2 = |M^2| \exp(i \Psi)$], i.e., the expression (\ref{ReBexp}) calculated with the central experimental spectral values $\langle \omega(\sigma_j)_{V+A} \rangle = \omega_{\rm exp}(\sigma_j)_{V+A}$. Square root of the diagonal elements of the above matrix
\be
\delta_B(M^2) = \left(  U_B(M^2,M^2) \right)^{1/2}
\label{dBM2}
\ee
gives the experimental standard deviation for the Borel transform ${\rm Re} B(M^2)$ at a given scale $M^2$. This standard deviation represents is the half width of the experimental grey bands in Figs.~\ref{FigPi64}-\ref{FigPsi0o26}, at given value of $M^2 = |M^2| \exp(i \Psi)$.

In order to fit the theoretical curves of ${\rm B}(M^2)$ to the central experimental curve, we have several possibilities. One is to simply minimize the sum of squared deviations
\be
\chi^2(\Psi) = \frac{1}{n} \sum_{\alpha=0}^n \left( {\rm Re} B_{\rm th} (M_{\alpha}^2) - {\rm Re} B_{\rm exp} (M_{\alpha}^2) \right)^2 ,
\label{chi2a}
\ee
where $M_{\alpha}^2 = |M_{\alpha}^2| \exp(i \Psi)$ with $\Psi$ fixed, and $|M_{\alpha}|^2$ are $n+1$ equidistant points in the considered $|M^2|$ interval. We choose in this work the interval $0.65 \ {\rm GeV}^2 \leq |M^2| \leq 1.50 \ {\rm GeV}^2$, and the number of subintervals $n=85$.

Another possibility is to weight the squared deviations by the squares of the standard deviations $\delta_B(M^2)$ of Eq.~(\ref{dBM2})
\be
\chi^2_{\delta}(\Psi) = \sum_{\alpha=0}^n \left( \frac{{\rm Re} B_{\rm th} (M_{\alpha}^2) - {\rm Re} B_{\rm exp} (M_{\alpha}^2)}{\delta_B(M^2_{\alpha})} \right)^2.
\label{chi2b}
\ee

Yet another possibility, a generalization of Eq.~(\ref{chi2b}), is to account for the entire covariance matrix (\ref{covMB})
\be
\chi^2_{\rm cov}(\Psi) = \sum_{\alpha=0}^n \sum_{\beta=0}^n \left( {\rm Re} B_{\rm th} (M_{\alpha}^2) - {\rm Re} B_{\rm exp} (M_{\alpha}^2) \right) (U_B(\Psi)^{-1})_{\alpha \beta} \left( {\rm Re} B_{\rm th} (M_{\beta}^2) - {\rm Re} B_{\rm exp} (M_{\beta}^2) \right).
\label{chi2c}
\ee
It turns out that the latter approach, i.e., the minimization of the quantity (\ref{chi2c}),  is not numerically stable when the number $n$ of subintervals of $|M^2|$ (of width $0.85/n \ {\rm GeV}^2$) increases. In such cases, the inversion $U_B(\Psi)^{-1}$ of the $(n+1) \times (n+1)$ matrix $U_B(\Psi)$ becomes in general numerically unstable and dependent on the number $n$. In the case of the $\A$-coupling framework, the instabilities occur for $n \geq 3$, and in the $\MSbar$ pQCD framework for $n \geq 2$. Nonetheless, we can use the values of the $D=4,6$ condensates obtained by the minimization of $\chi^2_{\rm cov}(\Psi)_{n=1}$ for $\Psi=\pi/6, \pi/4$. But then, in the OPAL case, at $\Psi=0$ the deviations from the experimental values of ${\rm Re}B(M^2)$ turn out to be larger than those when the central values of condensates Eq.~(\ref{cenres}) are used, the latter obtained by the minimization of $\chi^2(\Psi)_{n=85}$ of Eq.~(\ref{chi2a}). For numerical data on that, see the discussion below. This is the main reason why we do not use the minimization of $\chi^2_{\rm cov}(\Psi)_{n=1}$ for determination of the central values of the condensates in the OPAL case. On the other hand, in the ALEPH case the approaches (\ref{chi2a}) and (\ref{chi2c}) are comparably good, but we will use the approach (\ref{chi2a}) also in the ALEPH case.

On the other hand, the minimization of  $\chi^2_{\delta}$ (\ref{chi2b}) and $\chi^2$ (\ref{chi2a}) gives very similar results for the condensate values, the differences in these values being of the order of a per cent. In this work we will use the minimization of $\chi^2$, Eq.~(\ref{chi2a}), and not of $\chi^2_{\delta}$, for at least two reasons. One is the simplicity and the fact that the values of  $\delta_B(M^2)$ along the considered rays do not vary much within the considered interval $0.65 \ {\rm GeV}^2 \leq |M^2| \leq 1.50 \ {\rm GeV}^2$, as seen in Figs.~\ref{FigPi64}-\ref{FigPsi0o26}. The other reason is that the experimental uncertainties $\delta_B(M^2)$ are by about a factor of three smaller in the case of the ALEPH data (in comparison to the OPAL data), and therefore in such a case the comparison of $\chi^2_{\delta}$ between the OPAL with ALEPH cases will miss one important aspect. Namely, if the theoretical curve fits comparatively well both the central OPAL curve and the central ALEPH curve [for ${\rm Re} B(M^2)$ along a ray], then $\chi^2_{\delta}$ in the ALEPH case will be about ten times higher than in the OPAL case. However, we want to have a general indicator of quality of fit to the central experimental curves, and not only the fit quality as compared to the experimental bands. Therefore, we choose the minimization of $\chi^2$ of Eq.~(\ref{chi2a}) as our approach (with relatively large $n=85$). In Sec.~\ref{sec:BSR} we also present the corresponding values for the experimental bands
\be
\chi^2_{\rm exp}(\Psi) = \frac{1}{n} \sum_{\alpha=0}^n \delta_B(M^2_{\alpha})^2 .
\label{chi2aexp}
\ee
This quantity is in general ten times smaller in the ALEPH case, as compared to the OPAL case. The comparison of $\chi^2$ with the quantity (\ref{chi2aexp}) gives us a rough measure of quality of fit in comparison to the experimental band, i.e., the quality as quantified also by $\chi^2_{\delta}$ of Eq.~(\ref{chi2b}).

However, we will use the covariant matrix approach (\ref{chi2c}), with $n=1$ (i.e., with $|M_{\alpha}^2| = 0.65$ for $\alpha=0$ and $|M_{\alpha}^2| = 1.5 \ {\rm GeV}^2$ for $\alpha=1$), for the purpose of estimating the experimental uncertainties of the central values of condensates [we recall that the central values are obtained by minimizing $\chi^2(\Psi)_{n=85}$ of Eq.~(\ref{chi2a})]. Namely, in the OPAL case, the minimization of $\chi^2_{\rm cov,n=1}(\Psi)$ gives for $\Psi=\pi/6$ the value $\langle a GG \rangle = -0.0043 \ {\rm GeV}^4$, and for $\Psi=\pi/4$ the value $\langle O_6 \rangle_{V+A}=+0.0013 \ {\rm GeV}^6$ [differing somewhat from the central values Eq.~(\ref{cenres}) obtained by minimizing $\chi^2(\Psi)_{n=85}$ of Eq.~(\ref{chi2a})]. When using these central values in $\Psi=0$ case of ${\rm Re}B(|M^2| e^{i \Psi})$, we obtain for deviations $\chi^2(\Psi=0)_{n=85} = 1.4 \times 10^{-5}$, significantly larger than $\chi^2(\Psi=0)_{n=85} =2.1 \times 10^{-6}$ [Eq.~(\ref{xi2Psi0})] obtained when the central values Eqs.~(\ref{cenres}) are used. As mentioned, this is the main reason why we do not use for the central values of the condensates those obtained by the minimization of $\chi^2_{\rm cov}(\Psi)_{n=1}$, but rather those obtained by the minimization of $\chi^2(\Psi)_{n=85}$. On the other hand, the value of $\chi^2_{\rm cov}(\pi/6)_{n=1}$ increases from its minimal value $0.014$ by one unity (to $1.014$) when the value of the condensate $\langle a GG \rangle$ deviates from its value at the minimum by $\pm 0.0017 \ {\rm GeV}^4$. Analogously, the value of  $\chi^2_{\rm cov}(\pi/4)_{n=1}$ increases from its minimal value $0.006$ by one unity (to $1.006$) when the value of the condensate $\langle O_6 \rangle_{V+A}$ deviates from its value at the minimum by $\pm 0.0003 \ {\rm GeV}^6$. These deviations can be taken (and will be taken) as the estimates of the experimental uncertainties of the central values of the condensates, cf.~Eqs.~(\ref{cenres})-(\ref{cenresMS}). In the ALEPH case, completely analogous reasoning is applied. We refer to Refs.~\cite{PDG2014,PDG2016} on that point (Statistics).

\section{Massive variant of $\A$QCD}
\label{app:mAQCD}

Here we recapitulate, based on Refs.~\cite{Nest3a,Nest3b,NestBook}, the massive version for (any) analytic QCD, with a view of applying the procedure in $\A$QCD considered here for evaluations of the V-channel Adler function ${\cal D}_V(Q^2)$.

The production ratio for $e^+ e^- \to$ hadrons, $R(\sigma)$, which is a timelike quantity, and the Adler function $D_V(Q^2)$, which is a spacelike quantity, are related by the integral transformations
\bes
\label{DVRRDV}
\bea
D_V(Q^2) & = & Q^2 \int_{m^2}^{\infty} d \sigma \frac{R(\sigma)}{(\sigma + Q^2)^2},
\label{DVfromR}
\\
R(\sigma) & = & \frac{1}{2 \pi i} \int_{-\sigma - i \epsilon}^{-\sigma + i \epsilon} dQ^2 \frac{D_V(Q^2)}{Q^2},
\label{RfromDV}
\eea
\ees
where the last integral must avoid the cut $(-\infty,-m^2)$ in the $Q^2$ plane [$m^2 = (2 m_{\pi})^2$ with $m_{\pi}=0.13957$ GeV]. The kinematic threshold $\sigma \geq m^2$ in the production process $e^+ e^- \to$ hadrons is reflected in the production ratio $R(\sigma)$ \cite{Feynman,Akhiezer}
\be
R(\sigma) = R^{(0)}(\sigma)_m + \Theta(\sigma - m^2) r(\sigma),
\label{RTheta}
\ee
where the leading part is
\be
R^{(0)}(\sigma)_m= \Theta(\sigma - m^2) \left( 1 - \frac{m^2}{\sigma} \right)^{3/2},
\label{R0msig}
\ee
and $r(\sigma)$ is the QCD correction
\be
r(\sigma) = \frac{1}{\pi} \int_{\sigma}^{\infty} d \sigma' \frac{\rho_d(\sigma')}{\sigma'},
\label{rsig}
\ee
where  $\rho_d(\sigma) \equiv {\rm Im} \; d(-\sigma - i \epsilon; D=0)$ is the spectral function of the $D=0$ Adler function $d(Q^2; D=0)$ of Eqs.~(\ref{danc}) and (\ref{dBGan22}). In this approach, it is regarded that kinematic threshold $m^2$ is the major nonperturbative effect at low momenta or energies.
Application of the transformation (\ref{DVfromR}) to the function $R(\sigma)$ of Eq.~(\ref{RTheta}) then gives the V-channel Adler function in terms of $m^2$ and $\rho_d(\sigma)$
\be
   {\cal D}_V(Q^2) =  {\cal D}^{(0)}(Q^2)_m + d(Q^2)_m,
\label{DVmApp}
\ee
where the leading part comes from $R^{(0)}(\sigma)_m$ of Eq.~(\ref{R0msig})
\bes
\label{D0m}
\bea
{\cal D}^{(0)}(Q^2)_m &=& 1 + \frac{3}{z^2} \left[ 1 - \left( 1 + \frac{1}{z^2} \right)^{1/2} {\rm ArcSinh}(z) \right] {\bigg |}_{z=\sqrt{Q^2}/m}
\label{D0ma}
\\
& = & \frac{2}{5} z^2 - \frac{8}{35} z^4 + \frac{16}{105} z^6 + \ldots {\bigg |}_{z^2=Q^2/m^2},
\label{D0mb}
\eea
\ees
which is a holomorphic function of $Q^2$ for $Q^2 \in \mathbb{C} \backslash (-\infty, -m^2)$, with the cut $(-\infty, -m^2)$ in the complex $Q^2$-plane. Similarly, the QCD part $d(Q^2)_m$ in this approach is
\be
d(Q^2)_m = \frac{Q^2}{(Q^2+m^2)} \frac{1}{\pi} \int_{m^2}^{+\infty} d \sigma \left(1 - \frac{m^2}{\sigma} \right) \frac{\rho_d(\sigma)}{(\sigma + Q^2)} .
\label{dm}
\ee
In QCD with a holomorphic $\A(Q^2)$ (not necessarily the $\A$QCD considered in this work), direct algebra shows that this part can be expressed in terms of the mass-modified ('m') coupling $\A^{(m)}(Q^2)$
\be
d(Q^2)_m = - \frac{m^2}{(Q^2+m^2)} \A^{(m)}(0) + \A^{(m)}(Q^2) + {\td}_1  \tA_2^{(m)}(Q^2) + {\td}_2 \tA_3^{(m)}(Q^2) + {\td}_3 \tA_4^{(m)}(Q^2) + {\cal O}(\tA_5),
\label{dman}
\ee
where the mass-modified coupling is
\be
\A^{(m)}(Q^2) = \frac{1}{\pi} \int_{m^2}^{\infty} \frac{d \sigma \rho_{\A}(\sigma)}{(\sigma + Q^2)}
\label{Am}
\ee
and $\tA^{(m)}_{n+1}$ are logarithmic derivatives of this coupling
\be
\tA^{(m)}_{n+1}(Q^2)
\equiv \frac{(-1)^n}{\beta_0^n n!}
\frac{ \partial^n \A^{(m)}(Q^2)}{\partial (\ln Q^2)^n} \ .
\qquad (n=1,2,\ldots) \ ,
\label{tAnm}
\ee
in complete analogy with Eq.~(\ref{tAn}) for the original (non-mass-modified) holomorphic couplings $\tA_{n+1}$. The expansion coefficients $\td_n$ are given in Eqs.~(\ref{tdn}), cf.~also the analogous Eqs.~(\ref{dandtan}) for $d(Q^2;D=0)$ in the theory with the original holomorphic coupling $\A$. The function $\rho_{\A}(\sigma)$ appearing in Eq.~(\ref{Am}) is the spectral function of the original coupling $\A$: $\rho_{\A}(\sigma) \equiv {\rm Im} \A(-\sigma - i \epsilon)$. We recall that, in contrast with the integral (\ref{Am}) for $\A^{(m)}(Q^2)$, the original holomorphic coupling $\A(Q^2)$ has the dispersion integral running in principle across the entire positive semiaxis in $\sigma$
\be
\A(Q^2) = \frac{1}{\pi} \int_{0}^{\infty} \frac{d \sigma \rho_{\A}(\sigma)}{(\sigma + Q^2)}.
\label{Aorig}
\ee
In practice, the original spectral function $\rho_{\A}(\sigma)$ starts to be nonzero at some threshold value $M^2_{\rm thr}$, cf.~Eq.~(\ref{Adisp}). In the considered $\A$QCD model, $M^2_{\rm thr}=M_1^2$, cf.~Eq.~(\ref{rhoA}). Our $\A$QCD coupling gets modified by the kinematical threshold (\ref{Am}) [i.e., $\A^{(m)} \not= \A$] only when $m^2 > M_1^2$, which is the case in four of the seven considered cases of $\A$QCD here, cf.~Table \ref{tabres}.\footnote{
We recall that $M_1^2 = s_1 \Lambda_L^2$, so that the non-modification condition $(4 m_{\pi}^2 =)$ $m^2 < M_1^2$ is fulfilled only in the three cases: $\alpha_s(M_Z^2;\MSbar)=0.1185$ and $r^{(D=0)}_{\tau,{\rm th}}=0.199$;   $\alpha_s(M_Z^2;\MSbar)=0.1189$ and $r^{(D=0)}_{\tau,{\rm th}}=0.201$ or $0.203$. In the other four cases of Table \ref{tabres}, we have $m^2 > M_1^2$ and thus $\A^{(m)} \not= \A$.}

\end{document}